\documentclass[11pt,a4paper]{article}

\pdfoutput=1

\usepackage{jheppub}
\usepackage[T1]{fontenc}
\usepackage{epsfig}
\usepackage{amscd}
\usepackage{verbatim}
\usepackage{latexsym}
\usepackage{amsfonts,amsthm,amsmath,mathtools}
\usepackage{amssymb,stmaryrd,textcomp,enumerate,bbm}
\usepackage{color}
\usepackage{xcolor}
\usepackage{booktabs}
\usepackage{float}
\usepackage{graphicx}
\usepackage{lscape}
\usepackage{slashed}
\usepackage{subfigure}
\usepackage{multirow,makecell}

\numberwithin{equation}{section}

\def \be {\begin{equation}}
\def \ee {\end{equation}}
\def \ba {\begin{array}}
\def \ea {\end{array}}
\def \bea{\begin{eqnarray}}
\def \eea{\end{eqnarray}}
\def \nn {\nonumber}

\def \a {\alpha}
\def \b {\beta}
\def \g {\gamma}

\def \G {\Gamma}
\def \d {\delta}

\def \e {\epsilon}
\def \ve {\varepsilon}
\def \m {\mu}
\def \n {\nu}

\def \l {\lambda}

\def \s {\sigma}
\def \S {\Sigma}
\def \r {\rho}
\def \o {\omega}
\def \O {\Omega}
\def \th {\theta}
\def \vth {\vartheta}

\def \t {\tau}
\def \z {\zeta}

\def \mA {\mathcal A}
\def \mB {\mathcal B}

\def \mD {\mathcal D}

\def \mL {\mathcal L}

\def \mN {\mathcal N}

\def \mP {\mathcal P}

\def \p {\partial}
\def \f {\frac}

\def \mf {\mathfrak}

\def \lt {\left}
\def \rt {\right}

\def \ra {\rightarrow}

\def \sr {\sqrt}
\def \td {\tilde}

\def \inf {\infty}

\def \hi  {{\hat\imath}}
\def \hj  {{\hat\jmath}}
\def \hk  {{\hat k}}
\def \hl  {{\hat l}}

\def \tm {{\tilde\m}}
\def \ti {{\tilde\imath}}
\def \tj {{\tilde\jmath}}

\def \ho  {{\hat 1}}
\def \hw  {{\hat 2}}

\def \ep {\mathrm{e}}
\def \ii {\mathrm{i}}

\def \bg {\boldsymbol{\gamma}}
\def \bG {\boldsymbol{\Gamma}}
\def \bo {\boldsymbol{1}}

\def \Tr {{\textrm{Tr}}}

\def \diag {{\textrm{diag}}}

\def \AdS {{\textrm{AdS}}}

\def \X {{\textrm{X}}}
\def \rmS {{\textrm{S}}}

\def \gsym {{\textrm{AdS}_5\times\textrm{S}^5}}
\def \gabjm {{\textrm{AdS}_4\times\textrm{S}^7/\textrm{Z}_k}}
\def \goabjm {{\textrm{AdS}_4\times\textrm{S}^7/(\textrm{Z}_{rk}\times\textrm{Z}_{r})}}
\def \gtzt {{\textrm{AdS}_7\times\textrm{S}^4}}

\def \R {{\textrm{R}}}
\def \Z {{\textrm{Z}}}

\begin{document}

\title{String theory duals of Wilson loops from Higgsing}
\author[a]{Marco Lietti}
\author[a,b]{\!\!,~Andrea Mauri}
\author[a,b]{\!\!,~Silvia Penati}
\author[a,b]{\!\!,~Jia-ju Zhang}
\affiliation[a]{Dipartimento di Fisica, Universit\`a degli Studi di Milano-Bicocca,\\Piazza della Scienza 3, I-20126 Milano, Italy}
\affiliation[b]{INFN, Sezione di Milano-Bicocca, Piazza della Scienza 3, I-20126, Milano, Italy}
\emailAdd{m.lietti@campus.unimib.it, andrea.mauri@mi.infn.it, silvia.penati@mib.infn.it, jiaju.zhang@mib.infn.it}

\abstract{
For three-dimensional ABJ(M) theories and $\mathcal N=4$ Chern--Simons--matter quiver theories, we construct two sets of 1/2 BPS Wilson loop operators by applying the Higgsing procedure along independent directions of the moduli space, and choosing different massive modes. For theories whose dual M--theory description is known, we also determine the corresponding spectrum of 1/2 BPS M2--brane solutions. We identify the supercharges in M--theory and field theory, as well as the supercharges preserved by M2--/anti--M2--branes and 1/2 BPS Wilson loops. In particular, in $\mathcal N=4$ orbifold ABJM theory we find pairs of different 1/2 BPS Wilson loops that preserve exactly the same set of supercharges. In field theory they arise by Higgsing with the choice of either particles or antiparticles, whereas in the dual description they correspond to a pair of M2--/anti--M2--branes localized at different positions in the compact space. This result enlightens the origin of classical Wilson loop degeneracy in these theories, already discussed in arXiv:1506.07614. A discussion on possible scenarios that emerge by comparison with localization results is included.
}


\maketitle

\newpage

\section{Introduction}

Bogomol'nyi--Prasad--Sommerfield (BPS) Wilson loops (WLs) in supersymmetric gauge theories provide one of the main tools to test the AdS/CFT correspondence, being  non-protected operators that in many cases can be computed exactly at quantum level by using localization techniques. Matching the weak coupling expansion of the exact result with a field theory perturbative calculation and the strong coupling limit with the dual string or brane configuration in AdS provides in fact a strong check of the correspondence.

In this paper we focus on 1/2 BPS WLs in superconformal gauge theories and their string theory duals, in different realizations of the AdS/CFT correspondence. These WLs are gauge invariant non-local operators that preserve half of the original supersymmetry charges.

The prototype example of these operators is the 1/2 BPS WL in four-dimensional ${\cal N}=4$ SYM theory constructed in \cite{Maldacena:1998im} and dual to a fundamental string in $\gsym$ spacetime \cite{Maldacena:1998im, Rey:1998ik}.\footnote{In this paper we consider only WLs in fundamental representation. 1/2 BPS operators in more general representations are dual to D5--branes or D3--branes in $\gsym$\cite{Hartnoll:2006hr,Yamaguchi:2006tq,Gomis:2006sb,Gomis:2006im}.}
It corresponds to the holonomy of a generalized connection that includes also a coupling to the scalar fields of the theory. Using localization, the exact value for circular loops is given by a gaussian matrix model \cite{Drukker:2000rr,Pestun:2007rz}. At weak coupling it coincides with the perturbative result of \cite{Drukker:1999zq,Drukker:2000rr}, whereas at strong coupling it reproduces the type IIB fundamental string result \cite{Maldacena:1998im,Rey:1998ik,Drukker:1999zq,Drukker:2000rr}.

A similar approach has led to the construction of 1/2 BPS WLs in three-dimensional super Chern--Simons--matter (SCSM) theories. In particular, for the ABJ(M) models
\cite{Aharony:2008ug, Aharony:2008gk} this operator has been found in \cite{Drukker:2009hy} as the holonomy of a superconnection that contains, in addition to the gauge field, scalar and fermion matter fields in the bi--fundamental representation of the gauge group.
Less supersymmetric BPS WLs have been also constructed, which still contain additional scalars and/or fermions. In particular, the bosonic 1/6 BPS WL \cite{Drukker:2008zx,Chen:2008bp,Rey:2008bh}, dual to smeared fundamental strings or D-branes \cite{Drukker:2008zx}, plays an important role in the exact evaluation of 1/2 BPS WL, since the two operators only differ by a $Q$--term, where $Q$ is the charge used to localize the functional integral \cite{Drukker:2009hy}. The evaluation of the 1/2 BPS WL at weak coupling \cite{Bianchi:2013zda,Bianchi:2013rma,Griguolo:2013sma} and at strong coupling, via the M--theory  $\gabjm$ dual description \cite{Drukker:2008zx,Chen:2008bp,Rey:2008bh} matches the exact result from localization \cite{Kapustin:2009kz,Marino:2009jd,Drukker:2010nc}.

One important feature of 1/2 BPS WLs in four-dimensional ${\cal N}=4$ SYM and three-dimensional ABJ(M) theories is their uniqueness: for a specific set of preserved supercharges, there is at most
one single operator that is invariant under their action. This is true both at classical and quantum level, and it is consistent with the uniqueness of the localization result and the uniqueness of the string or M2--brane solutions in the corresponding dual description.

More recently, the construction of 1/2 BPS WLs in $\mN=4$ orbifold ABJM theory, and more generally in quiver $\mN=4$ SCSM theories with gauge group $\prod_{\ell=1}^{r}[U(N_{2\ell-1}) \times U(N_{2\ell})]$ and alternating levels \cite{Gaiotto:2008sd,Hosomichi:2008jd}, has been attacked \cite{Ouyang:2015qma,Cooke:2015ila}. 1/2 BPS operators can be defined locally for  each pair of adjacent quiver nodes. Referring to the $\ell$-- and $\ell+1$--nodes it is given by the holonomy of a superconnection that contains couplings to scalars and fermions in the bi--fundamental representation of the gauge groups $U(N_{2\ell-1}) \times U(N_{2\ell})$, or $U(N_{2\ell}) \times U(N_{2\ell+1})$, and adjacent nodes.

The novel feature that emerges for the first time in this context is the lack of uniqueness. In fact, at classical level two different WLs have been constructed that preserve the same set of four supercharges \cite{Cooke:2015ila}. The two operators, $\psi_1$--loop $W_{\psi_1}$ and $\psi_2$--loop $W_{\psi_2}$ in the language of \cite{Cooke:2015ila}, are evaluated along the same contour but differ for the matter couplings. Nevertheless, they are both cohomologically equivalent to the same 1/6 BPS WL, whose expectation value can be exactly computed using localization.

The existence of two different 1/2 BPS WLs seems to be in contrast with the M--theory dual description where in principle there should be one single M2--brane solution that is 1/2 BPS (see discussion in \cite{Cooke:2015ila}). It also seems to be puzzling when compared to the localization result that in principle provides a unique result, $\langle W_{\psi_1} \rangle = \langle W_{\psi_2} \rangle$, being both the operators $Q$--equivalent to the same 1/6 BPS WL.

In a perturbative setup, this puzzle has been solved in \cite{Bianchi:2016vvm} by computing the two WLs up to three loops.\footnote{Precisely, the result of \cite{Bianchi:2016vvm} holds for general $\mN=4$ SCSM $\prod_{\ell=1}^{r}[U(N_{2\ell-1})\times U(N_{2\ell})]$ quiver gauge theories with different ranks, but it cannot be extended to the case of equal ranks ($\mN=4$ orbifold ABJM theory).} While at one and two loops they have the same expectation value that coincides with the localization result \cite{Griguolo:2015swa}, at three loops they start being different and the localization result is matched only by the linear combination $\tfrac12 (W_{\psi_1} + W_{\psi_2})$ \cite{Bianchi:2016vvm}. Therefore, at weak coupling this combination seems to be the true BPS quantum operator.
However, this cannot be the end of the story. For ${\cal N}=4$ SCSM theories, a deeper comprehension of the physical mechanism that leads to two different WLs preserving the same set of supercharges would be desirable, as well as the construction of the corresponding duals in M--theory and the identification of the actual BPS operator at strong coupling.

Motivated by the above discussion, in this paper we perform a systematic construction of general 1/2 BPS WLs on the straight line and their string/M--theory duals, using  the heavy W--boson effective theory procedure and its dual counterpart in string/M--theory \cite{Maldacena:1998im,Rey:1998ik,Drukker:1999zq,Lee:2010hk}.%
\footnote{In this paper we consider the M--theory dual description of three-dimensional SCSM theories. Often it is more convenient to study SCSM theories in terms of a dual type IIB string theory, as done for example in \cite{Assel:2011xz,Assel:2012cj}.}
We begin by considering four-dimensional ${\cal N} =4$ SYM as a guideline, and then move to three-dimensional SCSM theories with decreasing amount of supersymmetry.

Different 1/2 BPS WL can be obtained by Higgsing along different (independent) directions in the scalar field space and/or choosing different massive modes corresponding to heavy particles or antiparticles. Different Higgsing directions correspond to different positions of the dual fundamental strings or M2--branes in the internal space and lead to WLs that are simply related by R--symmetry rotations and then correspond to the same quantum operator. Instead, choosing massive particle or antiparticle modes corresponds to choosing fundamental string/M2--brane or fundamental anti--string/anti--M2--brane solutions localized at the same position, and should lead to two physically distinguishable objects.

In all cases we construct two sets of independent WLs, one set ($W$ operators) obtained by Higgsing with heavy W--particles, the second one ($\tilde{W}$ operators) obtained by Higgsing with heavy W--antiparticles, both with the same mass. We study the overlapping of supercharges preserved by different WLs, as well as the overlapping of supercharges preserved by the dual fundamental strings or M2--branes. In all the cases we find perfect matching between field theory and string/M--theory results. In fact, we manage to identify the supercharges in string/M--theory with the supercharges in field theory, as well as the supercharges preserved by the M2--/anti--M2--branes with the supercharges preserved by the 1/2 BPS WLs.

While for four-dimensional ${\cal N}=4$ SYM, operators in the same set preserve different supercharges simply related by an internal rotation, and two WLs along the same line in different sets always preserve complementary sets of supercharges, for three-dimensional SCSM theories the overlapping configuration of preserved supercharges becomes more interesting.

In ABJM theory we find that any couple of WLs in the same set, let's say $W_I$ and $W_J$ with fixed $I,J= 1,2,3,4$ and $J \neq I$, always share two supercharges, $\theta_+^{IJ}, \e_{IJ KL}\theta_-^{KL}$. Operators belonging to two different sets, ${W}_I$ and $\tilde W_{J}$ with $J \neq I$, share four supercharges $\theta_+^{IK}, \theta_-^{JK}$ with $K \neq I,J$.

This overlapping becomes more stringent in ${\cal N}=4$ orbifold ABJM where it is possible to find one particle and one antiparticle configurations corresponding to different Higgsing directions in the scalar moduli space, which preserve exactly the same set of supercharges. These are the remnants of the overlapped ABJM spectrum after the orbifold projection.
In fact, under the R--symmetry breaking $SU(4) \to SU(2) \times SU(2)$ that implies the index relabeling $I \to (i , \hi)$, with $i=1,2$, $\hi= \hat{1}, \hat{2}$,
we find that four pairs of operators, say the pair $W_1, \tilde{W}_2$, the pair $\tilde{W}_1, W_2$, the pair $W_{\hat{1}}, \tilde{W}_{\hat{2}}$, and the pair $\tilde{W}_{\hat{1}}, W_{\hat{2}}$, preserve the same set of supercharges (see table \ref{tab4} and Figure \ref{f2a}). They are nothing but the $\psi_1$-- and $\psi_2$--type loops of \cite{Cooke:2015ila}.
Correspondingly, in M--theory in $\goabjm$ background we find that, contrary to the expectations, there exist pairs of M2-- and anti--M2-- branes at different positions that preserve the same set of supercharges (see Figure \ref{f2b}). They are the duals of $\psi_1$-- and $\psi_2$--loops.

We generalize this construction to $\mN=4$ SCSM quiver theories with gauge group and levels $\prod_{\ell=1}^{r}[U(N_{2\ell-1})_k \times U(N_{2\ell})_{-k}]$ and different group ranks. Again, we find that using  massive particles or antiparticles in the Higgsing procedure leads to the definition of two different classes of WL operators, with  special representatives that turn out to preserve the same set of supercharges. In this case, the dual M--theory description is not known.

\vskip 10pt

Our analysis enlightens the origin of the pairwise degeneracy of WL operators in $\mN=4$ SCSM theories, both from a (classical) field theory perspective and from a M--theory point of view. But at the same time it opens new questions.
In fact, on one side the existence in $\mN=4$ orbifold ABJM theory of pairwise M2--brane embeddings that preserve the same set of supercharges reconciles the WL degeneracy found in CFT with the AdS/CFT predictions, as both $\psi_1$-- and $\psi_2$-- WL operators have  distinct dual counterparts. On the other side, since the degeneracy persists at strong coupling, we cannot expect that the classical degeneracy gets lifted by quantum corrections, as previously suggested \cite{Cooke:2015ila,Bianchi:2016vvm}.
Rather, the present result seems to point towards the fact that both $\psi_1$-- and $\psi_2$-- WLs could be separately 1/2 BPS operators. As already mentioned, this should happen consistently with the localization result\footnote{Comparison with localization results makes sense only once we perform a Wick rotation to euclidean space and map the straight line to the circle by a conformal transformation.}
that predicts the same value for the two quantum operators. A perturbative calculation for 1/2 BPS WLs in $\mN=4$ orbifold ABJM could answer this question, but it is not available yet.
This is instead available for more general $\mN=4$ SCSM quiver theories with gauge group and levels $\prod_{\ell=1}^{r}[U(N_{2\ell-1})_k \times U(N_{2\ell})_{-k}]$ for which we know that the two results start  being different at three loops \cite{Bianchi:2016vvm}, and the localization result is matched by the unique BPS operator $\tfrac12 (W_{\psi_1} + W_{\psi_2})$. In this case it would be reasonable to expect that in the corresponding dual M--theory description no pairwise degeneracy of M2--brane embeddings would be present. Unfortunately, the M2--brane construction for this more general case has not been done yet. Therefore, in the absence of further indications it is difficult to clarify the whole picture and draw any definite conclusion.

\vskip 10pt

The rest of the paper is organized as follows.
In section~\ref{sec2} we briefly review the physical picture of the heavy W--boson effective theory obtained by Higgsing procedure and its string counterpart.
In section~\ref{sec3}, \ref{sec4}, \ref{sec5} we investigate 1/2 BPS WLs and their string theory or M--theory duals in, respectively, four-dimensional $\mN=4$ SYM theory, ABJM theory, and $\mN=4$ orbifold ABJM theory.
In section~\ref{sec6} we consider the 1/2 BPS WLs and Higgsing procedures in more general $\mN=4$ SCSM theories with alternating levels.
We conclude with a discussion of our results in section~\ref{sec7}.
In appendix~\ref{appA} we give spinor conventions and useful spinorial identities in three dimensions.
In appendix~\ref{appB} we collect the infinite mass limit for the relevant free field theories.
In appendix~\ref{appC} we give details to determine the Killing spinors in $\gsym$ spacetime.
In appendix~\ref{appD} we determine the Killing spinors in $\textrm{AdS}_4\times\textrm{S}^7$ spacetime.
Appendix \ref{appE} contains the detailed Higgsing procedure for general $\mN = 4$ SCSM theories.
Finally, in appendix \ref{appF} we first determine the Killing spinors in the $\gtzt$ spacetime and then use these results to construct 1/2 BPS M2-- and anti--M2--brane configurations that could be possibly dual to 1/2 BPS Wilson surfaces in six-dimensional ${\cal N}=(2,0)$ superconformal field theory.

\section{The Higgsing procedure}\label{sec2}

The guideline that we follow to identify a BPS WL and its dual in string or M--theory is the derivation of these operators through the Brout--Englert--Higgs mechanism applied on both sides of the AdS/CFT correspondence. The idea originates from \cite{Maldacena:1998im,Rey:1998ik}, and has been realized explicitly for four-dimensional ${\cal N}=4$ SYM theory in \cite{Drukker:1999zq} and for ABJM theory in \cite{Lee:2010hk}.  Here we briefly review this technique in the tantalizing example of four-dimensional $\mN=4$ SYM theory.\footnote{The procedures in \cite{Drukker:1999zq} and \cite{Lee:2010hk} are similar but not completely equivalent. In this paper we adopt the latter one.}

In a generic gauge theory, a WL along the timelike infinite straight line $x^\m=\t\d^\m_0$  corresponds to the phase associated to the semiclassical evolution of a very heavy quark in the gauge background. Since in $\mN=4$ SYM theory there are no massive particles, one can introduce them by a Higgsing procedure. Precisely, starting from the $\mN=4$ SYM theory with gauge group $SU(N+1)$, one breaks the gauge symmetry to $SU(N)\times U(1)$ by introducing an infinite expectation value for some of the scalar fields and  eventually gets  $\mN=4$ SYM theory with gauge group $SU(N)$ coupled to some infinitely massive particles. The corresponding Wilson operator turns out to be the holonomy of the actual gauge connection that appears in the resulting heavy particle effective lagrangian.

\begin{figure}[htbp]
  \centering
  \subfigure[A string stretching between a stack of $N$ D3--branes and one extra D3-brane]
            {\includegraphics[height=50mm]{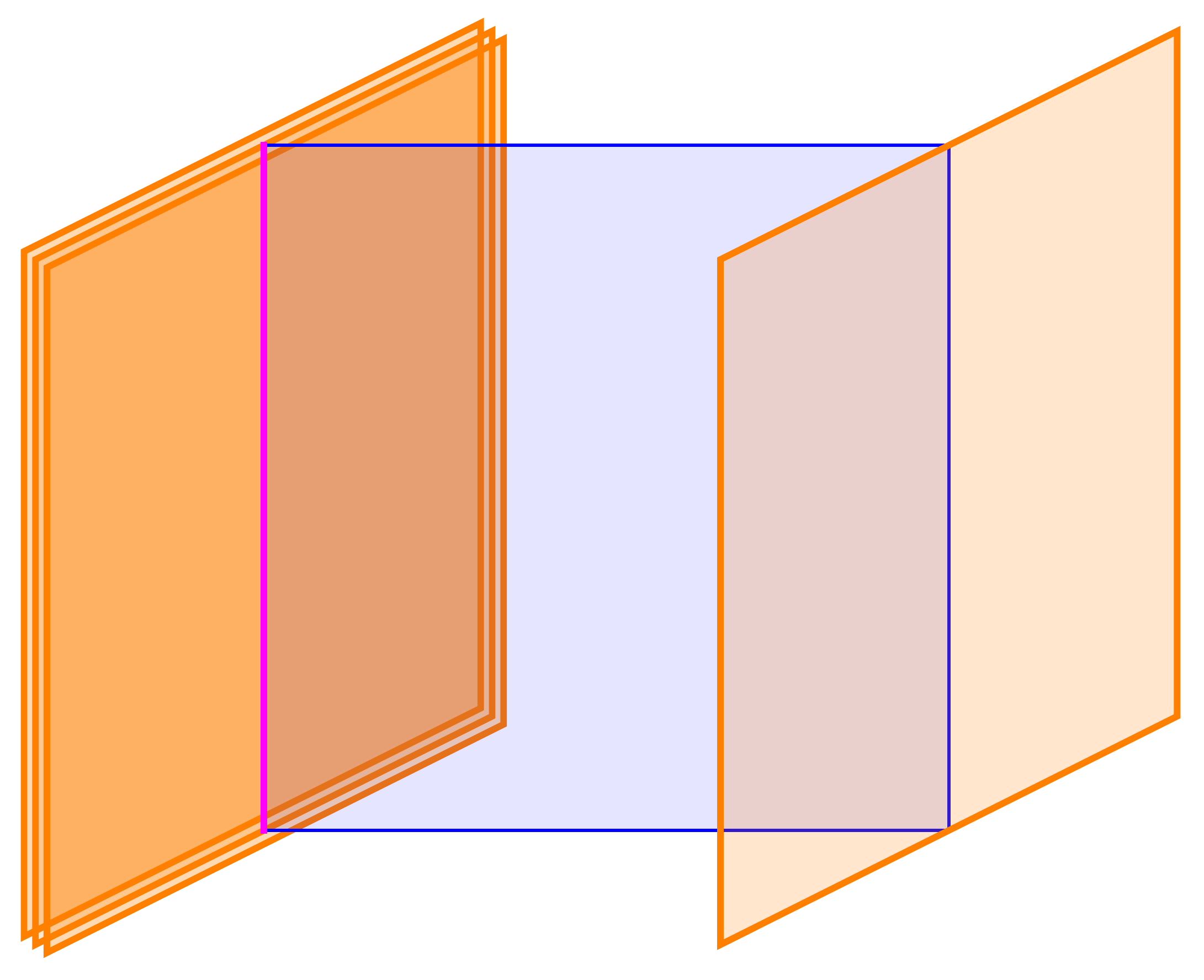}}
            ~~~~~~
  \subfigure[The string in the $\gsym$ geometry]
            {\includegraphics[height=50mm]{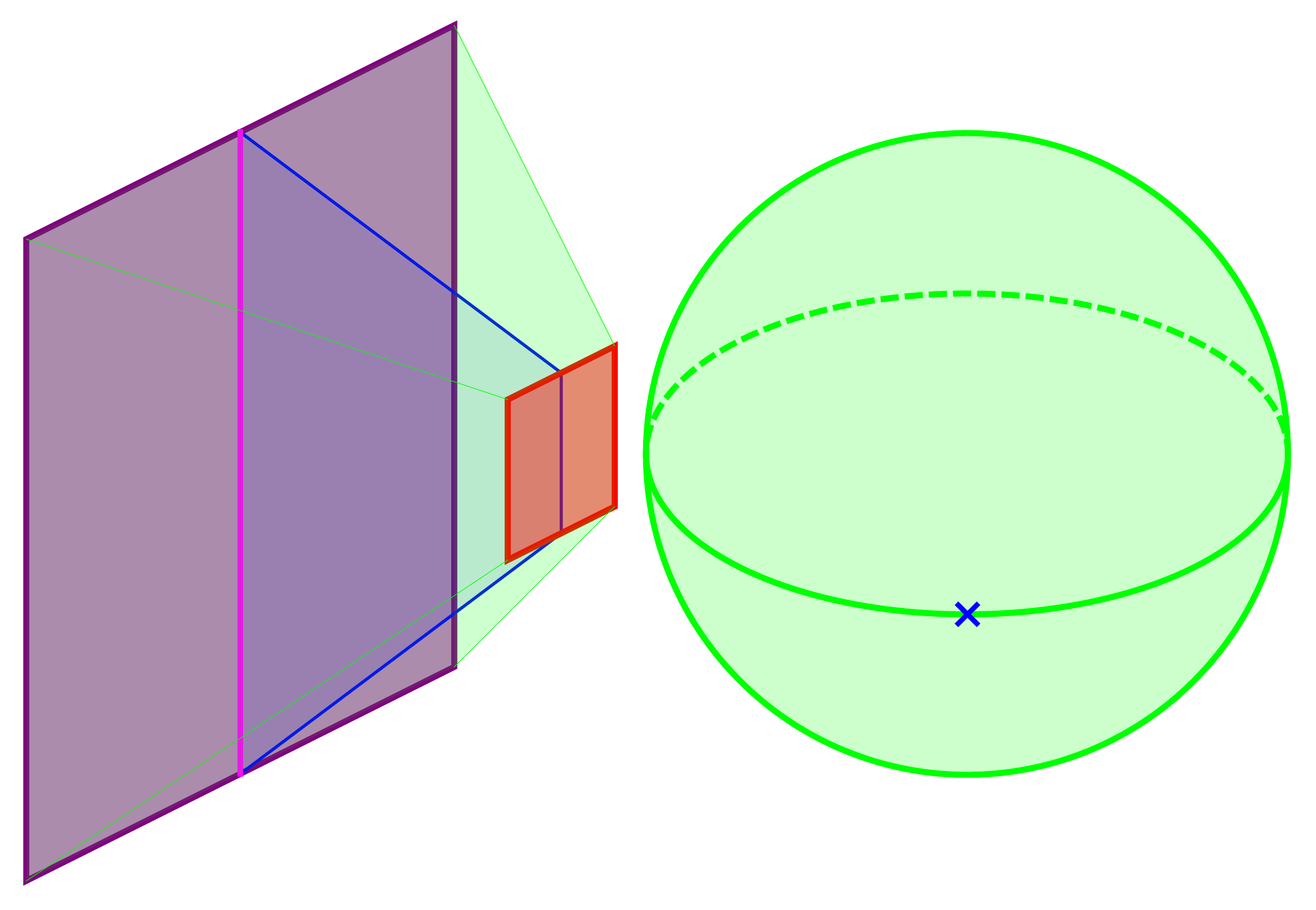}}\\
  \caption{The Higgsing procedure of four-dimensional $\mN=4$ SYM theory in the dual string theory description. In (a) the magenta line is the Wilson loop. In (b) the blue cross on S$^5$ represents the point where the string is localized in the internal space.}\label{hi}
\end{figure}

The string theory dual of the Higgsing procedure is shown in Figure~\ref{hi}. It corresponds to starting from a stack of $N+1$ coincident D3--branes and then moving one of the D3--branes to infinity in some particular direction. One can excite one fundamental string that connects the extra D3-brane with the remaining $N$ D3--branes. The worldline of the end-point of this string in the worldvolume of the $N$ D3--branes is precisely the Wilson loop.
By taking the near horizon limit of the $N$ D3--branes, we get the $\gsym$ geometry with a fundamental string stretching from the UV to IR in AdS$_5$ spacetime and being localized in the compact S$^5$ space.

There is a one-to-one correspondence between the Higgsing procedure in field theory and the dual string theory construction. In fact, the direction in which the extra D3-brane moves, and therefore its localization in the internal space, is related to the direction of the expectation value in the scalar field space of ${\cal N}=4$ SYM theory. Moreover, we have the freedom to excite 1/2 BPS fundamental strings or anti--strings between the stack of $N$ D3--branes and the extra probe brane. In the field theory language this corresponds to exciting different massive modes, that is massive particles or antiparticles. As we are going to show in the next section, fundamental strings and anti--strings localized at the same position in $\textrm{S}^5$ preserve complementary sets of supercharges, and they are dual to different WLs that also preserve complementary sets of supercharges.

This procedure can be easily generalized to M--theory in $\goabjm$ backgrounds where  different configurations of M2--branes or anti--M2--branes give rise to different sets of Wilson loop operators. A pair of  M2-- and anti--M2--branes localized at the same point preserve complementary sets of supercharges, and are dual to different WLs that also preserve non--overlapping sets of supercharges.

\section{Four-dimensional $\mN=4$ super Yang--Mills theory}\label{sec3}

As a warm-up, and also to fix our notations, in this section we review the Higgsing procedure for four-dimensional ${\cal N}=4$ SYM theory. We construct different (independent) 1/2 BPS WLs and focus on the spectrum of the corresponding preserved supercharges.

\subsection{1/2 BPS Wilson loops}\label{sec3.1}

Using ten-dimensional $\mN=1$ SYM theory formalism, the field content of the theory is given by one gauge field $A_\m$, six scalars $\phi_I$ with $I=4,5,\cdots,9$
and one ten-dimensional Majorana--Weyl spinor $\lambda$, all in the adjoint representation of the $SU(N)$ gauge group. The corresponding lagrangian is
\be \label{ld4ne4sy}
\mL = -\f1{4} \Tr ( F_{\m\n}F^{\m\n} + 2 D_\m\phi_I D^\m\phi_I -[\phi_I,\phi_J][\phi_I,\phi_J] )
      +\f{\ii}{2} \Tr [ \bar\l (\g^\m D_\m\l + \ii\g_I\phi_I\l) ]
\ee
with $F_{\m\n}= \p_\m A_\n - \p_\n A_\m + \ii [A_\m,A_\n]$, $D_\m\phi_I= \p_\m\phi_I + \ii [A_\m,\phi_I]$, $\bar\l=-\l^\dagger\g^0$ and $D_\m\l= \p_\m\l + \ii [A_\m,\l]$. The bosonic part of the supersymmetry (SUSY) transformations are
\be \label{e99}
\d A_\m=\ii\bar\l\g_\m ( \th + x^\m \g_\m \vth ), ~~ \d \phi_I=\ii\bar\l\g_I ( \th + x^\m \g_\m \vth )
\ee
Here $\th$ is the ten-dimensional Majorana--Weyl spinor with positive chirality associated to Poincar\'e supercharges, and $\vth$ is the Majorana--Weyl spinor with negative chirality associated to superconformal charges.

On a time-like infinite straight line $x^\m=(\t,0,0,0)$, we define the general 1/2 BPS WL \cite{Maldacena:1998im,Rey:1998ik,Gomis:2006sb}
\be
W = \mP \exp\Big( -\ii \int d\t \mA (\t) \Big)
\ee
with generalized connection
\be
 \label{e98-0}
\mA = A_0 - \phi_I n^I \nn\\
\ee
and preserved supercharges
\be
\label{e98}
\g_{0I}n^I \th=-\th, ~~ \g_{0I}n^I\vth=\vth
\ee
Here $n^I$ is a constant vector in R--symmetry space, with $\d_{IJ}n^I n^J =1$.

As particular cases, we consider a set $W$ of 1/2 BPS WLs with six independent representatives $W_I$ associated to
the generalized connections
\be
\label{eq:connection}
\mA_I = A_0 - \phi_I
\ee
The corresponding preserved supercharges are selected by
\be
\g_{0I}\th=-\th, ~~ \g_{0I}\vth=\vth
\ee
Similarly, we introduce a second set $\tilde{W}$ with representatives $\tilde{W}_I$  associated to
\be \label{a0pphi4}
\td\mA_I = A_0 + \phi_I
\ee
which lead to preserved supercharges
\be
\g_{0I}\th=\th, ~~ \g_{0I}\vth=-\vth
\ee

Although the six 1/2 BPS WLs in class $W$ are related by $SU(4)\cong SO(6)$ R--symmetry rotations, the relation among the corresponding preserved supercharges is interesting. Since matrices $\g_{0I}$, $\g_{0J}$ with $I\neq J$ do not commute, there is no overlapping among supercharges preserved by the $W$ loops.
The same is true for the six WLs in class $\tilde{W}$. Moreover, $W_I$ and $\td W_I$ with the same $I$--index always preserve complementary sets of supercharges.  In conclusion, there is no overlapping among the supercharges preserved by these WL representatives. This is a property that we will meet also in the gravity dual construction of section~\ref{subsecAdS5S5}.

\subsection{Wilson loops from Higgsing}\label{sec3.2}

Following the original idea of \cite{Maldacena:1998im,Rey:1998ik}, we now briefly review the Higgsing construction of WLs in ${\cal N}=4$ SYM theory.

Starting with $\mN=4$ $SU(N+1)$ SYM theory, we break the gauge group to $SU(N)$ by the following choice\footnote{We label gauge fields in $SU(N+1)$ and $SU(N)$ theories with the same letters, as long as this does not cause confusion.}
\be \label{amphii}
A_\m = \lt(\ba{cc} A_\m & W_\m \\ \bar W_\m & 0 \ea\rt),      ~~
\phi_J = \lt(\ba{cc} \phi_J & R_J \\ \bar R_J & v_J \ea\rt)
\ee
where $v_J = v n_J$, $v>0$, $\d^{IJ}n_I n_J =1$. To be definite we choose $n_J = \d_J^I$ with fixed $I=4, \cdots,9$. Taking $v \to \inf$ leads to particles with infinite mass, $m=v$.

The $v$--flux breaks half of the supersymmetries. The massive vector field $W_\m$ has three complex degrees of freedom, $W_\pm=W_1 \pm \ii W_2$, $W_3$, while working in unitary gauge ($R_I=0$), we are left with five scalars $R_i$, with $i\neq I$. These fields build up the bosonic part of the four-dimensional $\mN=2$ massive vector multiplet according to the following assignment
\begin{center}\begin{tabular}{r|c|c|c|c|c}
  {spin}       & 1     & 1/2      & 0            &$-1/2$    & $-1$ \\ \hline
  {degeneracy} & 1     & 4        & 6            & 4        & 1 \\ \hline
  {field}      & $W_+$ & $\cdots$ & $W_3$, $R_i$ & $\cdots$ & $W_-$
\end{tabular}\end{center}

Since we are interested in the low-energy dynamics of massive particles and their interactions with the $SU(N)$ SYM theory, we focus on terms in the lagrangian that are non-vanishing in the $v \to \inf$ limit. Inserting the ansatz (\ref{amphii}) into (\ref{ld4ne4sy}), we obtain the following lagrangian for the bosonic massive particles
\bea \label{j41}
&& \mL = -\f12 \bar W_{\m\n}W^{\m\n}-v^2\bar W_\m W^\m +2v\bar W_\m\phi_I W^\m \nn\\
&& \phantom{\mL =}
   -D_\m \bar R_i D^\m R_i -v^2 \bar R_i R_i + 2v \bar R_i\phi_I R_i
\eea
where $W_{\m\n} = D_\m W_\n-D_\n W_\m$, $D_\m W_\n=\p_\m W_\n+\ii A_\m W_\n$, $D_\m R_i=\p_\m R_i+\ii A_\m R_i$.

We now have two possibilities. If we use particle modes
\be \label{j1}
W_\m = \f{1}{\sqrt{2v}}(0,w_1,w_2,w_3)\ep^{-\ii m t} \quad , \quad
R_i = \f{1}{\sqrt{2v}}r_i\ep^{-\ii m t}
\ee
the non-relativistic lagrangian can be reduced to the following form (see appendix \ref{appB})
\be
\label{eq:nonrelL}
\mL = \ii \bar w_a \mD_0 w_a
     +\ii \bar r_i \mD_0 r_i
\ee
where $\mD_0 w_a=\p_0 w_a+\ii \mA_I w_a$, $\mD_0 r_i=\p_0 r_i+\ii \mA_I r_i$, and the new connection is
\be
\mA_I = A_0 - \phi_I
\ee
This is exactly the connection in (\ref{eq:connection}) that defines WLs in the $W$ set.

Alternatively, we can use antiparticle modes
\be\label{j2}
W_\m = \f{1}{\sqrt{2v}}(0,w_1,w_2,w_3)\ep^{\ii m t} \quad , \quad   R_i = \f{1}{\sqrt{2v}}r_i\ep^{\ii m t}
\ee
and we get the non-relativistic lagrangian
\be
\mL = \ii \Tr w_a \mD_0 \bar w_a
     +\ii \Tr r_i \mD_0 \bar r_i
\ee
where $\mD_0 \bar w_a=\p_0 \bar w_a-\ii \bar w_a \td\mA_I$, $\mD_0 \bar r_i=\p_0 \bar r_i-\ii\bar r_i \td\mA_I$, with the connection $\tilde{\cal A}$ being
\be
\td\mA_I = A_0 + \phi_I
\ee
This is indeed the connection (\ref{a0pphi4}) that enters the definition of 1/2 BPS WLs in the  $\td W$ set.

$W_I$ and $\td W_I$ preserve complementary sets of supercharges and they describe the evolution of massive particles and their antiparticles, as can be seen in (\ref{j1}) and (\ref{j2}).

An alternative, but equivalent procedure starts by Higgsing in the opposite direction in the scalar field space, that is choosing $v_J = -v \d_J^I$ with $v>0$ in ansatz (\ref{amphii}). In this case, by exciting modes (\ref{j1}) we get connection (\ref{a0pphi4}) that defines the $\td W_I$ loop, whereas exciting modes (\ref{j2}) we obtain connection (\ref{eq:connection}) that defines $W_I$.

Since we have two different, but equivalent ways to generate the two classes of WL operators, we will call them $W^{(1)}_I, W^{(2)}_I$ and $\tilde{W}_I^{(1)}, \tilde{W}_I^{(2)}$ although they represent the same operator. While this classification for the ${\cal N}=4$ SYM case seems quite redundant, it will become non-trivial when dealing with their string theory duals in the next subsection.

\subsection{Fundamental strings in AdS$_5\times$S$^5$ spacetime}\label{subsecAdS5S5}

We now determine the fundamental string solutions in $\gsym$ dual to the 1/2 BPS WL we have constructed.

Type IIB string $\gsym$ background with self-dual five-form flux is described by
\bea \label{ads5s5}
&& ds^2 = R^2 (ds^2_{\AdS_5} + ds^2_{\rmS^5}) \nn \\
&& F_{\td\m\td\n\td\r\td\s\td\l} = \f{4}{R} \ve_{\td\m\td\n\td\r\td\s\td\l}, ~~
F_{\ti\tj\td k\td l\td m} = -\f{4}{R} \ve_{\ti\tj\td k\td l\td m}
\eea
with $\ve_{\td\m\td\n\td\r\td\s\td\l}$ and $\ve_{\ti\tj\td k\td l\td m}$ being the volume forms of AdS$_5$ and S$^5$, respectively.

For the unit AdS$_5$ we choose the Poincar\'e coordinates
\be \label{ads5}
ds^2_{\AdS_5} = u^2 (-dt^2 + d x_1^2+ d x_2^2+ d x_3^2) + \f{du^2}{u^2}
\ee
with $u\to\inf$ being the boundary. Embedding S$^5$ in R$^6\cong$C$^3$ as
\bea
&& z_1 = \cos\th_1 \, \ep^{\ii\xi_1} = x_4 + \ii x_6             \nn \\
&& z_2 = \sin\th_1 \cos\th_2 \, \ep^{\ii\xi_2} = x_5 + \ii x_8        \nn\\
&& z_3 = \sin\th_1 \sin\th_2 \, \ep^{\ii\xi_3} = x_7 + \ii x_9
\eea
with $\th_{1,2}\in [0,\frac{\pi}{2}]$, $\xi_{1,2,3} \in [0,2\pi]$,
we get to the unit S$^5$ metric
\be \label{s5}
ds^2_{\rmS^5} = d\th_1^2+\cos^2\th_1 d\xi_1^2 + \sin^2\th_1 ( d \th_2^2 + \cos^2\th_2 d\xi_2^2 + \sin^2\b d\xi_3^2 )
\ee
Note that the R$^6\cong$C$^3$ is along the perpendicular directions of the stack of D3--branes before the near horizon limit is taken.

The Killing spinors for the $\gsym$ geometry are determined in appendix~\ref{appC}, following the procedure in \cite{Skenderis:2002vf}. They are given in  eqs. (\ref{ks1}), (\ref{e96}), (\ref{e95}), (\ref{e97}).

We now consider a fundamental string embedded in $\gsym$ spacetime as
\be
t=\s^0, ~~ x_{1,2,3}=0, ~~ u=\s^1
\ee
with $\s^{0,1}$ being the string worldsheet coordinates. We localize the string at some point on S$^5$, that is parametrized by a uniform vector $n^I$ in $\textrm{C}^3 \cong \textrm{R}^6$
\bea
&& n^4 = \cos\th_1\cos\xi_1, ~~ n^5 = \sin\th_1\cos\th_2\cos\xi_2, ~~  n^6 = \cos\th_1\sin\xi_1 \nn\\
&& n^7 = \sin\th_1\sin\th_2\cos\xi_3,~~ n^8 = \sin\th_1\cos\th_2\sin\xi_2,~~ n^9 = \sin\th_1\sin\th_2\sin\xi_3
\eea
The supercharges preserved by the fundamental string are given by\footnote{The names string and anti--string are interchangeable, and we choose the sign here for convenience of comparison to Wilson loops.}
\be \label{j49}
\g_{04}\e=-\e^c
\ee
with $\e^c$ being the charge conjugate of $\e$ \footnote{For a generic spinor $\e$, $\e^c$ is defined as the charge conjugate $\e^c = B^{-1}\e^*$, where $B$ is given in terms of gamma matrices, and satisfies the condition $B^{-1}\g_A^*B= \g_A$. The explicit form of $B$ can be found, for example, in \cite{Polchinski:1998rr}. In Majorana basis we have $\e^c = \e^\ast$.}.

Using the explicit expression (\ref{ks1}) for the Killing spinor we obtain
\be
h^{-1} \g_{04} h \e_1 = -\e_1^c, ~~ h^{-1} \g_{04} h \e_2 = \e_2^c
\ee
where $h$ has been defined in (\ref{e96}).
Expressing $\e_1$ and $\e_2$ as in (\ref{e97}), this is equivalent to
\be \label{j49prime}
h^{-1} \g_{04} h \th = -\th, ~~ h^{-1} \g_{04} h \vth =  \vth
\ee
where $\th$, $\vth$ are constant Majorana--Weyl spinors with respectively positive and negative chiralities. It turns out that \cite{Gomis:2006sb}
\be
h^{-1} \g_{04} h = \g_{0I}n^I
\ee
so that (\ref{j49prime}) becomes
\be \label{j49primeprime}
\g_{0I}n^I \th = -\th, ~~ \g_{0I}n^I \vth = \vth
\ee
These equations have exactly the same structure as the ones in eq. (\ref{e98}) defining the supercharges preserved by a general 1/2 BPS WL. Therefore, we are led to identify the Killing spinor components $\th$, $\vth$ in $\gsym$ with the Poincar\'e supercharges $\th$ and superconformal charges $\vth$ of four-dimensional SYM theory \cite{Gomis:2006sb}.
The spectrum of preserved supercharges depends on the particular string configuration, as we now describe.

We consider twelve different string configurations, $F_1^{(i)}, i=1, \cdots , 12$, localized at twelve different positions in the compact space. Their positions are explicitly listed in table~\ref{tab1}, both in terms of complex coordinates in $C^3\cong R^6$ and in terms of angular coordinates.
Solving constraint (\ref{j49primeprime}) for each specific string solution we obtain the corresponding preserved supercharges in the fourth column of table~\ref{tab1}.
In particular, we note that strings localized at opposite points in $\textrm{S}^5$, that are $F_1^{(i)}$ and $F_1^{(i+2)}$ solutions with $i=1,2,5,6,9,10$, preserve complementary sets of supercharges. In fact, the corresponding Killing spinor equations always differ by a sign on the r.h.s..

\begin{table}[htbp]\centering\begin{tabular}{|c|l|l|l|}\hline
  string      & \multicolumn{2}{c|}{position} & \multicolumn{1}{c|}{preserved supercharges} \\ \hline

  $F_1^{(1)}$ & $z_1=1$    & $\th_1=\xi_1=0$        & $\g_{04} \th = -\th$, $\g_{04} \vth = \vth$ \\ \hline
  $F_1^{(2)}$ & $z_1=\ii$  & $\th_1=0,\xi_1=\pi/2$  & $\g_{06} \th = -\th$, $\g_{06} \vth = \vth$ \\ \hline
  $F_1^{(3)}$ & $z_1=-1$   & $\th_1=0,\xi_1=\pi$    & $\g_{04} \th = \th$, $\g_{04} \vth = -\vth$ \\ \hline
  $F_1^{(4)}$ & $z_1=-\ii$ & $\th_1=0,\xi_1=3\pi/2$ & $\g_{06} \th = \th$, $\g_{06} \vth = -\vth$ \\ \hline

  $F_1^{(5)}$ & $z_2=1$    & $\th_1=\pi/2,\th_2=\xi_2=0$        & $\g_{05} \th = -\th$, $\g_{05} \vth = \vth$ \\ \hline
  $F_1^{(6)}$ & $z_2=\ii$  & $\th_1=\pi/2,\th_2=0,\xi_2=\pi/2$  & $\g_{08} \th = -\th$, $\g_{08} \vth = \vth$ \\ \hline
  $F_1^{(7)}$ & $z_2=-1$   & $\th_1=\pi/2,\th_2=0,\xi_2=\pi$    & $\g_{05} \th = \th$, $\g_{05} \vth = -\vth$ \\ \hline
  $F_1^{(8)}$ & $z_2=-\ii$ & $\th_1=\pi/2,\th_2=0,\xi_2=3\pi/2$ & $\g_{08} \th = \th$, $\g_{08} \vth = -\vth$ \\ \hline

  $F_1^{(9)}$  & $z_3=1$    & $\th_1=\th_2=\pi/2,\xi_3=0$      & $\g_{07} \th = -\th$, $\g_{07} \vth = \vth$ \\ \hline
  $F_1^{(10)}$ & $z_3=\ii$  & $\th_1=\th_2=\pi/2,\xi_3=\pi/2$  & $\g_{09} \th = -\th$, $\g_{09} \vth = \vth$ \\ \hline
  $F_1^{(11)}$ & $z_3=-1$   & $\th_1=\th_2=\pi/2,\xi_3=\pi$    & $\g_{07} \th = \th$, $\g_{07} \vth = -\vth$ \\ \hline
  $F_1^{(12)}$ & $z_3=-\ii$ & $\th_1=\th_2=\pi/2,\xi_3=3\pi/2$ & $\g_{09} \th = \th$, $\g_{09} \vth = -\vth$ \\ \hline
  \end{tabular}
\caption{The twelve different fundamental strings at different positions and their preserved supercharges. A similar table can be constructed for anti--string $\bar{F}_1^{(i)}$ solutions. The corresponding preserved charges are obtained by changing the sign of the r.h.s. of the Killing spinor equations.}\label{tab1}
\end{table}

Similarly, we can consider twelve fundamental anti--string configurations, $\bar F_1^{(i)}$, localized at the same internal points listed in table~\ref{tab1}. The corresponding preserved supercharges are obtained by solving the constraint
\be
\g_{0I}n^I \th = \th, ~~ \g_{0I}n^I \vth = -\vth
\ee
for each anti--string configuration. It turns out easily that the fundamental string and anti--string configurations localized at the same point preserve complementary sets of supercharges, whereas
$F_1^{(i)}$ and $\bar{F}_1^{(i+2)}$, or ${F}_1^{(i+2)}$ and $\bar F_1^{(i)}$, with $i=1,2,5,6,9,10$, located at opposite points always preserve the same set of supercharges.

Therefore, organizing the 12+12 (anti)string configurations in terms of the corresponding preserved supercharges, we find twelve pairs of fundamental string/anti--string solutions, such that each pair preserves the same set of supercharges. There is no overlapping of preserved supercharges between different pairs.
These pairs are in one-to-one correspondence with the twelve pairs of 1/2 BPS WLs $(W_I^{(1)}, W_I^{(2)})$ and  $(\td W_I^{(1)}, \td W_I^{(2)})$ discussed in section 3.1.

In conclusion, each 1/2 BPS operator can be obtained by two different Higgsing procedures in CFT, which in the dual description correspond to localize one fundamental string at some point in $\textrm{S}^5$ and one fundamental anti--string at the opposite point.

\section{ABJM theory}\label{sec4}

In the same spirit of the previous section, we now apply the Higgsing procedure in ABJM theory \cite{Aharony:2008ug} to build two different sets of 1/2 BPS WLs by assigning vev to different scalars and/or exciting different massive modes. Moreover, in the dual $\gabjm$ description we identify the corresponding  M2-- and anti--M2--brane solutions wrapping the M--theory circle and being localized
at different positions in the compact space. Both in field theory and in the dual constructions we discuss the spectra of preserved supercharges and their possible overlapping.

\vskip 10pt

The field content of $U(N)_k\times U(N)_{-k}$ ABJM theory is given by two gauge fields $A_\m$ and $B_\m$, four complex scalars $\phi^I$ and four Dirac fermions $\psi^I$, $I=1,2,3,4$, in the bi--fundamental representation $(N,\bar N)$ of the gauge group. The corresponding hermitian conjugates $\bar\phi_I=(\phi^I)^\dagger$ and $\bar\psi_I=(\psi^I)^\dagger$ belong to the bi--fundamental representation $(\bar N,N)$.

The ABJM lagrangian in components can be written as the sum of four terms
\bea \label{labjm}
&& \hspace{-2mm}
   \mL_{CS} = \f{k}{4\pi}\ve^{\m\n\r}\Tr \Big( A_\m\p_\n A_\r +\f{2\ii}{3}A_\m A_\n A_\r
                                             -B_\m\p_\n B_\r -\f{2\ii}{3}B_\m B_\n B_\r  \Big) \nn\\
&& \hspace{-2mm}
   \mL_k = \Tr( -D_\m\bar\phi^I D^\m\phi_I + \ii\bar\psi_I\bg^\m D_\m\psi^I ) \nn\\
&& \hspace{-2mm}
   \mL_p =\f{4\pi^2}{3k^2} \Tr(    \phi_I\bar\phi^I\phi_J\bar\phi^J\phi_K\bar\phi^K
                                +  \phi_I\bar\phi^J\phi_J\bar\phi^K\phi_K\bar\phi^I
                                +4 \phi_I\bar\phi^J\phi_K\bar\phi^I\phi_J\bar\phi^K
                                -6 \phi_I\bar\phi^J\phi_J\bar\phi^I\phi_K\bar\phi^K )  \nn\\
&& \hspace{-2mm}
   \mL_Y = \f{2\pi\ii}{k} \Tr(    \phi_I\bar\phi^I\psi^J\bar\psi_J
                               -2 \phi_I\bar\phi^J\psi^I\bar\psi_J
                               -  \bar\phi^I\phi_I\bar\psi_J\psi^J
                               +2 \bar\phi^I\phi_J\bar\psi_I\psi^J\\
&& \hspace{-2mm}
   \phantom{\mL_Y = \f{2\pi\ii}{k} \Tr(}
                               +\ve^{IJKL}\phi_I\bar\psi_J\phi_K\bar\psi_L
                               -\ve_{IJKL}\bar\phi^I\psi^J\bar\phi^K\psi^L )\nn
\eea
where $\ve^{IJKL}$, $\ve_{IJKL}$ are the totally anti-symmetric Levi--Civita tensors in four dimensions ($\ve^{1234}=\ve_{1234}=1$) and the  covariant derivatives are given by
\bea
&& D_\m\phi_I =\p_\m \phi_I + \ii A_\m \phi_I - \ii \phi_I B_\m \nn\\
&& D_\m\bar\phi^I =\p_\m \bar\phi^I - \ii \bar\phi^I A_\m + \ii B_\m \bar\phi^I  \nn\\
&& D_\m\psi^I =\p_\m \psi^I + \ii A_\m \psi^I - \ii \psi^I B_\m
\eea
The ABJM action is invariant under the following SUSY transformations \cite{Gaiotto:2008cg,Hosomichi:2008jb,Terashima:2008sy,Bandres:2008ry}
\bea \label{j68}
&& \d A_\m=-\f{2\pi}{k} \lt( \phi_I\bar\psi_J\bg_\m\e^{IJ} +\bar\e_{IJ}\bg_\m\psi^J\bar\phi^I \rt) \nn\\
&& \d B_\m=-\f{2\pi}{k} \lt( \bar\psi_J\phi_I\bg_\m\e^{IJ}+\bar\e_{IJ}\bg_\m\bar\phi^I\psi^J \rt) \nn\\
&& \d\phi_I=\ii\bar\e_{IJ}\psi^J, ~~ \d\bar\phi^I=\ii\bar\psi_J\e^{IJ}\\
&& \d\psi^I=\bg^\m\e^{IJ}D_\m\phi_J + \vth^{IJ}\phi_J
            +\f{2\pi}{k}\e^{IJ} \lt( \phi_J\bar\phi^K\phi_K-\phi_K\bar\phi^K\phi_J \rt)
            +\f{4\pi}{k}\e^{KL}\phi_K\bar\phi^I\phi_L \nn\\
&& \d\bar\psi_I=-\bar\e_{IJ}\bg^\m D_\m\bar\phi^J + \bar\vth_{IJ}\bar\phi^J
                -\f{2\pi}{k}\bar\e_{IJ} \lt( \bar\phi^J\phi_K\bar\phi^K-\bar\phi^K\phi_K\bar\phi^J \rt)
                -\f{4\pi}{k}\bar\e_{KL}\bar\phi^K\phi_I\bar\phi^L \nn
\eea
with the definitions $\e^{IJ}=\th^{IJ}+x^\m\bg_\m\vth^{IJ}$, $\bar\e_{IJ}=\bar\th_{IJ}-\bar\vth_{IJ} x^\m\bg_\m$, and $\th^{IJ}$ and $\vth^{IJ}$ denoting Poincar\'e and conformal supercharges respectively.
The SUSY parameters satisfy
\bea \label{e101}
&& \th^{IJ}=-\th^{JI}, ~~ (\th^{IJ})^*=\bar \th_{IJ}, ~~ \bar\th_{IJ}=\f{1}{2}\e_{IJKL}\th^{KL} \nn\\
&& \vth^{IJ}=-\vth^{JI}, ~~ (\vth^{IJ})^*=\bar \vth_{IJ}, ~~ \bar\vth_{IJ}=\f{1}{2}\e_{IJKL}\vth^{KL}
\eea

\vskip 10pt

\subsection{1/2 BPS Wilson loops} \label{WLABJM}

As in \cite{Drukker:2009hy}, one can construct the 1/2 BPS WLs along the straight line $\Gamma: x^\mu = \tau \delta^\mu_0$
\be
\label{WLdefinition}
W_I = \mP \exp\Big( -\ii \int_\Gamma d\t L_I(\t) \Big), ~~  I=1,2,3,4
\ee
as the holonomy of the superconnection\footnote{We use spinor decompositions (\ref{spinor1}) and (\ref{spinor2}). }
\be \label{L1}
L_I = \lt(\ba{cc} \mA &  \sqrt{\f{4\pi}{k}} \psi^I_+ \\ \sqrt{\f{4\pi}{k}} \bar \psi_{I-} & \mB \ea\rt) , ~~ \qquad
\ba{l} \mA=A_0 - \f{2\pi}{k}(\phi_I\bar\phi^I -\phi_i\bar\phi^i) \\
       \mB=B_0 - \f{2\pi}{k}(\bar\phi^I\phi_I - \bar\phi^i\phi_i) \ea
\ee
In the above formula the $I$ index is fixed and there is summation for index $i\neq I$.
The corresponding preserved Poincar\'e supercharges are (note that $\th^{IJ}$ and $\bar\th_{IJ}$ are not linearly independent)
\be \label{j3}
\th^{Ij}_+, ~ \th^{ij}_-, ~ \bar\th_{Ij-}, ~ \bar\th_{ij+} ~~\quad   i,j \neq I
\ee
For BPS WLs along infinite straight lines, the preserved Poincar\'e and conformal supercharges are similar, and in this paper we just consider the Poincar\'e supercharges, and the conformal supercharges can be inferred easily.
Due to relations (\ref{e101}), the preserved supercharges can be equivalently written  as
\be
\th^{Ij}_+, ~ \th^{ij}_-
~~\quad {\textrm{or}} ~~\quad
\th^{Ij}_+,  ~ \bar\th_{Ij-} ~~\quad   i,j \neq I
\ee
$W_I$ operators are class II 1/2 BPS WLs in \cite{Ouyang:2015iza,Ouyang:2015bmy}, up to some R--symmetry rotations.

Similarly, there are 1/2 BPS WLs $\td W_I$ still defined as in (\ref{WLdefinition}) but with superconnection
\be\label{tdL1}
\td L_I = \lt(\ba{cc} \td \mA & \sqrt{\f{4\pi}{k}} \psi^I_- \\ -\sqrt{\f{4\pi}{k}} \bar \psi_{I+} & \td \mB \ea\rt) , ~~ \qquad
\ba{l} \td \mA=A_0+\f{2\pi}{k}(\phi_I\bar\phi^I - \phi_i\bar\phi^i) \\
       \td \mB=B_0+\f{2\pi}{k}(\bar\phi^I\phi_I - \bar\phi^i\phi_i) \ea
\ee
They preserve the complementary set of Poincar\'e supercharges
\be
\th^{Ij}_-, ~ \th^{ij}_+, ~ \bar\th_{Ij+}, ~ \bar\th_{ij-} ~~ \quad  i,j \neq I
\ee
$\td W_I$ operators correspond to class I 1/2 BPS WLs in the classification of \cite{Ouyang:2015iza,Ouyang:2015bmy}, up to some R--symmetry rotations.

\begin{figure}[htbp]
  \centering
  \subfigure[Overlapping supercharges of 1/2 BPS Wilson loops]
            {\includegraphics[height=0.4\textwidth]{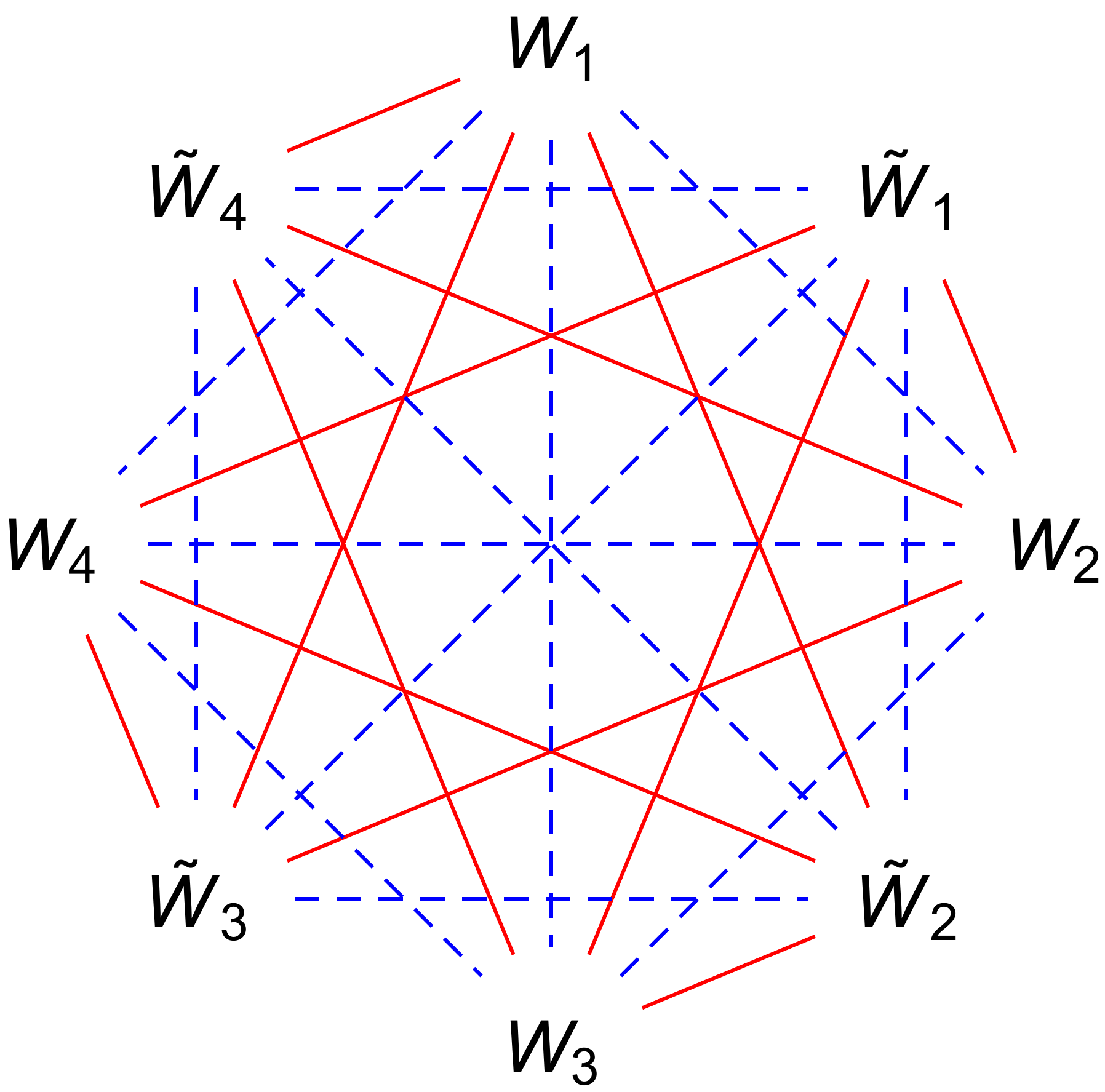}\label{f1a}}
            ~~~~~~
  \subfigure[Overlapping supercharges of M2-- and anti--M2--branes]
            {\includegraphics[height=0.4\textwidth]{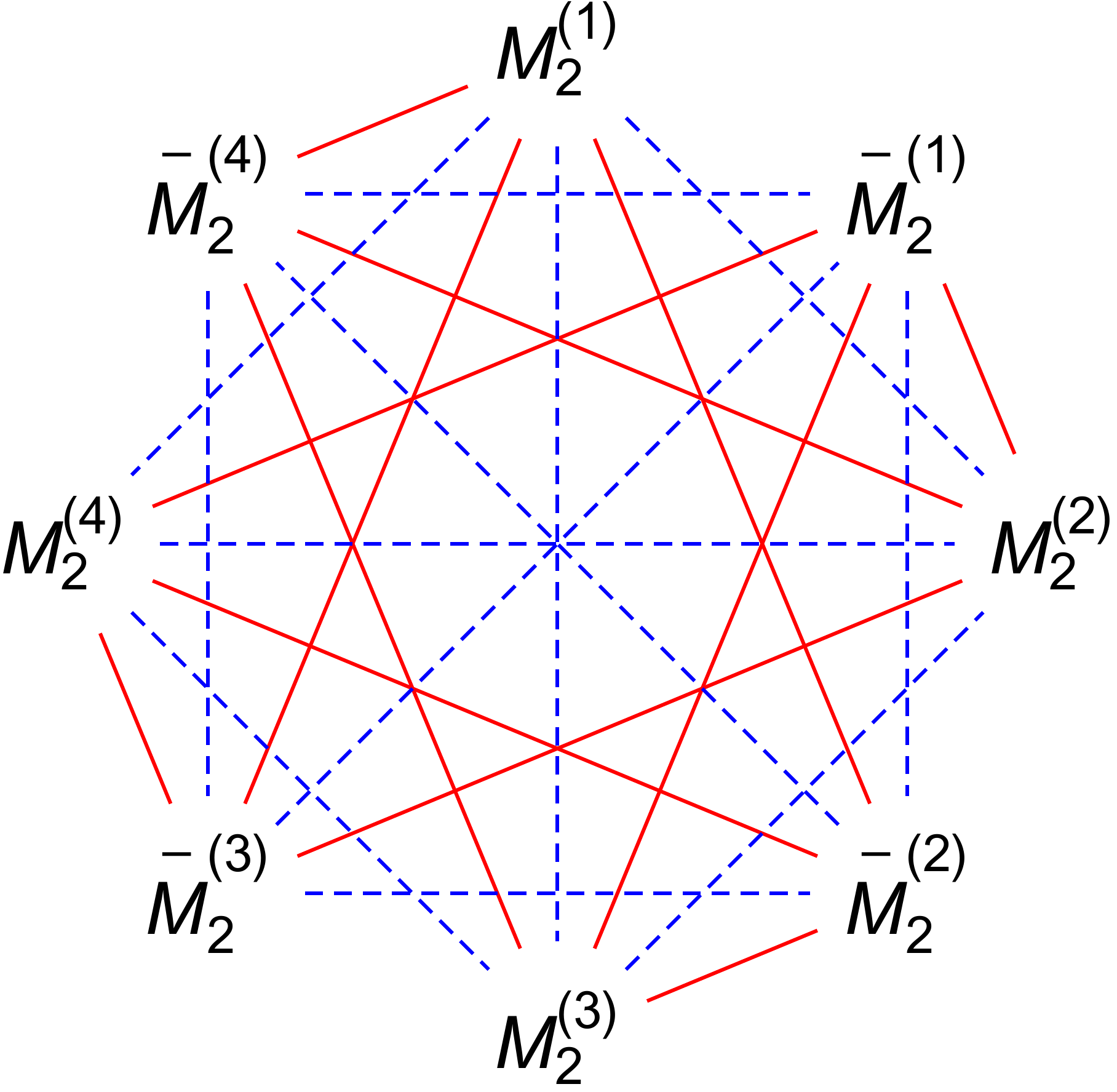}\label{f1b}}\\
  \caption{Amount of overlapping supercharges between each pair of 1/2 BPS WLs in ABJM theory and between each pair of M2-- and anti--M2--branes in $\gabjm$. A red solid line means that the two WLs or branes share 2/3 of preserved supercharges. A blue dashed line means that the two WLs or branes share 1/3 of preserved supercharges. Two WLs or branes that are not directly connected by any line have no common preserved supercharges.}
\end{figure}

Table \ref{tab2} summarizes the ``natural'' representatives of the two classes of 1/2 BPS WLs and their preserved supercharges.
Each WL preserves six real Poincar\'e plus six real superconformal charges, and so a total of twelve real supercharges. $W_I$ and $\tilde{W}_I$ for a fixed $I$--index preserve complementary sets of supercharges.
It is important to note that there are non--trivial overlappings among the supercharges preserved by different WLs. Precisely, any couple of WLs in the same set, let's say $W_I$ and $W_J$ with $I \neq J$, always share two Poincar\'e supercharges $\theta_+^{IJ}, \e_{IJKL}\theta_-^{KL}$.  Operators belonging to two different sets, $\tilde{W}_I$ and $W_{J \neq I}$, share four Poincar\'e supercharges $\theta_-^{IK}, \theta_+^{JK}$, $K\neq I,J$. The amount of the overlapping supercharges between each pair of WLs are shown in Figure~\ref{f1a}.

\begin{table}[htbp]
  \centering
  \begin{tabular}{|c|c|}\hline
   Wilson loop & preserved supercharges \\ \hline
   $W_1$       & $\th^{12}_+$, $\th^{13}_+$, $\th^{14}_+$, $\th^{23}_-$, $\th^{24}_-$, $\th^{34}_-$ \\ \hline
   $\td W_1$   & $\th^{12}_-$, $\th^{13}_-$, $\th^{14}_-$, $\th^{23}_+$, $\th^{24}_+$, $\th^{34}_+$ \\ \hline

  $W_2$       & $\th^{12}_+$, $\th^{23}_+$, $\th^{24}_+$, $\th^{13}_-$, $\th^{14}_-$, $\th^{34}_-$ \\ \hline
  $\td W_2$   & $\th^{12}_-$, $\th^{23}_-$, $\th^{24}_-$, $\th^{13}_+$, $\th^{14}_+$, $\th^{34}_+$ \\ \hline

  $W_3$       & $\th^{13}_+$, $\th^{23}_+$, $\th^{34}_+$, $\th^{12}_-$, $\th^{14}_-$, $\th^{24}_-$ \\ \hline
  $\td W_3$   & $\th^{13}_-$, $\th^{23}_-$, $\th^{34}_-$, $\th^{12}_+$, $\th^{14}_+$, $\th^{24}_+$ \\ \hline

  $W_4$       & $\th^{14}_+$, $\th^{24}_+$, $\th^{34}_+$, $\th^{12}_-$, $\th^{13}_-$, $\th^{23}_-$ \\ \hline
  $\td W_4$   & $\th^{14}_-$, $\th^{24}_-$, $\th^{34}_-$, $\th^{12}_+$, $\th^{13}_+$, $\th^{23}_+$ \\ \hline

  \end{tabular}
  \caption{The 1/2 BPS WLs in ABJM theory and the preserved Poincar\'e supercharges. We have not included $\bar\th_{IJ}$ supercharges, since they are not linearly independent from $\th^{IJ}$.}\label{tab2}
\end{table}

The 1/2 BPS WLs introduced above are special cases of a general 1/2 BPS Wilson loop $W$ with superconnection
\be \label{z128}
L = \lt( \ba{cc}
A_0 + \f{2\pi}{k} (\d^I_J - 2\a^I\bar\a_J) \phi_I\bar\phi^J & \sqrt{\f{4\pi}{k}} \bar\a_I\psi^I_+ \\
\sqrt{\f{4\pi}{k}}\bar\psi_{I-} \a^I & B_0 + \f{2\pi}{k} (\d^I_J - 2\a^I\bar\a_J) \bar\phi^J\phi_I
\ea \rt)
\ee
where $\bar\a_I=(\bar\a_1,\bar\a_2,\bar\a_3,\bar\a_4)$, $\a^I\equiv(\bar\a_I)^*$, $|\a|^2\equiv\bar\a_I\a^I=1$. The corresponding preserved Poincar\'e and conformal supercharges are
\be \label{z13}
\bar\a_I\th^{IJ}_+, ~~ \a^I \bar\th_{IJ-}, ~~ \bar\a_I\vth^{IJ}_+, ~~ \a^I \bar\vth_{IJ-}
\ee
Similarly, we can introduce a 1/2 BPS Wilson loop $\td W$ with superconnection
\be \label{z129}
\td L = \lt( \ba{cc}
A_0 + \f{2\pi}{k} (- \d^I_J+2\a^I\bar\a_J) \phi_I\bar\phi^J & \sqrt{\f{4\pi}{k}} \bar\a_I\psi^I_- \\
-\sqrt{\f{4\pi}{k}}\bar\psi_{I+} \a^I & B_0 + \f{2\pi}{k} (-\d^I_J+2\a^I\bar\a_J) \bar\phi^J\phi_I
\ea \rt)
\ee
where $|\a|^2=1$, and preserved supercharges
\be \label{z14}
\bar\a_I\th^{IJ}_-, ~~ \a^I \bar\th_{IJ+}, ~~ \bar\a_I\vth^{IJ}_-, ~~ \a^I \bar\vth_{IJ+}
\ee
When $\bar\a_I = \d_I^J$ with fixed $J=1,2,3,4$, the general operator $W$ coincides with $W_J$ in (\ref{L1}), while $\td W$ is exactly $\td W_J$ in (\ref{tdL1}).

\subsection{Wilson loops from Higgsing}\label{sec4.2}

The Higgs construction of fermionic 1/2 BPS WL in ABJM theory has been proposed in \cite{Lee:2010hk}. We now review this construction by generalizing it in order to obtain both $W$ and $\tilde{W}$ kinds of operators.

We start with $U(N+1)\times U(N+1)$ ABJM theory and break it to $U(N)\times U(N)$ by choosing the following field configurations
\bea \label{abphipsi}
&& A_\m = \lt(\ba{cc} A_\m & W_\m \\ \bar W_\m & 0 \ea\rt),      ~~
   B_\m = \lt(\ba{cc} B_\m & Z_\m \\ \bar Z_\m  & 0 \ea\rt)     \nn\\
&& \phi_J = \lt(\ba{cc} \phi_J & R_J \\ \bar S_J & v_J \ea\rt),  ~~
   \bar\phi^J = \lt(\ba{cc} \bar\phi^J & S^J \\ \bar R^J & \bar v^J \ea\rt) \nn\\
&& \psi^J = \lt(\ba{cc} \psi^J & \O^J \\ \bar\S^J & 0 \ea\rt),  ~~
   \bar\psi_J = \lt(\ba{cc} \bar\psi_J & \S_J \\ \bar\O_J & 0 \ea\rt)
\eea
with $v_J=v\bar\a_J$, $|\a|^2=1$, $v>0$. In principle, we could perform Higgsing along this general direction. However, in order to be definite and avoid clutter of symbols, we consider the special case $\bar\a_I=\d_I^1$.

It is convenient to work in the unitary gauge where we set $R_1=\bar R^1=S^1=\bar S_1=0$. We are then left with three-dimensional massive $\mN=3$ vector multiplets
\be
\label{j32}
(W_\m, \O^i, R_i, \O^1) ~ {\rm  and}  ~ (Z_\m, \S_i, S^i, \S_1), ~~  i = 2,3,4
 \ee
with mass $m = \f{2\pi v^2}{k}$.

Inserting this ansatz into the ABJM lagrangian and taking the limit $v\to\inf$ the terms relevant for the dynamics of the massive particles can be organized as
\be
\label{leeaabjm}
\mL = \mL_v + \mL_s + \mL_{f1}+\mL_{f2}+\mL_{f3}
\ee
where we have defined
\bea \label{leeaabjm1}
&& \mL_v =  \f{k}{2\pi}\e^{\m\n\r}( \bar W_\m D_\n W_\r - \bar Z_\m D_\n Z_\r )
           -\bar W_\m (v^2+\phi_I\bar\phi^I)W^\m - \bar Z_\m (v^2+\bar\phi^I\phi_I)Z^\m \nn \\
&& \phantom{\mL_v =}
           +2v\bar W_\m\phi_1Z^\m +2v\bar Z_\m\bar\phi^1W^\m
\eea
\bea \label{leeaabjm2}
&& \mL_s = -D_\m\bar R^i D^\m R_i - D_\m\bar S_i D^\m S^i - \f{4\pi^2v^4}{k^2} (\bar R^i R_i + \bar S_i S^i)\\
&&       -\f{4\pi^2v^2}{k^2}\big( 2\bar R^i (-\phi_1\bar\phi^1+\phi_j\bar\phi^j)R_i - \bar R^i\phi_i\bar\phi^j R_j
                                   +2\bar S_i (-\bar\phi^1\phi_1+\bar\phi^j\phi_j)S^i - \bar S_i\bar\phi^i\phi_j S^j \big) \nn
\eea
\bea \label{leeaabjm3}
\hspace{-0.8cm}
&& \hspace{-0.2cm} \mL_{f1}=\ii\bar\O_1\bg^\m D_\m\O^1 + \f{2\pi\ii v^2}{k}\bar\O_1\O^1
            +\f{2\pi\ii}{k}\bar\O_1 (-\phi_1\bar\phi^1+\phi_i\bar\phi^i)\O^1
            +\bar\O_I\bg^\m\psi^I Z_\m + \bar Z_\m \bar\psi_I\bg^\m\O^I \nn\\
&&     \hspace{-0.4cm}   +\ii\bar\S^1\bg^\m D_\m\S_1 - \f{2\pi\ii v^2}{k}\bar\S^1\S_1
            -\f{2\pi\ii}{k}\bar\S^1 (-\bar\phi^1\phi_1+\bar\phi^i\phi_i)\S_1
            +\bar\S^I\bg^\m\bar\psi_I W_\m + \bar W_\m \psi^I\bg^\m\S_I
\eea
\bea \label{leeaabjm4}
&& \mL_{f2}=\ii\bar\O_i\bg^\m D_\m\O^i - \f{2\pi\ii v^2}{k}\bar\O_i\O^i
            +\f{2\pi\ii}{k}( \bar\O_i\phi_J\bar\phi^J\O^i - 2\bar\O_i\phi_j\bar\phi^i\O^j
                            +2v\bar\O_i\psi^1S^i+2v\bar S_i\bar\psi_1\O^i )\nn\\
&&       +\ii\bar\S^i\bg^\m D_\m\S_i + \f{2\pi\ii v^2}{k}\bar\S^i\S_i
            -\f{2\pi\ii}{k}( \bar\S^i \bar\phi^J\phi_J\S_i - 2\bar\S^i \bar\phi^j\phi_i\S_j
                            +2v\bar\S^i\bar\psi_1R_i + 2v\bar R^i\psi^1\S_i ) \nn\\
&&        +\f{4\pi\ii v}{k} ( \ve^{ijk}\bar\O_i\phi_j\S_ k - \ve_{ijk}\bar\S^i\bar\phi^j\O^k )
\eea
\bea \label{leeaabjm5}
&& \mL_{f3}=-\f{4\pi\ii}{k} ( \bar\O_1\phi_i\bar\phi^1\O^i + \bar\O_i\phi_1\bar\phi^i\O^1
                             -\bar\S^1\bar\phi^i\phi_1\S_i - \bar\S^i\bar\phi^1\phi_i\S_1 )
\eea

Before choosing the non-relativistic modes, one has to redefine the subleading orders of fields \cite{Lee:2010hk}. There is some freedom in doing it. We choose the following field redefinitions
\bea
&& W_\m \to \Big( 1+\f{\phi_1\bar\phi^1}{2v^2} \Big)W_\m + \f{\phi_1}{v}Z_\m \quad , \quad
   Z_\m \to \Big( 1+\f{\bar\phi^1\phi_1}{2v^2} \Big)Z_\m + \f{\bar\phi^1}{v}W_\m                \nn\\
&& R_i \to R_i + \f{\phi_i\bar\phi^j}{2v^2}R_j \quad , \quad
   S^i \to S^i + \f{\bar\phi^i\phi_j}{2v^2}S^j                                                   \nn\\
&& \O^1 \to \O^1 \quad , \quad
   \O^i \to \O^i + \f{1}{v}\ve^{ijk}\phi_j\S_k + \f{1}{2v^2}\phi_j(\bar\phi^i\O^j-\bar\phi^j\O^i)  \nn\\
&& \S_1 \to \S_1 \quad , \quad
   \S_i \to \S_i + \f{1}{v}\ve_{ijk}\bar\phi^j\O^k + \f{1}{2v^2}\bar\phi^j(\phi_i\S_j-\phi_j\S_i)
\eea
which are slightly different from but equivalent to that in \cite{Lee:2010hk}.

As for the ${\cal N}=4$ SYM theory, we can now choose the modes either corresponding to particle or antiparticle excitations.
Exciting particles amounts to choose
\bea\label{ABJMpartciles}
W_\m = \sqrt{\f{\pi}{k}}(0,1,-\ii) w\,  \ep^{-\ii mt} \quad ,     && Z_\m = \sqrt{\f{\pi}{k}}(0,1,\ii) z \,\ep^{-\ii mt} \nn\\
\O^i = u_- \o^i \, \ep^{-\ii mt} \quad ,                          && \S_i = u_+ \s_i \, \ep^{-\ii mt} \nn\\
R_i  = \sqrt{\f{k}{4\pi}} \f{1}{v} r_i \, \ep^{-\ii mt} \quad ,   && S^i  = \sqrt{\f{k}{4\pi}} \f{1}{v} s^i \, \ep^{-\ii mt} \nn\\
\O^1 = u_+ \o^1 \, \ep^{-\ii mt} \quad ,                          && \S_1 = u_- \s_1 \, \ep^{-\ii mt}
\eea
where $u_\pm$ are bosonic spinors defined in (\ref{bspinors}).

The Higgsing procedure breaks half of the supersymmetries. The non-relativistic mode excitations organize themselves in ${\cal N}=3$ SUSY multiplets as follows \cite{Lee:2010hk}
\begin{center}\begin{tabular}{r|c|c|c|ccr|c|c|c|c}
  spin       & 1   & 1/2          & 0           & $-1/2$ && spin       & $-1$ & $-1/2$       & 0           & 1/2 \\ \cline{1-5}\cline{7-11}
  degeneracy & 1   & 3            & 3           & 1      && degeneracy & 1    & 3            & 3           & 1   \\ 
  mode       & $w$ & $\o^{2,3,4}$ & $r_{2,3,4}$ & $\o^1$ && mode       & $z$  & $\s_{2,3,4}$ & $s^{2,3,4}$ & $\s_1$
\end{tabular}\end{center}

\vskip 7pt
Inserting expressions (\ref{ABJMpartciles}) in the previous lagrangian, after some long but straightforward calculation, we obtain the non-relativistic lagrangian
\bea
&& \mL = \ii (  \bar w \mD_0 w + \bar r^i \mD_0 r_i + \bar \o_I \mD_0 \o^I
               +\bar z \mD_0 z + \bar s_i \mD_0 s^i + \bar \s^I \mD_0 \s_I ) \\
&& +\sqrt{\f{4\pi}{k}}
(- \bar w \psi^1_+\s_1 - \bar \o_1 \psi^1_+z  - \bar \s^1\bar \psi_{1-}w - \bar z\bar\psi_{1-}\o^1
 + \bar r^i \psi^1_+\s_i + \bar \o_i \psi^1_+s^i  + \bar \s^i\bar \psi_{1-}r_i + \bar s_i\bar\psi_{1-}\o^i ) \nn
\eea
where for $w$, $r_i$ and $\o^I$ we have defined $\mD_0=\p_0+\ii\mA$, whereas for $z$, $s^i$ and $\s_I$ we have $\mD_0=\p_0+\ii\mB$, with $\mA$ and $\mB$ defined in (\ref{L1}) and acting on the left. Defining
\bea \label{j7}
&& \Psi_1 = \lt(\ba{cc} w & \o^1 \\ \s_1 & z \ea\rt) ,  ~~
  \bar\Psi^1 = \lt(\ba{cc} \bar w & \bar\s^1 \\ \bar\o_1 & \bar z \ea\rt) \nn\\
&& \Psi_i = \lt(\ba{cc} r_i & -\o^i \\ -\s_i & s^i \ea\rt), ~~
   \bar\Psi^i = \lt(\ba{cc} \bar r^i & -\bar\s^i \\ -\bar\o_i & \bar s_i \ea\rt)
\eea
the previous result can be written in the compact form
\be
\mL = \ii \Tr \bar\Psi^I \mf D_0 \Psi_I
\ee
with $\mf D_0 \Psi_I = \p_0 \Psi_I + \ii L_1 \Psi_I$ and $L_1$ being just the connection (\ref{L1}).

Similarly, we can choose antiparticle modes
\bea \label{ABJMantiparticles}
W_\m = \sqrt{\f{\pi}{k}}(0,1,\ii) w \, \ep^{\ii mt} \quad  ,     && Z_\m = \sqrt{\f{\pi}{k}}(0,1,-\ii) z \, \ep^{\ii mt} \nn\\
\O^i = u_+ \o^i \, \ep^{\ii mt} \quad ,                          && \S_i = u_- \s_i \, \ep^{\ii mt} \nn\\
R_i  = \sqrt{\f{k}{4\pi}} \f{1}{v} r_i \, \ep^{\ii mt} \quad ,   && S^i  = \sqrt{\f{k}{4\pi}} \f{1}{v} s^i \, \ep^{\ii mt} \nn\\
\O^1 = u_- \o^1 \, \ep^{\ii mt} \quad  ,                          && \S_1 = u_+ \s_1  \,\ep^{\ii mt}
\eea
Inserting these expressions in lagrangian (\ref{leeaabjm}), in the limit $v \to \infty$ we obtain
\bea
&& \mL = \ii \Tr(  w \mD_0 \bar w + r_i\mD_0\bar r^i + \o^I\mD_0\bar \o_I
               + z \mD_0 \bar z + s^i \mD_0 \bar s_i + \s_I\mD_0 \bar \s^I ) \\
&&   +\sqrt{\f{4\pi}{k}}\Tr
(  \s_1\bar w \psi^1_- - z\bar \o_1 \psi^1_-  - w\bar \s^1\bar \psi_{1+} + \o^1\bar z\bar\psi_{1+}
 + \s_i\bar r^i \psi^1_- - s^i\bar \o_i \psi^1_-  - r_i\bar \s^i\bar \psi_{1+} + \o^i\bar s_i\bar\psi_{1+} ) \nn
\eea
where we have defined $\mD_0=\p_0-\ii\td\mA$ for $\bar w$, $\bar r^i$ and $\bar\o_I$ , and $\mD_0=\p_0-\ii\td\mB$ for $\bar z$, $\bar s_i$ and $\bar \s^I$ , with $\td\mA$ and $\td\mB$ given in (\ref{tdL1}) and acting on the right. With definitions
\bea \label{j8}
&& \Psi_1 = \lt(\ba{cc} w & -\o^1 \\ \s_1 & z \ea\rt), ~~
   \bar\Psi^1 = \lt(\ba{cc} \bar w & \bar\s^1 \\ -\bar\o_1 & \bar z \ea\rt) \nn\\
&& \Psi_i = \lt(\ba{cc} r_i & -\o^i \\ \s_i & s^i \ea\rt) , ~~
   \bar\Psi^i = \lt(\ba{cc} \bar r^i & \bar\s^i \\ -\bar\o_i & \bar s_i \ea\rt)
\eea
the previous result can be written in the following compact form
\be
\mL = \ii \Tr \Psi_I \mf D_0 \bar\Psi^I
\ee
with $\mf D_0 \bar\Psi^I = \p_0 \bar\Psi^I - \ii \bar\Psi^I \td L_1 $, and $\td L_1$ being exactly the connection in (\ref{tdL1}).

Applying the same procedure with $v_J = v \d_J^i, i=2,3,4$, or equivalently applying R--symmetry rotations, we generate all $W_I$ and $\tilde{W}_I$ previously defined in section~\ref{WLABJM}. Furthermore, Higgsing in the general direction with $v_J = v \bar\a_J$ we could get the general 1/2 BPS Wilson loops $W$ and $\td W$ corresponding to superconnections (\ref{z128}) and (\ref{z129}), respectively.

\vskip 10pt
An analogue procedure can be used to construct 1/2 BPS WLs in the more general $U(N)_k \times U(M)_{-k}$ Aharony-Bergman-Jafferis (ABJ) theory with $N \neq M$ \cite{Hosomichi:2008jb,Aharony:2008gk}. The general structure of the operators is still the one in (\ref{WLdefinition}), (\ref{L1}), (\ref{tdL1}) with the matter fields now in the bi--fundamental representation of $U(N) \times U(M)$. The Higgsing procedure works exactly as for the ABJM theory and we can classify WLs in two main sets, depending whether we excite particle or antiparticle modes. The configuration of preserved supercharges can be still read in table \ref{tab2} and the overlapping of the preserved supercharges can be seen in Figure \ref{f1a}.

\subsection{M2--branes in AdS$_4\times$S$^7$/Z$_k$ spacetime}\label{sec4.3}

For the ABJM theory we now investigate the gravity dual of the Higgsing procedure by constructing different M2--brane embeddings that correspond to the previous 1/2 BPS WLs. In particular, we will be interested in classifying M2--brane configurations in terms of their sets of preserved supercharges.

ABJM theory is dual to M--theory on the $\gabjm$ background with self-dual  four-form flux, described by
\bea\label{ads4s7zk}
&& ds^2=R^2 \lt( \frac14 ds^2_{\AdS_4}+ds^2_{ {\rm S}^7/\Z_k} \rt) \nn \\
&& F_{\td\m\td\n\td\r\td\s} = \f{6}{R} \ve_{\td\m\td\n\td\r\td\s}
\eea
with $\ve_{\td\m\td\n\td\r\td\s}$ being the $\AdS_4$ volume form.

We use the $\AdS_4$ metric in the form
\be \label{ads4}
ds^2_{\AdS_4}= u^2 (-dt^2 + dx_1^2 + dx_2^2) + \f{du^2}{u^2}
\ee
whereas, in order to write the unit $\rm{S}^7$ metric, following \cite{Drukker:2008zx}, we embed it in ${\rm{C}^4}\cong\rm{R}^8$ with coordinates $z_i$, $i=1,2,3,4$, parametrized as
\bea \label{z1234}
&& z_1 = \cos\frac{\b}{2}\cos\frac{\theta_1}{2} \ep^{\ii\xi_1}, ~~
   \xi_1 = -\frac14(2\phi_1+\chi+\zeta)                              \nn\\
&& z_2 = \cos\frac{\b}{2}\sin\frac{\theta_1}{2} \ep^{\ii\xi_2}, ~~
   \xi_2 = -\frac14(-2\phi_1+\chi+\zeta)                             \nn\\
&& z_3 = \sin\frac{\b}{2}\cos\frac{\theta_2}{2} \ep^{\ii\xi_3}, ~~
   \xi_3 = -\frac14(2\phi_2-\chi+\zeta)                              \nn\\
&& z_4 = \sin\frac{\b}{2}\sin\frac{\theta_2}{2} \ep^{\ii\xi_4}, ~~
   \xi_4 = -\frac14(-2\phi_2-\chi+\zeta)
\eea
with $\b,\th_{1,2} \in [0, \pi]$, $\xi_{1,2,3,4} \in [0,2\pi]$, and so $\phi_{1,2} \in [0, 2\pi]$, $\chi\in [0, 4\pi]$, $\zeta \in [0, 8\pi]$. The $\z$ direction is the M--theory circle.
The metric of unit $\rm{S}^7$ is then
\bea \label{s7}
&& ds^2_{\rm{S}^7}=\frac14 \Big[ d\b^2+\cos^2\frac{\b}{2} \big( d\theta_1^2+\sin^2\theta_1 d\varphi_1^2 \big)
                                +\sin^2\frac{\b}{2} \big( d\theta_2^2+\sin^2\theta_2 d\varphi_2^2 \big) \nn\\
&& \phantom{ds^2_{S^7}=}
+\sin^2\frac{\b}{2}\cos^2\frac{\b}{2} (d\chi+\cos\theta_1d\varphi_1-\cos\theta_2d\varphi_2)^2 \nn\\
&& \phantom{ds^2_{S^7}=}
+ \Big( \frac12d\zeta+\cos^2\frac{\b}2\cos\theta_1d\varphi_1+\sin^2\frac{\b}2\cos\theta_2d\varphi_2+\frac12\cos\b d\chi \Big)^2 \Big]
\eea
The quotient space ${\rm S}^7/\Z_k$ is generated by the identification $z_i \sim \exp\Big( \f{2\pi\ii}{k} \Big) z_i$, or equivalently
\be \label{quotient1}
\z \sim \z - \f{8\pi}{k}
\ee

We now study M2-- and anti--M2--brane configurations and the corresponding preserved supercharges.

In appendix~\ref{appD} we provide the Killing spinors of $\AdS_4\times\rmS^7$, eq. (\ref{e105}), in terms of two constant spinors $\e_1$ and $\e_2$ that can be decomposed in two different ways, one way given in (\ref{e106}), and the second way given in (\ref{z2}).
The first decomposition is more suitable when constructing explicitly M2-- and anti--M2--brane configurations that have the same properties of the 1/2 BPS Wilson loops $W_I$ and
$\td{W}_I$ obtained by Higgsing.
The second way is instead more suitable to perform the correct identification of Killing spinors in M--theory with Poincar\'e and conformal supercharges in field theory. It is also useful for identifying the supercharges preserved by the M2-- and anti--M2--branes at a general position with the supercharges preserved by the general 1/2 BPS Wilson loops $W$ and $\td{W}$ corresponding to superconnections (\ref{z128}, \ref{z129}). Therefore, it is worth analyzing the two decompositions separately.

For Killing spinors in $\gabjm$, the quotient (\ref{quotient1}) leads to
\be \label{z5}
(\g_{3\natural} + \g_{58} + \g_{47} + \g_{69})\e_1 = (\g_{3\natural} + \g_{58} + \g_{47} + \g_{69})\e_2=0
\ee
Decomposing $\e_1$ and $\e_2$ as in  (\ref{e106}) this constraint corresponds to
\be
s_1+s_2+s_3+s_4=0
\ee
so that only six of the eight states in (\ref{e106}) survive
\be \label{z7}
(s_1,s_2,s_3,s_4) = (++--), (+-+-), (+--+),
                    (-++-), (-+-+), (--++)
\ee
This is consistent with the fact that there are 24 real supercharges in ABJM theory, with 12 real Poincar\'e supercharges and 12 real conformal supercharges.

We want to realize M2--brane embeddings preserving half of the supersymmetries, which are dual to the 1/2 BPS WL operators $W_I$, $\td W_I$ constructed in section~\ref{sec4.2}.
To this end, we consider a M2--brane with coordinates  $(\sigma^{0}, \sigma^{1},\sigma^{2})$ embedded in the $\gabjm$ spacetime (\ref{ads4s7zk}) as
\be \label{embedding}
t=\sigma^0, ~~x_1=x_2=0, ~~ u=\sigma^1, ~~ \zeta=\sigma^2
\ee
and localized in $\rmS^7/{\Z}_k$.

In the presence of this M2--brane supersymmetry is broken by the condition \cite{Drukker:2008zx}
\be \label{z12}
\g_{03\natural} \e = \e ~ \implies ~
h^{-1} \g_{03\natural} h \e_1 = \e_1, ~~ h^{-1} \g_{03\natural} h \e_2 = \e_2
\ee
with $h$ being defined in (\ref{gh2}).
Explicitly, we have
\bea \label{z8}
&& h^{-1}\g_{03\natural}h = \cos^2\f{\b}{2}\Big( \g_{03\natural}\cos^2\f{\th_1}{2} + \g_{058}\sin^2\f{\th_1}{2} \Big)
                          + \sin^2\f{\b}{2}\Big( \g_{047}\cos^2\f{\th_2}{2} + \g_{069}\sin^2\f{\th_2}{2} \Big) \nn\\
&& ~~~
   +\cos^2\f{\b}{2}\cos\f{\th_1}{2}\sin\f{\th_1}{2}\big[ (\g_{038}+\g_{05\natural})\cos(\xi_1-\xi_2)
                                                       + (\g_{035}-\g_{08\natural})\sin(\xi_1-\xi_2) \big] \nn\\
&& ~~~
   +\sin^2\f{\b}{2}\cos\f{\th_2}{2}\sin\f{\th_2}{2}\big[ (\g_{049}+\g_{067})\cos(\xi_3-\xi_4)
                                                       + (\g_{046}+\g_{079})\sin(\xi_3-\xi_4) \big] \nn\\
&& ~~~
   +\cos\f\b2\sin\f\b2\Big\{ \cos\f{\th_1}{2}\cos\f{\th_2}{2}\big[ (\g_{037}+\g_{04\natural})\cos(\xi_1-\xi_3)
                                                                 + (\g_{034}-\g_{07\natural})\sin(\xi_1-\xi_3) \big] \nn\\
&& ~~~
                            +\cos\f{\th_1}{2}\sin\f{\th_2}{2}\big[ (\g_{039}+\g_{06\natural})\cos(\xi_1-\xi_4)
                                                                 + (\g_{036}-\g_{09\natural})\sin(\xi_1-\xi_4) \big] \nn\\
&& ~~~
                            +\sin\f{\th_1}{2}\cos\f{\th_2}{2}\big[ (\g_{048}+\g_{057})\cos(\xi_2-\xi_3)
                                                                 - (\g_{045}+\g_{078})\sin(\xi_2-\xi_3) \big] \nn\\
&& ~~~
                            +\sin\f{\th_1}{2}\sin\f{\th_2}{2}\big[ (\g_{059}+\g_{068})\cos(\xi_2-\xi_4)
                                                                 + (\g_{056}+\g_{089})\sin(\xi_2-\xi_4) \big] \Big\}
\eea

In general, a M2--brane localized in  ${\rm S}^7/\Z_k$ except for the M--theory circle is 1/2 BPS. In order to make the discussion more explicit, we consider four special configurations and classify the corresponding preserved supercharges.
\begin{enumerate}
\item[1)] For a M2--brane at position $|z_1|=1$, $z_{2,3,4}=0$ ($\b=\th_1=0$), we have
\be\label{c1}
\g_{03\natural}\e_1 = \e_1, ~~ \g_{03\natural}\e_2 = \e_2
\ee
which leads to the constraint $s_1=+$. According to (\ref{z7}), it preserves three states. The M2--brane is 1/2 BPS, and we call it $M_2^{(1)}$.
\item[2)] Similarly, for a M2--brane $M_2^{(2)}$ localized at $|z_2|=1$, $z_{1,3,4}=0$ ($\b=0,$ $\th_1=\pi$) we have
\be\label{c2}
\g_{058}\e_1 = \e_1, ~~ \g_{058}\e_2 =  \e_2
\ee
and this leads to the constraint $s_2=+$. This is still compatible with three states in (\ref{z7}). We call this 1/2 BPS solution $M_2^{(2)}$.
\item[3)] A M2--brane at position $|z_3|=1$, $z_{1,2,4}=0$ ($\b=\pi,$ $\th_2=0$) corresponds to the condition
\be\label{c3}
\g_{047}\e_1 = \e_1, ~~ \g_{047}\e_2 = \e_2
\ee
which is solved by $s_3=+$. We call this 1/2 BPS solution $M_2^{(3)}$.
\item[4)] Finally, we consider a M2--brane localized at $|z_4|=1$, $z_{1,2,3}=0$ ($\b=\th_2=\pi$), which gives
\be\label{c4}
\g_{069}\e_1 = \e_1, ~~ \g_{069}\e_2 = \e_2
\ee
This is solved  by $s_4=+$. We will call it $M_2^{(4)}$ solution.
\end{enumerate}

In addition, we can consider anti--M2--brane solutions. In the presence of an anti--M2--brane supersymmetry is broken by
\be
\g_{03\natural} \e = -\e ~ \implies ~
h^{-1} \g_{03\natural} h \e_1 = -\e_1, ~~ h^{-1} \g_{03\natural} h \e_2 = -\e_2
\ee
The classification of solutions works as before with all the plus signs on the r.h.s. of (\ref{c1}--\ref{c4}) replaced by minus signs.
We can then construct four 1/2 BPS anti--brane solutions localized at the same positions as the previous brane solutions. We will call these solutions $\bar{M}_2^{(I)}$, $I=1, 2,3,4$.

In table~\ref{tab3} we summarize the eight different M2--brane/anti--M2--brane solutions together with their positions and preserved supercharges, i.e., components of the Killing spinors.
It turns out that, while $M_2^{(I)}$ and $\bar M_2^{(I)}$ solutions always preserve complementary sets of supercharges, there are non--trivial overlappings of supercharges corresponding to M2-- and anti--M2--branes located at different points. The precise structure of this overlapping is shown in Figure~\ref{f1b}. Notably, this reproduces exactly the same configuration of overlappings for $W_I$ and $\td{W}_I$ WLs in ABJM theory, given in Figure~\ref{f1a}.
The fact that the two pictures are identical strongly supports the conjecture that $W_{1,2,3,4}$ operators are respectively dual to $M_2^{(1,2,3,4)}$ M2--branes, and $\td W_{1,2,3,4}$ WLs are dual to $\bar M_2^{(1,2,3,4)}$ anti--M2--branes. As a further confirmation, in section~\ref{sec4.2} it was shown that the pair ($W_I$, $\td W_I$) emerges from Higgsing in the $\phi_I$ direction, and correspondingly here we have shown that the pair ($M_2^{(I)}$, $\bar M_2^{(I)}$) is localized at the same position $|z_I|=1$, $z_j=0$, $j \neq I$. Therefore, there is one-to-one correspondence between the Higgsing direction in the scalar field space in the superconformal field theory and the position where the M2--/anti--M2--brane resides.

\begin{table}[htbp]
  \centering
  \begin{tabular}{|c|l|l|l|l|}\hline
  brane             & \multicolumn{2}{c|}{position} & \multicolumn{2}{c|}{preserved supercharges} \\ \hline

  $M_2^{(1)}$       & \multirow{2}*{$|z_1|=1$} & \multirow{2}*{$\b=\th_1=0$} & $s_1=+$     & $(++--)$, $(+-+-)$, $(+--+)$ \\ \cline{1-1}\cline{4-5}
  $\bar M_2^{(1)}$  &                          &                             & $s_1=-$     & $(-++-)$, $(-+-+)$, $(--++)$ \\ \hline

  $M_2^{(2)}$       & \multirow{2}*{$|z_2|=1$} & \multirow{2}*{$\b=0$, $\th_1=\pi$} & $s_2=+$ & $(++--)$, $(-++-)$, $(-+-+)$ \\ \cline{1-1}\cline{4-5}
  $\bar M_2^{(2)}$  &                          &                                    & $s_2=-$ & $(+-+-)$, $(+--+)$, $(--++)$ \\ \hline

  $M_2^{(3)}$       & \multirow{2}*{$|z_3|=1$} & \multirow{2}*{$\b=\pi$, $\th_2=0$} & $s_3=+$     & $(+-+-)$, $(-++-)$, $(--++)$ \\ \cline{1-1}\cline{4-5}
  $\bar M_2^{(3)}$  &                          &                                    & $s_3=-$     & $(++--)$, $(+--+)$, $(-+-+)$ \\ \hline

  $M_2^{(4)}$       & \multirow{2}*{$|z_4|=1$} & \multirow{2}*{$\b=\th_2=\pi$} & $s_4=+$ & $(+--+)$, $(-+-+)$, $(--++)$ \\ \cline{1-1}\cline{4-5}
  $\bar M_2^{(4)}$  &                          &                               & $s_4=-$ & $(++--)$, $(+-+-)$, $(-++-)$ \\ \hline

  \end{tabular}
  \caption{The 1/2 BPS M2-- and anti--M2--branes in $\gabjm$ spacetime, their positions, and the supercharges they preserve.}\label{tab3}
\end{table}

Supported by this first evidence, we now investigate the identification of supercharges in gravity and field theory for more general configurations \footnote{We thank the JHEP referee for suggesting the possibility to perform this general analysis.}.  To this end, it is worth using the second way of decomposing Killing spinors,  given in (\ref{z2}). Using decompositions (\ref{e92}) and (\ref{z11}), constraints (\ref{z5}) lead to
\be
(\bG_{3\natural} + \bG_{58} + \bG_{47} + \bG_{69})\eta=0
\ee
In terms of the eigenstates (\ref{z1}) this amounts to
\be
t_1+t_2+t_3+t_4=0
\ee
and only six of the eight states (\ref{z6}) for the $\eta$ spinor survive
\be
(t_1,t_2,t_3,t_4) = (++--), (+-+-), (+--+),
                    (-++-), (-+-+), (--++)
\ee
In the present order we call them $\eta_i$, $i=2,3,4,5,6,7$. We rename $\eta_i$, $\th^i$ and $\vth^i$ in (\ref{z2}) as
\bea
&& \eta_2 = \eta_{12} = -\eta_{21}, ~~
   \eta_3 = \eta_{13} = -\eta_{31}, ~~
   \eta_4 = \eta_{14} = -\eta_{41} \nn\\
&& \eta_5 = \eta_{23} = -\eta_{32}, ~~
   \eta_6 = -\eta_{24} = \eta_{42}, ~~
   \eta_7 = \eta_{34} = -\eta_{43} \nn\\
&& \th^2 = \th^{12} = -\th^{21}, ~~
   \th^3 = \th^{13} = -\th^{31}, ~~
   \th^4 = \th^{14} = -\th^{41} \nn\\
&& \th^5 = \th^{23} = -\th^{32}, ~~
   \th^6 = -\th^{24} = \th^{42}, ~~
   \th^7 = \th^{34} = -\th^{43} \nn\\
&& \vth^2 = \vth^{12} = -\vth^{21}, ~~
   \vth^3 = \vth^{13} = -\vth^{31}, ~~
   \vth^4 = \vth^{14} = -\vth^{41} \nn\\
&& \vth^5 = \vth^{23} = -\vth^{32}, ~~
   \vth^6 = -\vth^{24} = \vth^{42}, ~~
   \vth^7 = \vth^{34} = -\vth^{43}
\eea
Then, defining $\bar\eta^{IJ} = \eta_{IJ}^c$, $\bar\th_{IJ} = \th^{IJc}=(\th^{IJ})^*$ and $\bar\vth_{IJ} = \vth^{IJc}=(\vth^{IJ})^*$ we write $\e_1, \e_2$ as
\bea \label{z9}
&& \e_1 = \sum_{i=2}^4 ( \th^i \otimes \eta_i + \bar\th_i \otimes \bar\eta^{i} )
     = \f12 \th^{IJ} \otimes \eta_{IJ}
     = \f12 \bar\th_{IJ} \otimes \bar\eta^{IJ} \nn \\
&&     \e_2 = \sum_{i=2}^4 ( \vth^i \otimes \eta_i + \bar\vth_i \otimes \bar\eta^{i} )
     = \f12 \vth^{IJ} \otimes \eta_{IJ}
     = \f12 \bar\vth_{IJ} \otimes \bar\eta^{IJ}
\eea
where $\th^{IJ}$, $\bar\th_{IJ}$ satisfy relations (\ref{e101}) as a consequence of  (\ref{z4}). It is therefore tempting to identify $\th^{IJ}$, $\bar\th_{IJ}$, $\vth^{IJ}$, $\bar\vth_{IJ}$ components of the Killing spinors in $\gabjm$ with the supercharges in ABJM theory.

To perform the exact identification, in $\rm{C}^4 \cong \rm{R}^8$ we use complex coordinates
\bea
&& z^1 = x_3 + \ii x_\natural, ~~
   \bar z_1 = \bar z^{\bar 1} =  x_3 - \ii x_\natural \nn\\
&& z^2 = x_5 + \ii x_8, ~~
   \bar z_2 = \bar z^{\bar 2} =  x_5 - \ii x_8 \nn\\
&& z^3 = x_4 + \ii x_7, ~~
   \bar z_3 = \bar z^{\bar 3} =  x_4 - \ii x_7 \nn\\
&& z^4  = x_6+ \ii x_9, ~~
   \bar z_4 = \bar z^{\bar 4} =  x_6- \ii x_9
\eea
The metric then reads  $ds^2_{\rm{C}^4} = dz^I d\bar z_I = g_{I\bar J} d z^I d \bar z^{\bar J}$
with non-vanishing components $g_{1\bar 1}=g_{2\bar 2}=g_{3\bar 3}=g_{4\bar 4}=1$.
Correspondingly, we introduce gamma matrices
\bea
&& g_1 = \f{1}{\sr{2}} ( \g_3 -\ii \g_\natural ), ~~ g_{\bar 1} = \f{1}{\sr{2}} ( \g_3 + \ii \g_\natural ) \nn\\
&& g_2 = \f{1}{\sr{2}} ( \g_5 -\ii \g_8 ), ~~ g_{\bar 2} = \f{1}{\sr{2}} ( \g_5 + \ii \g_8 ) \nn\\
&& g_3 = \f{1}{\sr{2}} ( \g_4 -\ii \g_7 ), ~~ g_{\bar 3} = \f{1}{\sr{2}} ( \g_4 + \ii \g_7 ) \nn\\
&& g_4 = \f{1}{\sr{2}} ( \g_6 -\ii \g_9 ), ~~ g_{\bar 4} = \f{1}{\sr{2}} ( \g_6 + \ii \g_9 )
\eea
that satisfy the algebra
$ \{g_I, g_J\}=\{g_{\bar I},g_{\bar J}\}=0, ~~ \{g_I,g_{\bar J}\}=2g_{I\bar J}$. For later convenience, we also define $g_0=\g_0$.

Considering the decomposition (\ref{z11}), we also define
\bea
&& \textbf{G}_1 = \f{1}{\sr{2}} ( \bG_3 -\ii \bG_\natural ), ~~ \textbf{G}_{\bar 1} = \f{1}{\sr{2}} ( \bG_3 + \ii \bG_\natural ) \nn\\
&& \textbf{G}_2 = \f{1}{\sr{2}} ( \bG_5 -\ii \bG_8 ), ~~ \textbf{G}_{\bar 2} = \f{1}{\sr{2}} ( \bG_5 + \ii \bG_8 ) \nn\\
&& \textbf{G}_3 = \f{1}{\sr{2}} ( \bG_4 -\ii \bG_7 ), ~~ \textbf{G}_{\bar 3} = \f{1}{\sr{2}} ( \bG_4 + \ii \bG_7 ) \nn\\
&& \textbf{G}_4 = \f{1}{\sr{2}} ( \bG_6 -\ii \bG_9 ), ~~ \textbf{G}_{\bar 4} = \f{1}{\sr{2}} ( \bG_6 + \ii \bG_9 )
\eea
In $\rm{C}^4$ we introduce the unit vector
\bea\label{unit}
&& \a^I = (\cos\frac{\b}{2}\cos\frac{\theta_1}{2} \ep^{\ii\xi_1},
        \cos\frac{\b}{2}\sin\frac{\theta_1}{2} \ep^{\ii\xi_2},
        \sin\frac{\b}{2}\cos\frac{\theta_2}{2} \ep^{\ii\xi_3},
        \sin\frac{\b}{2}\sin\frac{\theta_2}{2} \ep^{\ii\xi_4})\\
&& \bar\a_I \equiv (\a^I)^* = \bar\a^{\bar I} =(\cos\frac{\b}{2}\cos\frac{\theta_1}{2} \ep^{-\ii\xi_1},
        \cos\frac{\b}{2}\sin\frac{\theta_1}{2} \ep^{-\ii\xi_2},
        \sin\frac{\b}{2}\cos\frac{\theta_2}{2} \ep^{-\ii\xi_3},
        \sin\frac{\b}{2}\sin\frac{\theta_2}{2} \ep^{-\ii\xi_4})\nn
\eea
that satisfies $\a^I\bar\a_I = g_{I\bar J}\a^I\bar\a^{\bar J}=1$. Localizing the M2-- or anti--M2--brane in the compact space at the point described by this vector,
it turns out that (\ref{z8}) can be written as
\be
h^{-1}\g_{03\natural}h = -\ii g_{0I\bar J} \a^I \bar\a^{\bar J} = \ii\bg_0 \otimes \bG \textbf{G}_{I\bar J}\a^I\bar\a^{\bar J}
\ee
whereas (\ref{z9}) becomes
\bea
&& \e_1 =  \bar\a_I \th^{IK} \otimes \a^J \eta_{JK} +  \a^I\bar\th_{IK} \otimes \bar\a_J \bar\eta^{JK} \nn\\
&& \e_2 =  \bar\a_I \vth^{IK} \otimes \a^J \eta_{JK} +  \a^I\bar\vth_{IK} \otimes \bar\a_J \bar\eta^{JK}
\eea
Inserting in (\ref{z12}) and using
\bea
&& (\bG \textbf{G}_{I\bar J}\a^I\bar\a^{\bar J})(\a^K\eta_{KL}) = (\a^K\eta_{KL})\nn\\
&& (\bG \textbf{G}_{I\bar J}\a^I\bar\a^{\bar J})(\bar\a_K \bar\eta^{KL}) = -(\bar\a_K \bar\eta^{KL})
\eea
we find that the $(\th^{IJ}, \vth^{IJ})$ supercharges preserved by a generic M2--brane satisfy
\bea \label{e122}
&& \bg_0 \bar\a_I \th^{IJ} = -\ii \bar\a_I \th^{IJ}, ~~ \bg_0 \a^I\bar\th_{IJ}=\ii\a^I\bar\th_{IJ} \nn\\
&& \bg_0 \bar\a_I \vth^{IJ} = -\ii \bar\a_I \vth^{IJ}, ~~ \bg_0 \a^I\bar\vth_{IJ}=\ii\a^I\bar\vth_{IJ}
\eea
These are indeed supercharges (\ref{z13}) preserved by a general Wilson loop $W$. Similarly, a general anti--M2--brane at the position specified by $\a^I$ preserves supercharges satisfying
\bea
&& \bg_0 \bar\a_I \th^{IJ} = \ii \bar\a_I \th^{IJ}, ~~ \bg_0 \a^I\bar\th_{IJ}=-\ii\a^I\bar\th_{IJ} \nn\\
&& \bg_0 \bar\a_I \vth^{IJ} = \ii \bar\a_I \vth^{IJ}, ~~ \bg_0 \a^I\bar\vth_{IJ}=-\ii\a^I\bar\vth_{IJ}
\eea
which are supercharges (\ref{z14}) preserved by a general $\td W$ operator.

In summary, we have proved that the supercharges in $\gabjm$ preserved by a M2-- or anti--M2--brane embedded as in (\ref{embedding}) and localized in the internal space at a point described by vector (\ref{unit}) can be identified with the Poincar\'e and conformal supercharges in ABJM theory preserved by general $W$ or $\td{W}$ 1/2 BPS operators.

\section{$\mN=4$ orbifold ABJM theory}\label{sec5}

In all the previous examples, we have given evidence of the fact that different, independent 1/2 BPS WL operators can share at most a subset of preserved supercharges. Therefore, for each configuration of 1/2 conserved supersymmetries there is at most one  WL operator that is invariant under that set. The same property emerges in the spectrum of string/M2--brane solutions dual to these operators.

We now consider $\mN=4$ SCSM theories where, as we will discuss, such a uniqueness property is lost and one can find pairs of different WL operators or dual brane configurations sharing exactly the same preserved supersymmetries. We begin by considering the $\mN=4$ orbifold ABJM theory and postpone to the next section the discussion for more general  $\mN=4$ SCSM theories.

\vskip 10pt
The $\mN=4$ orbifold ABJM theory with gauge group and levels $[U(N)_k \times U(N)_{-k}]^r$ can be obtained from the $U(r N)_k \times U(r N)_{-k}$ ABJM theory by performing a $Z_r$ quotient \cite{Benna:2008zy}.
To begin with, the field content is given by $rN \times rN$ matrix fields $A_\m$, $B_\m$, $\phi_I$, $\psi^I$ with $I=1,2,3,4$. Under the $Z_r$ projection each matrix is decomposed into $r \times r$ blocks and each block is an $N\times N$ matrix. Moreover, the R--symmetry group $SU(4)\cong SO(6)$ is broken to $SU(2)\times SU(2)\cong SO(4)$, and consequently the $I$ index is decomposed as
\be \label{e120}
I=1,2,4,3 \to i=1,2  , ~\hi=\hat 1,\hat 2
\ee
In particular, the SUSY parameters are now labeled as Poincar\'e supercharges $\th^{i\hi}$, $\bar\th_{i\hi}$ and superconformal charges  $\vth^{i\hi}$, $\bar\vth_{i\hi}$, and they are subject to the constraints
\bea \label{e121}
&& (\th^{i\hi})^*=\bar \th_{i\hi}, ~~ \bar\th_{i\hi}=\ve_{ij}\ve_{\hi\hj}\th^{j\hj} \nn\\
&& (\vth^{i\hi})^*=\bar \vth_{i\hi}, ~~ \bar\vth_{i\hi}=\ve_{ij}\ve_{\hi\hj}\vth^{j\hj}
\eea
where the antisymmetric tensors are defined as $\ve_{12}=\ve_{\hat 1 \hat 2}=1$.

\begin{figure}[htbp]
  \centering
  \includegraphics[width=0.8\textwidth]{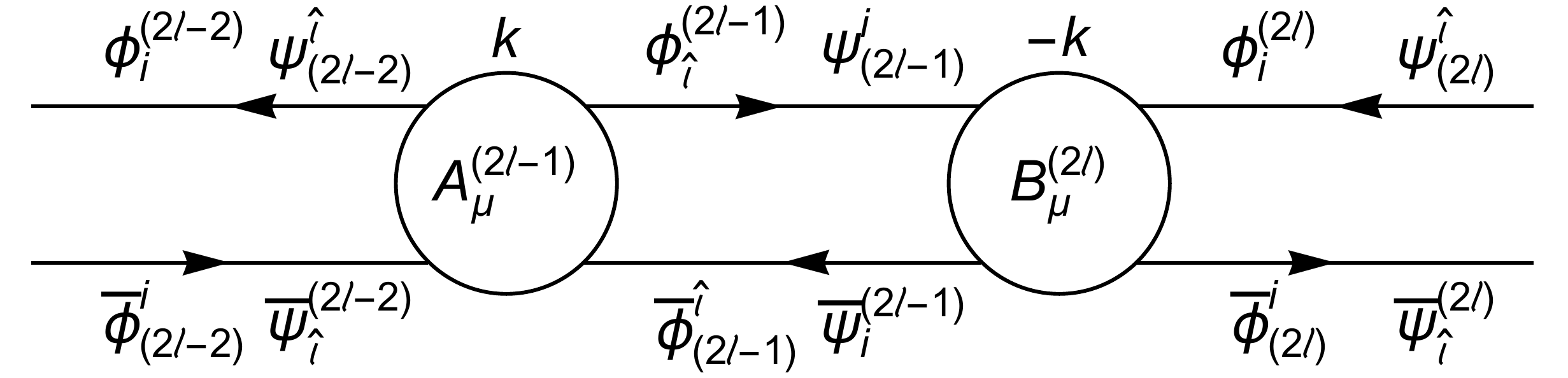}
  \caption{A slice of the quiver diagram of $\mN=4$ orbifold ABJM theory with gauge group and levels $[U(N)_k\times U(N)_{-k}]^r$. The quiver diagram is closed, so index identifications $(2r+1)= (1)$ and  $(2r) = (0)$ are understood.}\label{necklace}
\end{figure}

Explicitly, the original ABJM fields are decomposed as
\bea
&& A_\m = \diag(A_\m^{(1)},A_\m^{(3)},\cdots,A_\m^{(2r-1)}) , ~~
   B_\m = \diag(B_\m^{(0)},B_\m^{(2)},\cdots,B_\m^{(2r-2)})         \nn\\
&& \phi_i = \diag(\phi_i^{(0)},\phi_i^{(2)},\cdots,\phi_i^{(2r-2)}) , ~~~
   \bar\phi^i = \diag(\bar\phi^i_{(0)},\bar\phi^i_{(2)},\cdots,\bar\phi^i_{(2r-2)}) \nn
\eea\bea
&& \phi_\hi=\lt( \ba{ccccc}
              0&\phi_\hi^{(1)}&&& \\
              &0&\phi_\hi^{(3)}&& \\
              &&\ddots&\ddots& \\
              &&&0&\phi_\hi^{(2r-3)} \\
              \phi_\hi^{(2r-1)}&&&&0 \ea \rt),                   ~~
   \bar\phi^\hi=\lt( \ba{ccccc}
              0&&&&\bar\phi^\hi_{(2r-1)} \\
              \bar\phi^\hi_{(1)}&0&&& \\
              &\bar\phi^\hi_{(3)}&\ddots&& \\
              &&\ddots&0& \\
              &&&\bar\phi^\hi_{(2r-3)}&0 \ea \rt)                     \nn\\
&& \psi^i=\lt( \ba{ccccc}
                0&\psi^i_{(1)}&&& \\
                &0&\psi^i_{(3)}&& \\
                &&\ddots&\ddots& \\
                &&&0&\psi^i_{(2r-3)} \\
                \psi^i_{(2r-1)}&&&&0  \ea \rt)  , ~~
   \bar\psi_i=\lt( \ba{ccccc}
                0&&&&\bar\psi_i^{(2r-1)} \\
                \bar\psi_i^{(1)}&0&&& \\
                &\bar\psi_i^{(3)}&\ddots&& \\
                &&\ddots&0& \\
                &&&\bar\psi_i^{(2r-3)}&0  \ea \rt) \nn\\
&& \psi^\hi = \diag(\psi^\hi_{(0)},\psi^\hi_{(2)},\cdots,\psi^\hi_{(2r-2)})  ,  ~~
   \bar\psi_\hi = \diag(\bar\psi_\hi^{(0)},\bar\psi_\hi^{(2)},\cdots,\bar\psi_\hi^{(2r-2)})  \label{j78}
\eea
A slice of the corresponding necklace quiver diagram is shown in Figure \ref{necklace}, where arrows indicate that matter fields are in the fundamental representation of one gauge group (outgoing arrow) and in the anti--fundamental of the next one (incoming arrow).

\subsection{1/2 BPS Wilson loops}\label{sec5.1}

1/2 BPS WLs in $\mN=4$ orbifold ABJM theory can be easily obtained by taking the $\Z_r$ quotient of ABJM 1/2 BPS WLs constructed in section~\ref{WLABJM}.

We start by considering $W_1$ operator, i.e., (\ref{WLdefinition}) with $I=1$. Its connection (\ref{L1}) decomposes as
\bea \label{LL1}
L_1 = \lt(\ba{cccccccccc}
          \mA^{(1)}&&&&&0&f_1^{(1)}&&&\\
          &\mA^{(3)}&&&&&0&f_1^{(3)}&&\\
          &&\ddots&&&&&\ddots&\ddots&\\
          &&&\mA^{(2r-3)}&&&&&0&f_1^{(2r-3)}\\
          &&&&\mA^{(2r-1)}&f_1^{(2r-1)}&&&&0\\
          0&&&&f_2^{(2r-1)}&\mB^{(0)}&&&&\\
          f_2^{(1)}&0&&&&&\mB^{(2)}&&&\\
          &f_2^{(3)}&\ddots&&&&&\ddots&&\\
          &&\ddots&0&&&&&\mB^{(2r-4)}&\\
          &&&f_2^{(2r-3)}&0&&&&&\mB^{(2r-2)}\\
\ea\rt)
\eea
with the definitions
\bea
&& \mA^{(2\ell-1)}=A_0^{(2\ell-1)}+\f{2\pi}{k}( -\phi_1^{(2\ell-2)}\bar\phi^1_{(2\ell-2)}
                                                +\phi_2^{(2\ell-2)}\bar\phi^2_{(2\ell-2)}
                                                +\phi_\hi^{(2\ell-1)}\bar\phi^\hi_{(2\ell-1)} ) \nn\\
&& \mB^{(2\ell)}=B_0^{(2\ell)}+\f{2\pi}{k}( -\bar\phi^1_{(2\ell)}\phi_1^{(2\ell)}
                                            +\bar\phi^2_{(2\ell)}\phi_2^{(2\ell)}
                                            +\bar\phi^\hi_{(2\ell-1)}\phi_\hi^{(2\ell-1)} )  \nn\\
&& f_1^{(2\ell-1)} = \sqrt{\f{4\pi}{k}} \psi^1_{(2\ell-1)+} , ~~
   f_2^{(2\ell-1)} = \sqrt{\f{4\pi}{k}} \bar \psi_{1-}^{(2\ell-1)}
\eea
The connection can be re-organized as
\be \label{LLL1}
L_1 = \diag( L_1^{(1)} , L_1^{(2)}, \cdots, L_1^{(r)} ) ~~  {\rm with} ~~
L_1^{(\ell)} = \lt(\ba{cc} \mA^{(2\ell-1)} & f_1^{(2\ell-1)} \\ f_2^{(2\ell-1)} & \mB^{(2\ell)} \ea\rt)
\ee
This time we have the freedom to define double--node operators $W_1^{(\ell)}$, with $\ell=1,2,\cdots,r$, corresponding to the $L_1^{(\ell)}$ superconnection localized at quiver nodes $2\ell-1$ and $2\ell$. One can easily show that all these WLs preserve Poincar\'e supercharges
$(\th^{1\hi}_+, ~ \th^{2\hi}_-, ~ \bar\th_{1\hi-}, ~ \bar\th_{2\hi+})$.  Therefore, we can define a ``global'' $W_1$ operator as the holonomy of the complete $L_1$ superconnection. This is nothing but $W_1 = \sum_{\ell=1}^r W_1^{(\ell)}$, and preserves the same supercharges.

With a similar procedure, but starting from $W_2$ in eq. (\ref{WLdefinition}) we can construct 1/2 BPS WL $W_2 = \sum_{\ell=1}^r W_2^{(\ell)}$ preserving Poincar\'e supercharges $(\th^{2\hi}_+, ~ \th^{1\hi}_-, ~ \bar\th_{2\hi-}, ~ \bar\th_{1\hi+})$.
From $W_4$ operator in  eq. (\ref{WLdefinition}) we construct $W_\ho = \sum_{\ell=1}^r  W_\ho^{(\ell)}$ with preserved Poincar\'e supercharges
$(\th^{i\ho}_+, ~ \th^{i\hw}_-, ~ \bar\th_{i\ho-}, ~ \bar\th_{i\hw+})$.
Finally, from $W_3$ we obtain $W_\hw = \sum_{\ell=1}^r  W_\hw^{(\ell)}$ preserving Poincar\'e supercharges
$(\th^{i\hw}_+, ~ \th^{i\ho}_-, ~ \bar\th_{i\hw-}, ~ \bar\th_{i\ho+})$.

Alternatively, we can do the orbifold projection starting from the ABJM superconnection $\tilde{L}_1$, i.e., (\ref{tdL1}) with $I=1$. The corresponding superconnection in $\mN=4$ SCSM theory then reads
\be
\td L_1 = \diag(\td  L_1^{(1)} ,\td  L_1^{(2)}, \cdots,\td  L_1^{(r)} )
\ee
where
\bea
&& \td L_1^{(\ell)} = \lt(\ba{cc} \td \mA^{(2\ell-1)} & \td f_1^{(2\ell-1)} \\ \td f_2^{(2\ell-1)} & \td \mB^{(2\ell)} \ea\rt), ~~
   \td f_1^{(2\ell-1)} = \sqrt{\f{4\pi}{k}} \psi^1_{(2\ell-1)-}, ~~
   \td f_2^{(2\ell-1)} = -\sqrt{\f{4\pi}{k}} \bar \psi_{1+}^{(2\ell-1)} \nn\\
&& \td \mA^{(2\ell-1)}=A_0^{(2\ell-1)}+\f{2\pi}{k}( \phi_1^{(2\ell-2)}\bar\phi^1_{(2\ell-2)}
                                                   -\phi_2^{(2\ell-2)}\bar\phi^2_{(2\ell-2)}
                                                   -\phi_\hi^{(2\ell-1)}\bar\phi^\hi_{(2\ell-1)} )  \nn\\
&& \td \mB^{(2\ell)}=B_0^{(2\ell)}+\f{2\pi}{k}(  \bar\phi^1_{(2\ell)}\phi_1^{(2\ell)}
                                                -\bar\phi^2_{(2\ell)}\phi_2^{(2\ell)}
                                                -\bar\phi^\hi_{(2\ell-1)}\phi_\hi^{(2\ell-1)} )
\eea
We then define double--node WLs $\td W_1^{(\ell)}$ with $\ell=1,2,\cdots,r$ as the holonomy of the $\td L_1^{(\ell)}$ superconnections, and the ``global'' operator $\td W_1 = \sum_{\ell=1}^r \td W_1^{(\ell)}$. They all preserve supercharges $(\th^{1\hi}_-, ~ \th^{2\hi}_+, ~ \bar\th_{1\hi+}, \bar\th_{2\hi-})$.

From WLs $\td W_{2,4,3}$ of the ABJM theory, we obtain 1/2 BPS operators $\td W_2 = \sum_{\ell=1}^r \td W_2^{(\ell)}$, $\td W_\ho =\sum_{\ell=1}^r \td W_\ho^{(\ell)}$ and $\td W_\hw =\sum_{\ell=1}^r \td W_\hw^{(\ell)}$ respectively, with corresponding preserved Poincar\'e supercharges given in the summarizing table \ref{tab4}.

According to the classification of \cite{Ouyang:2015iza,Ouyang:2015bmy}, $W_1$ and $W_2$ operators (and the corresponding double--node operators) belong to class II, up to some R--symmetry rotations; $\td W_1$ and $\td W_2$ belong to class I, whereas WLs $W_\ho$, $W_\hw$, $\td W_\ho$ and  $\td W_\hw$ were not considered therein. In particular, $W_1$ (or the double--node version $W_1^{(\ell)}$) is the $\psi_1$-loop that was constructed in \cite{Ouyang:2015qma,Cooke:2015ila}. Wilson loop $\td W_2$ (or $\td W_2^{(\ell)}$) corresponds to the $\psi_2$--loop of \cite{Cooke:2015ila}.

\begin{figure}[htbp]
  \centering
  \subfigure[Overlapping supercharges of 1/2 BPS Wilson loops]
            {\includegraphics[height=0.4\textwidth]{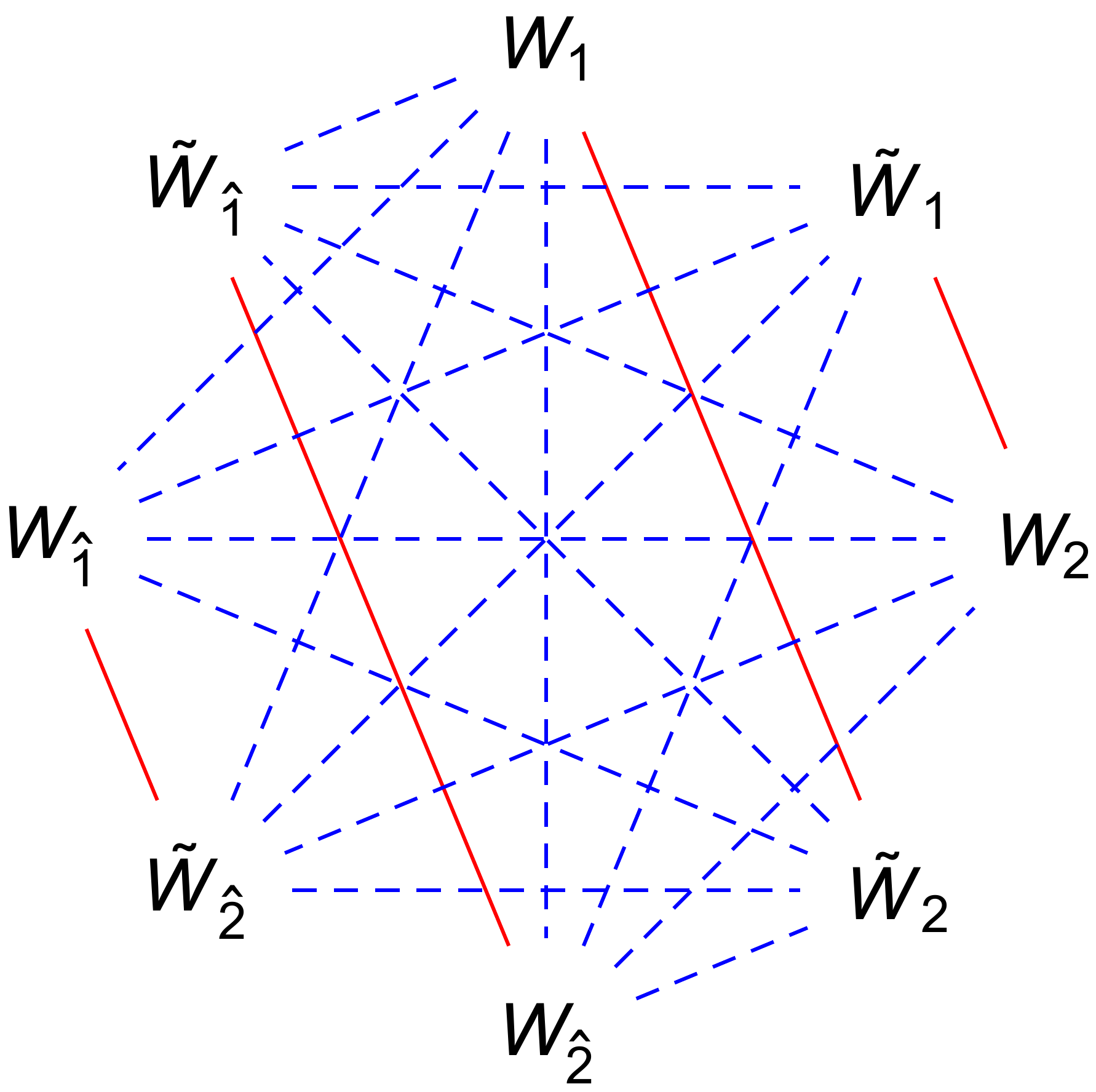}\label{f2a}}
            ~~~~~~
  \subfigure[Overlapping supercharges of M2-- and anti--M2--branes]
            {\includegraphics[height=0.4\textwidth]{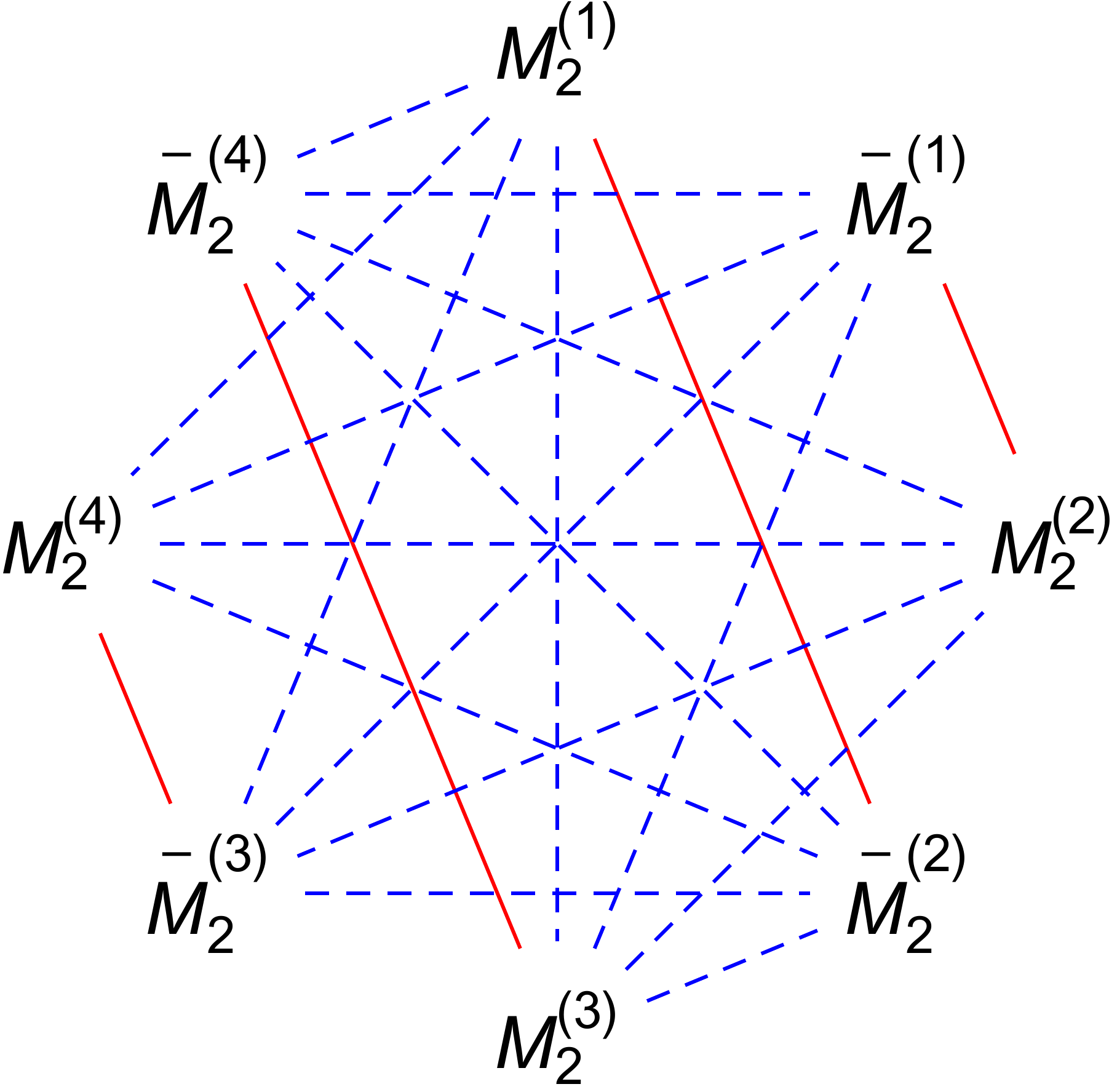}\label{f2b}}\\
  \caption{Amount of overlapping supercharges between each pair of 1/2 BPS WLs in $\mN=4$ orbifold ABJM theory and between each pair of M2-- and anti--M2--branes in $\goabjm$. A red solid line means that the two WLs or branes preserve exactly the same supercharges. A blue dashed line means that the two WLs or branes share 1/2 of preserved supercharges. Two WLs or branes that are not directly connected by any line have no common preserved supercharges.}
\end{figure}

Each WL preserves four real Poincar\'e plus four real superconformal charges. Therefore, they are all 1/2 BPS operators.
From table \ref{tab4} it is easy to realize that there is non-trivial overlapping of preserved supercharges among different WLs, as shown in Figure~\ref{f2a}.
In particular, we see that there are four pairs of WLs that preserve exactly the same set of supercharges (WLs connected by a red line in Figure \ref{f2a}). Therefore, as already mentioned, in the ${\cal N}=4$ orbifold ABJM theory the uniqueness property of WLs corresponding to a given set of preserved supercharges is no longer valid. This is in fact the result already found in \cite{Cooke:2015ila} for the ($\psi_1$--loop, $\psi_2$--loop) pair.

\begin{table}[htbp]
  \centering
  \begin{tabular}{|c|c|}\hline
   Wilson loop & preserved supercharges \\ \hline
   $W_1$       & $\th^{1 \hat 1}_+$, $\th^{1 \hat 2}_+$, $\th^{2 \hat 1}_-$, $\th^{2 \hat 2}_-$ \\ \hline
   $\td W_1$   & $\th^{1 \hat 1}_-$, $\th^{1 \hat 2}_-$, $\th^{2 \hat 1}_+$, $\th^{2 \hat 2}_+$ \\ \hline
  $W_2$        & $\th^{2 \hat 1}_+$, $\th^{2 \hat 2}_+$, $\th^{1 \hat 1}_-$, $\th^{1 \hat 2}_-$ \\ \hline
  $\td W_2$    & $\th^{2 \hat 1}_-$, $\th^{2 \hat 2}_-$, $\th^{1 \hat 1}_+$, $\th^{1 \hat 2}_+$ \\ \hline
  $W_\ho$      & $\th^{1 \hat 1}_+$, $\th^{2 \hat 1}_+$, $\th^{1 \hat 2}_-$, $\th^{2 \hat 2}_-$ \\ \hline
  $\td W_\ho$  & $\th^{1 \hat 1}_-$, $\th^{2 \hat 1}_-$, $\th^{1 \hat 2}_+$, $\th^{2 \hat 2}_+$ \\ \hline
  $W_\hw$      & $\th^{1 \hat 2}_+$, $\th^{2 \hat 2}_+$, $\th^{1 \hat 1}_-$, $\th^{2 \hat 1}_-$ \\ \hline
  $\td W_\hw$  & $\th^{1 \hat 2}_-$, $\th^{2 \hat 2}_-$, $\th^{1 \hat 1}_+$, $\th^{2 \hat 1}_+$ \\ \hline
  \end{tabular}
  \caption{The 1/2 BPS WLs in $\mN=4$ orbifold ABJM theory and the supercharges they preserve. We have not shown $\bar{\theta}_{i\hi}$ supercharges, since they are not independent. }\label{tab4}
\label{table4}
\end{table}

Starting from $W_1$ or $W_2$ operators defined above, we can apply a R--symmetry rotation and obtain a 1/2 BPS Wilson loop $W$ with connection
\bea \label{e127}
&& L = \diag( L^{(1)} , L^{(2)}, \cdots, L^{(r)} ), ~~
   L^{(\ell)} = \lt(\ba{cc} \mA^{(2\ell-1)} & f_1^{(2\ell-1)} \\ f_2^{(2\ell-1)} & \mB^{(2\ell)} \ea\rt) \nn\\
&& \mA^{(2\ell-1)}=A_0^{(2\ell-1)}+\f{2\pi}{k}\Big[ \big( \d^i_j - {2\a^i\bar\a_j} \big)
                                                 \phi_i^{(2\ell-2)}\bar\phi^j_{(2\ell-2)}
                                                +\phi_\hi^{(2\ell-1)}\bar\phi^\hi_{(2\ell-1)} \Big] \nn\\
&& \mB^{(2\ell)}=B_0^{(2\ell)}+\f{2\pi}{k}\Big[ \big( \d^i_j - {2\a^i\bar\a_j} \big)
                                                \bar\phi^j_{(2\ell)}\phi_i^{(2\ell)}
                                               +\bar\phi^\hi_{(2\ell-1)}\phi_\hi^{(2\ell-1)} \Big]  \nn\\
&& f_1^{(2\ell-1)} = \sqrt{\f{4\pi}{k}} \bar\a_i\psi^i_{(2\ell-1)+} , ~~
   f_2^{(2\ell-1)} = \sqrt{\f{4\pi}{k}} \bar \psi_{i-}^{(2\ell-1)} {\a^i}
\eea
where $\bar\a_i=(\bar\a_1,\bar\a_2)$, $\a^i=(\bar\a_i)^*$, $|\a|^2 = \bar\a_i\a^i =1$. The corresponding preserved supercharges are
\be \label{sc1}
\bar\a_i\th^{i\hi}_+, ~~ \a^i\bar\th_{i\hi-}, ~~ \bar\a_i\vth^{i\hi}_+, ~~ \a^i\bar\vth_{i\hi-}
\ee
Similarly, we can construct the 1/2 BPS Wilson loop $W_\wedge$ with connection
\bea \label{e133}
&& L_\wedge = \diag( L_\wedge^{(1)} , L_\wedge^{(2)}, \cdots, L_\wedge^{(r)} ), ~~
   L_\wedge^{(\ell)} = \lt(\ba{cc} \mB^{(2\ell)} & f_1^{(2\ell)} \\ f_2^{(2\ell)} & \mA^{(2\ell+1)} \ea\rt) \nn\\
&& \mB^{(2\ell)} =  B_0^{(2\ell)}
   + \f{2\pi}{k} \Big[ \bar\phi^i_{(2\ell)}\phi_i^{(2\ell)}
                      +\big( \d^\hi_\hj - {2\a^\hi\bar\a_\hj} \big)\bar\phi^\hj_{(2\ell-1)}\phi_\hi^{(2\ell-1)}
                 \Big] \nn\\
&& \mA^{(2\ell+1)} =  A_0^{(2\ell+1)}
   + \f{2\pi}{k} \Big[ \phi_i^{(2\ell)}\bar\phi^i_{(2\ell)}
                      +\big( \d^\hi_\hj - {2\a^\hi\bar\a_\hj} \big) \phi_\hi^{(2\ell+1)}\bar\phi^\hj_{(2\ell+1)}
                 \Big] \nn\\
&& f_1^{(2\ell)} = \sqrt{\f{4\pi}{k}} \bar \psi_{\hi-}^{(2\ell)} {\a^\hi} , ~~
   f_2^{(2\ell)} = \sqrt{\f{4\pi}{k}} \bar\a_\hi\psi^\hi_{(2\ell)+}
\eea
where $\bar\a_\hi=(\bar\a_\ho,\bar\a_\hw)$, $\a^\hi = (\bar\a_\hi)^*$, $|\a|^2=\bar\a_\hi\a^\hi =1$, and preserved supercharges
\be \label{sc2}
\bar\a_\hi\th^{i\hi}_+, ~~ \a^\hi\bar\th_{i\hi-}, ~~ \bar\a_\hi\vth^{i\hi}_+, ~~ \a^\hi\bar\vth_{i\hi-}
\ee
Furthermore, we can obtain the 1/2 BPS Wilson loop $\td W$ with parameters $\bar\a_i$, $\a^i$ and preserved supercharges
\be \label{sc3}
\bar\a_i\th^{i\hi}_-, ~~ \a^i\bar\th_{i\hi+}, ~~ \bar\a_i\vth^{i\hi}_-, ~~ \a^i\bar\vth_{i\hi+}
\ee
as well the 1/2 BPS Wilson loop $\td W_\wedge$ with parameters $\bar\a_\hi$, $\a^\hi$ and preserved supercharges
\be \label{sc4}
\bar\a_\hi\th^{i\hi}_-, ~~ \a^\hi\bar\th_{i\hi+}, ~~ \bar\a_\hi\vth^{i\hi}_-, ~~ \a^\hi\bar\vth_{i\hi+}
\ee
The corresponding connections can be easily figured out, and we will not bother writing them out.

It is interesting to note that if we apply the orbifold projection directly to the general 1/2 BPS WLs in ABJM theory corresponding to connections (\ref{z128}) and (\ref{z129}), we obtain new fermionic 1/4 BPS operators. We will report the results, as well as their M2--/anti--M2--brane duals, elsewhere \cite{work-newWL}.

\subsection{Wilson loops from Higgsing} \label{sec5.2}

The easiest way to obtain the previous WLs via the Higgsing procedure is to perform the orbifold projection of the construction done for the ABJM theory. In fact, orbifolding the Higgsing reduction of $U(r N+r)_k \times U(rN+r)_{-k}$ ABJM theory to $U(r N)_k \times U(r M)_{-k}$ ABJM theory, is equivalent to directly Higgsing a $[U(N+1)_k \times U(N+1)_{-k}]^r$ $\mN=4$ orbifold ABJM theory to a $[U(N)_k \times U(N)_{-k}]^r$ $\mN=4$ orbifold ABJM theory.
Since the procedure is similar for all the WLs, we will show it explicitly only for the $W_1$ operator.

We consider the low energy non-relativistic particle modes of the ABJM theory given in eqs. (\ref{ABJMpartciles}) and (\ref{j7}) and write them in terms of fields in $\mN=4$ orbifold ABJM theory
\be \label{j30}
\Psi_1 = \lt(\ba{cc} w & \o^1 \\ \s_1 & z \ea\rt), ~~
\Psi_2 = \lt(\ba{cc} r_2 & -\o^2 \\ -\s_2 & s^2 \ea\rt), ~~
\Psi_\hi = \lt(\ba{cc} r_\hi & -\o^\hi \\ -\s_\hi & s^\hi \ea\rt)
\ee
where we have defined
\bea
&& w = \diag(w^{(1)},w^{(3)},\cdots,w^{(2r-1)}) , ~~
   z = \diag(z^{(0)},z^{(2)},\cdots,z^{(2r-2)})                          \nn\\
&& r_2 = \diag(r_2^{(0)},r_2^{(2)},\cdots,r_2^{(2r-2)}) , ~~
   s^2 = \diag(s^2_{(0)},s^2_{(2)},\cdots,s^2_{(2r-2)})                 \nn\\
&& \o^\hi = \diag(\o^\hi_{(0)},\o^\hi_{(2)},\cdots,\o^\hi_{(2r-2)}) , ~~
   \s_\hi = \diag(\s_\hi^{(0)},\s_\hi^{(2)},\cdots,\s_\hi^{(2r-2)})   \nn \\
&& r_\hi=\lt( \ba{ccccc}
              0&r_\hi^{(1)}&&& \\
              &0&r_\hi^{(3)}&& \\
              &&\ddots&\ddots& \\
              &&&0&r_\hi^{(2r-3)} \\
              r_\hi^{(2r-1)}&&&&0 \ea \rt) , ~~
   s^\hi=\lt( \ba{ccccc}
              0&&&&s^\hi_{(2r-1)} \\
              s^\hi_{(1)}&0&&& \\
              &s^\hi_{(3)}&\ddots&& \\
              &&\ddots&0& \\
              &&&s^\hi_{(2r-3)}&0 \ea \rt)          
\eea\bea
&& \o^i=\lt( \ba{ccccc}
                0&\o^i_{(1)}&&& \\
                &0&\o^i_{(3)}&& \\
                &&\ddots&\ddots& \\
                &&&0&\o^i_{(2r-3)} \\
                \o^i_{(2r-1)}&&&&0  \ea \rt) , ~~
  \s_i=\lt( \ba{ccccc}
              0&&&&\s_i^{(2r-1)} \\
              \s_i^{(1)}&0&&& \\
              &\s_i^{(3)}&\ddots&& \\
              &&\ddots&0& \\
              &&&\s_i^{(2r-3)}&0 \ea \rt)          \nn
\eea
The non-relativistic lagrangian then becomes
\be \label{j31}
\mL = \ii \Tr \bar\Psi^i \mf D_0 \Psi_i + \ii \Tr \bar\Psi^\hi \mf D_0 \Psi_\hi
\ee
with
\be
\mf D_0 \Psi_i = \p_0 \Psi_i + \ii L_1 \Psi_i \qquad , \qquad \mf D_0 \Psi_\hi = \p_0 \Psi_\hi + \ii L_1 \Psi_\hi
\ee
and $L_1$ being the connection in eq. (\ref{LL1}).

It is convenient to re-organize the $L_1$ connection as in (\ref{LLL1}) and modes (\ref{j30}) as
\bea
&& \Psi_i = \diag( \Psi_i^{(1)},\Psi_i^{(2)},\cdots,\Psi_i^{(r)}) \nn \\
&& \Psi_1^{(\ell)} = \lt(\ba{cc} w^{(2\ell-1)} & \o^1_{(2\ell-1)} \\ \s_1^{(2\ell-1)} & z^{(2\ell)} \ea\rt), ~~
   \Psi_2^{(\ell)} = \lt(\ba{cc} r_2^{(2\ell-2)} & -\o^2_{(2\ell-1)} \\ -\s_2^{(2\ell-1)} & s^2_{(2\ell)} \ea\rt)
\eea
and
\be
\Psi_\hi = \diag( \Psi_\hi^{(1)},\Psi_\hi^{(2)},\cdots,\Psi_\hi^{(r)}),  ~~
\Psi_\hi^{(\ell)} = \lt(\ba{cc} r_\hi^{(2\ell-1)} & -\o^\hi_{(2\ell-2)} \\ -\s_\hi^{(2\ell)} & s^\hi_{(2\ell-1)} \ea\rt) \nn\\
\ee
so that we can write
\be
\Tr \bar \Psi^i L_1 \Psi_i = \sum_{\ell=1}^{r} \Tr \bar \Psi_{(\ell)}^i L_1^{(\ell)} \Psi_i^{(\ell)}  , ~~ \Tr \bar \Psi^\hi L_1 \Psi_\hi = \sum_{\ell=1}^{r} \Tr \bar \Psi_{(\ell)}^\hi L_1^{(\ell)} \Psi_\hi^{(\ell)}
\ee
Therefore, using these new definitions we can rewrite lagrangian (\ref{j31}) as
\be
\mL     = \ii \sum_{\ell=1}^r \Tr \lt( \bar\Psi^i_{(\ell)} \mf D_0 \Psi_i^{(\ell)} +
                                   \bar\Psi^\hi_{(\ell)} \mf D_0 \Psi_\hi^{(\ell)} \rt)
\ee
where the covariant derivatives are given by
\be
 \mf D_0 \Psi_i^{(\ell)} = \p_0 \Psi_i^{(\ell)}+ \ii L_1^{(\ell)} \Psi_i^{(\ell)}  \qquad , \qquad
\mf D_0 \Psi_\hi^{(\ell)} = \p_0 \Psi_\hi^{(\ell)}+ \ii L_1^{(\ell)} \Psi_\hi^{(\ell)}
\ee
We have then obtained the generalized connections $L_1$, $L_1^{(\ell)}$ that need to be used to define the 1/2 BPS Wilson loops $W_1$, $W_1^{(\ell)}$.

Replacing particle excitations with antiparticle ones in eqs. (\ref{ABJMantiparticles}), (\ref{j8}) and performing the orbifold quotient we obtain a non-relativistic lagrangian with derivatives covariantized by generalized connections $\td L_1, \td L_1^{(\ell)}$, which enter the definitions of $\td W_1, \td W_1^{(\ell)}$ operators.

\vskip 7pt
The Higgsing procedure breaks half of the supersymmetries. It is then interesting to analyze how the non-relativistic modes organize themselves in ${\cal N}=2$ SUSY multiplets.
Exploiting the fact that in three-dimensions a $\mN=3$ massive vector multiplet can be written as a $\mN=2$ massive vector multiplet plus a $\mN=2$ massive fermion multiplet, in $\mN=4$ orbifold ABJM theory the non-relativistic modes of the original ABJM theory can be re-organized in $\mN=2$ massive super multiplets as follows
\begin{center}
\begin{tabular}{r|c|c|c|c}
  spin       & 1               & 1/2                        & 0                         & $-1/2$             \\ \cline{1-5}
  degeneracy & 1               & 2                          & 1                         &                    \\
  mode       & $w^{(2\ell-1)}$ & $\o_{(2\ell-2)}^{\ho,\hw}$ & $r_2^{(2\ell-2)}$         &                    \\ \cline{1-5}
  degeneracy &                 & 1                          & 2                         & 1                  \\
  mode       &                 & $\o_{(2\ell-1)}^2$         & $r_{\ho,\hw}^{(2\ell-1)}$ & $\o^1_{(2\ell-1)}$
\end{tabular}
\end{center}

\vskip 5pt
\begin{center}
\begin{tabular}{r|c|c|c|c}
  spin       & $-1$          & $-1/2$                   & 0                         & 1/2 \\ \cline{1-5}
  degeneracy & 1             & 2                        & 1                         &     \\
  mode       & $z^{(2\ell)}$ & $\s_{\ho,\hw}^{(2\ell)}$ & $s^2_{(2\ell)}$           &     \\ \cline{1-5}
  degeneracy &               & 1                        & 2                         & 1   \\
  mode       &               & $\s_2^{(2\ell-1)}$       & $s^{\ho,\hw}_{(2\ell-1)}$ & $\s_1^{(2\ell-1)}$
\end{tabular}
\end{center}

\vskip 10pt
Therefore, 1/2 BPS WLs in $\mN=4$ orbifold ABJM theory emerge from the low energy dynamics of $\mN=2$ massive supermultiplets.

\vskip 20pt
\subsection{M2--branes in AdS$_4\times$S$^7/$(Z$_{rk}\times$Z$_r$) spacetime}\label{sec5.3}

The $\mN=4$ orbifold ABJM theory is dual to M--theory in $\goabjm$ spacetime \cite{Benna:2008zy,Imamura:2008nn,Fuji:2008yj}. We use the ${\rm AdS}_4$ metric in (\ref{ads4}) and parametrize the $\rmS^7$ unit sphere with the $z_i$  complex coordinates given in (\ref{z1234}). The quotient $\Z_{rk}\times\Z_r$ is obtained by identifying
\bea
&& (z_1, z_2, z_3, z_4) \sim \ep^{\f{2\pi\ii}{r k}} ( z_1,  z_2, z_3, z_4) \nn\\
&& (z_1, z_2, z_3, z_4) \sim (\ep^{ \f{2\pi\ii}{r}} z_1,  \ep^{ \f{2\pi\ii}{r}} z_2, z_3, z_4)
\eea
or equivalently, in terms of the angular coordinates
\be
\zeta\sim \zeta - \frac{8\pi}{rk} , ~~\chi\sim \chi - \frac{4\pi}{r} , ~~ \zeta\sim\zeta - \frac{4\pi}{r}
\ee
Note that this quotient convention is consistent with conventions on the R--symmetry indices decomposition (\ref{e120}).
Acting with the orbifold projection on the $\AdS_4\times\rmS^7$ Killing spinors (\ref{e105}) we obtain the following constraints
\bea \label{z15}
&& (\g_{3\natural}+\g_{58})\e_1=0 , ~~ (\g_{47}+\g_{69})\e_1=0 \nn\\
&& (\g_{3\natural}+\g_{58})\e_2=0 , ~~ (\g_{47}+\g_{69})\e_2=0
\eea
Using decomposition (\ref{e107}), we get
\bea
s_1+s_2=0, ~~ s_3+s_4=0
\eea
Therefore, only four of the eight states (\ref{e106}) survive
\be
(s_1,s_2,s_3,s_4) = (+-+-), (-++-), (+--+), (-+-+)
\ee
The Killing spinors in $\goabjm$ spacetime have 16 real degrees of freedom, and this is consistent with the fact the $\mN=4$ orbifold ABJM theory has eight real Poincar\'e supercharges plus eight real superconformal charges.

Following what has been done in section~\ref{sec4.3} for the ABJM theory, we construct 1/2 BPS M2-- and anti--M2--brane solutions preserving  eight real supersymmetries. These configurations wrap the M--theory circle and are embedded in ${\rm AdS}_4$ as in (\ref{embedding}). Different positions in the internal space lead to different M2--brane configurations that preserve different sets of supercharges.

A set of independent solutions is listed in table \ref{tab5}.
For $M_2^{(I)}$, $I = 1, \cdots , 4$ solutions localized at $|z_I|=1$, constraints in (\ref{z12}) give respectively $s_I=+$. For the anti--M2--branes $\bar M_2^{(I)}$,  localized at $|z_I|=1$,  the constraints give respectively $s_I=-$.

\begin{table}[htbp]
  \centering
  \begin{tabular}{|c|l|l|l|l|}\hline

  brane & \multicolumn{2}{c|}{position} & \multicolumn{2}{c|}{preserved supercharges} \\ \hline

  $M_2^{(1)}$       & \multirow{2}*{$|z_1|=1$} & \multirow{2}*{$\b=\th_1=0$} & $s_1=+$ & $(+-+-)$, $(+--+)$ \\ \cline{1-1}\cline{4-5}
  $\bar M_2^{(1)}$  &                          &                             & $s_1=-$ & $(-++-)$, $(-+-+)$ \\ \hline

  $M_2^{(2)}$       & \multirow{2}*{$|z_2|=1$} & \multirow{2}*{$\b=0$, $\th_1=\pi$} & $s_2=+$ & $(-++-)$, $(-+-+)$ \\ \cline{1-1}\cline{4-5}
  $\bar M_2^{(2)}$  &                          &                                    & $s_2=-$ & $(+-+-)$, $(+--+)$ \\ \hline

  $M_2^{(3)}$       & \multirow{2}*{$|z_3|=1$} & \multirow{2}*{$\b=\pi$, $\th_2=0$} & $s_3=+$ & $(+-+-)$, $(-++-)$ \\ \cline{1-1}\cline{4-5}
  $\bar M_2^{(3)}$  &                          &                                    & $s_3=-$ & $(+--+)$, $(-+-+)$ \\ \hline

  $M_2^{(4)}$       & \multirow{2}*{$|z_4|=1$} & \multirow{2}*{$\b=\th_2=\pi$} & $s_4=+$ & $(+--+)$, $(-+-+)$ \\ \cline{1-1}\cline{4-5}
  $\bar M_2^{(4)}$  &                          &                               & $s_4=-$ & $(+-+-)$, $(-++-)$ \\ \hline

  \end{tabular}
  \caption{The 1/2 BPS M2-- and anti--M2--branes in $\goabjm$ spacetime, their positions and the preserved supercharges.}\label{tab5}
\end{table}

As it turns out from this table,  there is non--trivial overlapping among the sets of preserved supercharges. In particular, there are four pairs of M2--branes and M2--anti--branes localized at different positions, which preserve exactly the same supercharges. This is shown in Figure~\ref{f2b} where red solid lines connect elements of the same pair.

It is important to note that Figure~\ref{f2a} showing the overlapping scheme of supercharges preserved by the 1/2 BPS WLs in table \ref{tab4} is exactly the same as Figure~\ref{f2b} representing the overlapping scheme of supercharges preserved by M2-- and anti--M2--branes in table \ref{tab5}.
Precisely, to each pair ($W_1$, $\td W_2$),  ($W_2$, $\td W_1$), ($W_\hw$, $\td W_\ho$) and ($W_\ho$, $\td W_\hw$) of BPS WLs preserving the same set of supercharges correspond pairs ($M_2^{(1)}$, $\bar M_2^{(2)}$), ($M_2^{(2)}$, $\bar M_2^{(1)}$), ($M_2^{(3)}$, $\bar M_2^{(4)}$) and ($M_2^{(4)}$, $\bar M_2^{(3)}$) of M2--/anti--M2--branes that preserve the same supercharges. In each pair of WLs, one operator is dual to an M2--brane configuration, while the other one is dual to an anti--M2--brane at a different position.

In particular, it follows that if $W_1$, the $\psi_1$-loop in \cite{Ouyang:2015qma,Cooke:2015ila}, is made dual to the $M_2^{(1)}$ brane localized at  $|z_1|=1$, then $\td W_2$, the $\psi_2$--loop in \cite{Cooke:2015ila}, is dual to the $\bar M_2^{(2)}$ anti--brane at position $|z_2|=1$. The $\psi_1$-- and $\psi_2$--loops are different operators that happen to preserve the same supercharges. Correspondingly, they have a dual description in terms of  two different M2-- and anti--M2--brane configurations located at different positions.

Therefore, the WL degeneracy found in \cite{Cooke:2015ila} for the $\psi_1$-- and $\psi_2$--loops is also present in their dual description and no contradiction with the AdS/CFT correspondence emerges. In particular, our construction of dual M2--, anti--M2--brane pairs seems to indicate that no degeneracy uplifting should occur at quantum level and points towards
the possibility for both $\psi_1$-- and $\psi_2$--loops to be separately BPS operators.
However, as already mentioned in the introduction, this may have problematic consequences, in particular when compared with the localization result that seems to be unique. We will  come back to this point in the conclusions.

Using decomposition (\ref{z2}) we can identify the supercharges in M--theory and field theory. For $\eta$ in (\ref{z1}), (\ref{z6}), the orbifold constraints (\ref{z15}) lead to
\be
t_1+t_2 = t_3+t_4 =0
\ee
so that only $\eta_3$, $\eta_4$, $\eta_5$, $\eta_6$ survive. In (\ref{z2}) we redefine
\bea
&& \eta_3 = \eta_{1\hat2}=-\bar\eta^{2\hat1}, ~~
   \eta_4 = \eta_{2\hat2}=\bar\eta^{1\hat1}, ~~
   \eta_5 = \eta_{1\hat1}=\bar\eta^{2\hat2}, ~~
   \eta_6 = -\eta_{2\hat1}=\bar\eta^{1\hat2} \nn\\
&& \th^3 = \th^{1\hat2}=-\bar\th_{2\hat1}, ~~
   \th^4 = \th^{2\hat2}=\bar\th_{1\hat1}, ~~
   \th^5 = \th^{1\hat1}=\bar\th_{2\hat2}, ~~
   \th^6 = -\th^{2\hat1}=\bar\th_{1\hat2}
\eea
and rewrite the Killing spinor decompositions as
\bea
&& \e_1 = \sum_{i=3,4} ( \th^i \otimes \eta_i + \bar\th_i \otimes \bar\eta^i ) = \th^{i\hi}\otimes\eta_{i\hi} = \bar\th_{i\hi}\otimes\bar\eta^{i\hi} \nn\\
&& \e_2 = \sum_{i=3,4} ( \vth^i \otimes \eta_i + \bar\vth_i \otimes \bar\eta^i ) = \vth^{i\hi}\otimes\eta_{i\hi} = \bar\vth_{i\hi}\otimes\bar\eta^{i\hi}
\eea
Since $\th^{i\hi}$, $\bar\th_{i\hi}$, $\vth^{i\hi}$, $\bar\vth_{i\hi}$ satisfy (\ref{e121}), we can identify them with the Poincar\'e and conformal supercharges of the $\mN=4$ orbifold ABJM theory.

In fact, the analysis in section~\ref{sec4.3} of the spectrum of conserved supercharges for a generic M2-- or anti--M2--brane configuration can be easily applied to the present case, simply setting
\be
\th^{12}=\th^{34}=\bar\th_{12}=\bar\th_{34}=\vth^{12}=\vth^{34}=\bar\vth_{12}=\bar\vth_{34}=0
\ee
and using the redefinition of compact space indices as in (\ref{e120}).

We first consider a M2--brane solution. Using complex coordinates ($z^1$, $z^2$, $z^{\hat 1}$, $z^{\hat 2}$) for the $\rm{C}^2 \times \rm{C}^2$ internal space, parametrized as in
(\ref{z1234}), we choose a M2--brane configuration determined by the constant vectors in $\rmS^7/(\Z_{rk}\times\Z_r)$
\bea
&& \a^i = ( \cos\frac{\b}{2}\cos\frac{\theta_1}{2} \ep^{\ii\xi_1},
            \cos\frac{\b}{2}\sin\frac{\theta_1}{2} \ep^{\ii\xi_2}), ~~
   \a^\hi =( \sin\frac{\b}{2}\sin\frac{\theta_2}{2} \ep^{\ii\xi_4},
             \sin\frac{\b}{2}\cos\frac{\theta_2}{2} \ep^{\ii\xi_3})    \\
&& \bar\a_i =( \cos\frac{\b}{2}\cos\frac{\theta_1}{2} \ep^{-\ii\xi_1},
               \cos\frac{\b}{2}\sin\frac{\theta_1}{2} \ep^{-\ii\xi_2}), ~~
   \bar\a_\hi = ( \sin\frac{\b}{2}\sin\frac{\theta_2}{2} \ep^{-\ii\xi_4},
\sin\frac{\b}{2}\cos\frac{\theta_2}{2} \ep^{-\ii\xi_3}) \nn
\eea
satisfying $\bar\a_i\a^i+\bar\a_\hi\a^\hi=1$.
From (\ref{e122}), the corresponding preserved supercharges are given by
\bea \label{e128}
&& \bg_0 \bar\a_i\th^{i\hi} = -\ii \bar\a_i \th^{i\hi}, ~~
   \bg_0 \a^i\bar\th_{i\hi} =  \ii \a^i\bar\th_{i\hi}, ~~
   \bg_0 \bar\a_\hi\th^{i\hi} = -\ii \bar\a_\hi \th^{i\hi}, ~~
   \bg_0 \a^\hi\bar\th_{i\hi} =  \ii \a^\hi\bar\th_{i\hi} \nn\\
&& \bg_0 \bar\a_i\vth^{i\hi} = -\ii \bar\a_i \vth^{i\hi}, ~~
   \bg_0 \a^i\bar\vth_{i\hi} =  \ii \a^i\bar\vth_{i\hi}, ~~
   \bg_0 \bar\a_\hi\vth^{i\hi} = -\ii \bar\a_\hi \vth^{i\hi}, ~~
   \bg_0 \a^\hi\bar\vth_{i\hi} =  \ii \a^\hi\bar\vth_{i\hi}
\eea
We discuss three different configurations.
\begin{enumerate}
  \item[1)] When $\b=0$, we have $\bar\a_\hi=0$ and the M2--brane wraps only the first $\rm{C}^2$. The preserved supercharges are
  \be
  \bg_0 \bar\a_i\th^{i\hi} = -\ii \bar\a_i \th^{i\hi}, ~~
  \bg_0 \a^i\bar\th_{i\hi} =  \ii \a^i\bar\th_{i\hi}, ~~
  \bg_0 \bar\a_i\vth^{i\hi} = -\ii \bar\a_i \vth^{i\hi}, ~~
  \bg_0 \a^i\bar\vth_{i\hi} =  \ii \a^i\bar\vth_{i\hi}
  \ee
These are exactly supercharges (\ref{sc1}) preserved by the $W$ operator with superconnection (\ref{e127}).
  \item[2)] When $\b=\pi$, we have $\bar\a_i=0$ and the M2--brane wraps only the second $\rm{C}^2$. The preserved supercharges are
  \be
  \bg_0 \bar\a_\hi\th^{i\hi} = -\ii \bar\a_\hi \th^{i\hi}, ~~
  \bg_0 \a^\hi\bar\th_{i\hi} =  \ii \a^\hi\bar\th_{i\hi},~~
  \bg_0 \bar\a_\hi\vth^{i\hi} = -\ii \bar\a_\hi \vth^{i\hi}, ~~
  \bg_0 \a^\hi\bar\vth_{i\hi} =  \ii \a^\hi\bar\vth_{i\hi}
  \ee
 These are exactly supercharges (\ref{sc2}) preserved by the  $W_\wedge$ operator with superconnection (\ref{e133}).
 \item[3)] When $\b\neq0$ and $\b\neq\pi$, we have $\bar\a_i\a^i\neq0$ and $\bar\a_\hi\a^\hi\neq0$. Such M2--branes are 1/4 BPS, and they are dual to the new 1/4 BPS WLs \cite{work-newWL}.
\end{enumerate}

For an anti--M2--brane, the analysis is similar. When $\b=0$, it preserves the same supercharges (\ref{sc3}) as the $\td W$ operator. When $\b=\pi$, it preserves the same supercharges (\ref{sc4}) as the $\td W_\wedge$ operator. For generic $\b$, it is 1/4 BPS and it is dual to a 1/4 BPS WL \cite{work-newWL}.

\section{General $\mN=4$ SCSM theories with alternating levels}\label{sec6}

Finally, we study 1/2 BPS WL operators in more general $\mN=4$ SCSM theories with gauge group and levels $\prod_{\ell=1}^{r}[U(N_{2\ell-1})_k\times U(N_{2\ell})_{-k}]$, where the $N_1,N_2,\cdots N_{2r}$ integers are generically different from each other \cite{Gaiotto:2008sd,Hosomichi:2008jd}. The quiver diagram is the same as the one for the $\mN=4$ orbifold ABJM theory in Figure~\ref{necklace} with the boundary identification $N_{2r+1}=N_1$ and $N_{2r}=N_0$.\footnote{$\mN=4$ SCSM theories with vanishing Chern--Simons levels have been introduced in \cite{Imamura:2008dt} and the BPS WLs  were studied in \cite{Cooke:2015ila}. They turn out to be very different from the ones considered in this paper.}

In order to apply the Higgsing procedure to construct 1/2 BPS WLs we can follow two different strategies.

The first strategy is based on the initial observation that a general $\mN=4$ SCSM theory with gauge group $\prod_{\ell=1}^{r}[U(N_{2\ell-1})_k\times U(N_{2\ell})_{-k}]$ can be obtained by a quotient of the $U(N)_k \times U(M)_{-k}$ ABJ theory where we decompose $N=N_1+N_3+\cdots+N_{2r-1}$ and $M=N_2+N_4+\cdots+N_{2r}$. As a consequence, WL operators can be easily obtained from the ones for the ABJ theory (see section~\ref{sec5.2}) by performing the orbifold projection on the excited non-relativistic modes. This is exactly the procedure we have used in the previous section to obtain WLs in the $\mN=4$ orbifold ABJM theory from the ones of the ABJM theory. Therefore, we will not repeat it here.

The second strategy consists instead in applying the Higgsing procedure directly on the lagrangian of the $\mN=4$ SCSM theory along the lines of what we have done in section~\ref{sec4.2} for the ABJ(M) theory. The calculation is straightforward but tedious, and we report it in appendix~\ref{appE} only for $W_1$, $W_1^{(\ell)}$, $\td W_1$, $\td W_1^{(\ell)}$ operators.

As for the $\mN=4$ orbifold ABJM theory, we can define double--node operators $W^{(\ell)}_{i=1,2}$,  $W^{(\ell)}_{\hi= \hat{1}, \hat{2}}$,  and the corresponding global WLs
\be
W_{i=1,2} = \sum_{\ell=1}^{r} W^{(\ell)}_{i=1,2} \qquad  , \qquad  W_{\hi= \hat{1}, \hat{2}} = \sum_{\ell=1}^{r} W^{(\ell)}_{\hi= \hat{1}, \hat{2}}
\ee
They are given by the holonomy of superconnections in eqs. (\ref{j76}), (\ref{j77}), (\ref{j91}), (\ref{j92}), and the superconnections that can be got from by R--symmetry rotations, and these superconnections contain gauge fields corresponding to the nodes of the quiver diagram plus scalar and fermion matter fields that coupled to them. The spectrum of the preserved supercharges is still given in table \ref{table4}.
As for the orbifold case there is a pairwise degeneracy of WL operators that preserve exactly the same set of supercharges.
Since we do not know the M--theory dual of  general $\mN=4$ SCSM theories with alternating levels, we cannot identify the gravity duals of these WLs and discuss this degeneracy at strong coupling. We will be back to this point briefly in section~\ref{sec7}.

\section{Conclusions and discussion}\label{sec7}

For superconformal gauge theories in three and four dimensions we have investigated 1/2 BPS Wilson loops and their string theory or M--theory duals. Using the Higgsing procedure, for each theory we have constructed two sets of WLs, $W$ and $\td W$,  that can be obtained by exciting particle and antiparticle modes, respectively. Correspondingly, each WL in the $W$ set has a string or M2--brane dual, whereas each WL in the $\td W$ set has a dual description in terms of an anti--string or anti--M2--brane.

In general, different WLs may share some preserved supercharges. We have studied the configuration of overlappings of preserved supercharges both for the operators and for the corresponding dual objects. In all cases there is a perfect matching between the two configurations.
In particular, we have found confirmation that in ${\cal N}=4$ SYM theory in four dimensions and three-dimensional ABJM theory different WLs have only a partial overlapping of preserved supercharges, so that for a given set of supercharges there is at most one single operator that is invariant under their action. For three-dimensional ${\cal N}=4$ orbifold ABJM theory  we have solved the degeneracy problem raised in \cite{Cooke:2015ila} concerning the existence of two different WLs,  $\psi_1$-- and $\psi_2$--loops, preserving exactly the same set of supercharges, apparently in contrast with the expectation that there should be only one 1/2 BPS M2--brane dual solution. In fact, we have found that the two operators are respectively dual to a M2--brane and an anti--M2--brane localized at different positions in $\goabjm$ but preserving exactly the same set of supercharges.

This WL degeneracy may have problematic consequences when compared with localization predictions. We then devote a careful discussion to this point, focusing on ABJM theory first and then on its orbifold projection.

ABJM theory can be localized to a matrix model \cite{Kapustin:2009kz}, and using this approach one can compute the expectation values of bosonic BPS WLs exactly  \cite{Kapustin:2009kz,Drukker:2009hy,Marino:2009jd,Drukker:2010nc}. In particular, since classically 1/2 BPS WLs differ from bosonic 1/6 BPS WLs by a $Q$--exact term where $Q$ is the supercharge used to localize the model \cite{Drukker:2009hy}, localization predicts the same vacuum expectation value for all 1/2 BPS and 1/6 BPS operators (note that we have to consider circular BPS WLs in euclidean space to have non-trivial expectation values).
At weak coupling, expanding the exact result one obtains total agreement with the perturbative calculations, both for the 1/6 BPS WLs \cite{Drukker:2008zx,Chen:2008bp,Rey:2008bh} and 1/2 BPS WLs \cite{Bianchi:2013zda,Bianchi:2013rma,Griguolo:2013sma}, once the framing factor is appropriately subtracted.%
\footnote{In fact, even for the bosonic 1/6 BPS WLs the framing factor is non-trivial at high orders \cite{Bianchi:2016yzj}.}
Regarding the two sets of WL operators that we have constructed in ABJM, 1/2 BPS Wilson loops $W_{I=1,2,3,4}$ are expected to have the same expectation value, being related by R--symmetry rotations. In the same way, 1/2 BPS Wilson loops $\td W_{I=1,2,3,4}$ should have the same expectation value. Using results in \cite{Bianchi:2013zda,Bianchi:2013rma,Griguolo:2013sma}, one can easily see that $W_{I}$ and $\td W_{I}$ have the same expectation value up to two loops. More generally, from the results in \cite{Griguolo:2015swa,Bianchi:2016vvm}, we may infer that $\langle W_{I} \rangle$ and $\langle \td W_{I} \rangle$ should be the same at any even order in perturbation theory, while they should be opposite at any odd order. Therefore, consistency with the matrix model result implies that odd order terms should be identically vanishing. Unfortunately, this has not been directly checked in perturbation theory yet.
At strong coupling, $W_{I}$ and $\td W_{I}$ operators are dual, respectively, to a M2--brane and an anti--M2--brane localized at the same position.
The corresponding classical actions in euclidean space have the same Dirac--Born--Infeld (DBI) term and the opposite Chern--Simons (CS) terms
\be \label{j72}  I_{M2} \sim   \int_\S d^3\s \sqrt{g} + \ii \int_\S H  \qquad , \qquad
I_{\overline{M2}} \sim   \int_\S d^3\s \sqrt{g} - \ii \int_\S H
\ee
where $H$ is the three-form field in M--theory. Classically the CS term is vanishing, but it may be no longer true when including quantum corrections. This may be related to the possibility that $W_{I}$ and $\td W_{I}$ operators have opposite expectation values at odd orders.

For the ${\cal N}=4$ orbifold ABJM theory the situation is even more unclear, given the appearance of WL degeneracy. In fact, both $\psi_1$-- and $\psi_2$--loops are cohomologically equivalent to a 1/4 BPS bosonic WL
\cite{Ouyang:2015qma,Cooke:2015ila} for which we know the matrix model result \cite{Herzog:2010hf,Marino:2011eh,Ouyang:2015hta,Griguolo:2015swa,Bianchi:2016vvm}. Therefore, we should expect $\langle W_{\psi_1} \rangle = \langle W_{\psi_2} \rangle$ at any perturbative order. However, even in this case, using the results in \cite{Griguolo:2015swa,Bianchi:2016vvm} we conclude that this identity certainly breaks down at odd orders where the two expectation values should have opposite sign, unless they vanish. In \cite{Drukker:2009hy} it was proposed  that the failure for the two operators to separately match the matrix model result could be an indication that the actual quantum BPS operator should be given by a suitable linear combination of the two, matching the matrix model result. However, our present result about the existence of different M2--brane configurations dual to the two operators gives strong indication that the two WLs are different BPS operators also at quantum level and no degeneracy lifting should be expected from quantum corrections. If this is true, the only possibility for being consistent with the matrix model prediction is that the two expectation values vanish at any odd order. An explicit calculation to check this prediction at three loops would be desirable. If this were not the case, then the interesting question about the validity of the cohomological equivalence of the two operators at quantum level should be addressed.

\vskip 10pt
There are several interesting generalizations of our results both in field theory and gravity sides. In field theory the generalizations are straightforward.
From the $U(N)_k\times U(N)_{-k}$ ABJM theory we can easily obtain results for the $U(N)_k\times U(M)_{-k}$ ABJ theory with $N \neq M$ \cite{Hosomichi:2008jb,Aharony:2008gk}. Similarly, results for $\mN=4$ orbifold ABJ theory, as well as for a general $\mN=4$ SCSM theory with alternating levels are obtained using techniques close to the ones used for $\mN=4$ orbifold ABJM theory, as we have discussed in section~\ref{sec6} and appendix~\ref{appE}.

The gravity generalizations are instead less trivial.
The ABJ theory is dual to M--theory in $\gabjm$ background with additional torsion flux \cite{Aharony:2008gk}, and so it is possible that the $\mN=4$ orbifold ABJ theory is dual to M--theory in $\goabjm$ background with some possibly more complex torsion flux.
We do not know the M--theory dual of a general $\mN=4$ SCSM theory with alternating levels. It is supposed to be dual to M--theory in $\AdS_4 \times \X^7$ spacetime, with $\X^7$ being some non-trivial deformation of $\rmS^7$, and with some nontrivial flux turned on. It would be very interesting to investigate the supercharges preserved by M2--brane  BPS configurations in these backgrounds.
In particular, it would be interesting to construct the gravity duals of 1/2 BPS WLs $W_{i=1,2}$, $W_{\hat{i}= \hat{1}, \hat{2}}$, $\td W_{i=1,2}$ and $\td W_{\hat{i}= \hat{1}, \hat{2}}$ that we have discussed in
$\mN=4$ orbifold ABJ theory and more general $\mN=4$ SCSM theories with alternating levels. Since the pairwise degeneracy problem of WLs is present also in these theories, it would be crucial to establish whether a similar pattern is also present in the dual description. At the moment, comparison between the matrix model result and the perturbative calculation \cite{Griguolo:2015swa,Bianchi:2016vvm} shows that
at quantum level there should exist only one 1/2 BPS WL given by the linear combination $\tfrac12 (W_1 + \td W_1)$. This should be reflected by the appearance at strong coupling of one single 1/2 BPS M2--brane configuration. If this were not the case, it would mean that the cohomological equivalence may be broken quantum mechanically. We hope to come back to this interesting problem in the future.

In $\mN=6$ SCSM theories fermionic 1/6 BPS WLs have been also constructed \cite{Ouyang:2015iza,Ouyang:2015bmy}, which depend on continuous parameters and interpolate between the bosonic 1/6 BPS WL and the fermionic 1/2 BPS operator. Similarly, in $\mN=4$ SCSM theories there are also fermionic 1/4 BPS WLs \cite{Ouyang:2015iza,Ouyang:2015bmy}. In both cases, it would be nice to investigate whether these less supersymmetric fermionic WLs can be obtained using the Higgsing procedure and whether their string theory or M--theory duals can be identified. This is a project we are currently working on \cite{work-quantum}.

\acknowledgments

We would like to thank Marco Bianchi, Nadav Drukker, Luca Griguolo, Matias Leoni, Noppadol Mekareeya, Domenico Seminara and Jun-Bao Wu for helpful discussions. S.P. thanks the Galileo Galilei Institute for Theoretical Physics (GGI) for the hospitality and INFN for partial support during the completion of this work, within the program ``New Developments in AdS3/CFT2 Holography''.
This work has been supported in part by Italian Ministero dell'Istruzione, Universit\`a e Ricerca (MIUR) and Istituto
Nazionale di Fisica Nucleare (INFN) through the ``Gauge Theories, Strings, Supergravity'' (GSS) research project.
J.-j.Z. is supported by the ERC Starting Grant 637844-HBQFTNCER.

\appendix

\section{Spinor conventions in three-dimensional spacetime}\label{appA}

In three-dimensional Minkowski spacetime we use $(-++)$ signature and gamma matrices
\be \label{gm}
\bg^\m{}_\a{}^\b = (\ii\s^2,\s^1,\s^3)
\ee
with $\s^{1,2,3}$ being the usual Pauli matrices. Throughout the paper we use boldface font to indicate gamma  matrices in three dimensions. They satisfy
\be
\bg^\m\bg^\n=\eta^{\m\n}+\ve^{\m\n\r}\bg_\r
\ee
with $\ve^{012}=-\ve_{012}=1$.
We have a two-component spinor and its complex conjugate
\be
\th_\a = \left( \ba{cc} \th_1 \\ \th_2 \ea \right), ~~
\th^*_\a = \left( \ba{cc} \th^*_1 \\ \th^*_2 \ea \right)
\ee
The spinor indices are raised and lowered as
\be
\th^\a=\ve^{\a\b}\th_\b \qquad , \qquad  \th_\a=\ve_{\a\b}\th^\b
\ee
where $\ve^{12} = - \ve_{12} = 1$. We use the following shortening notation
\be
\th\psi = \th^\a\psi_\a, ~~
\th\bg^\m\psi = \th^\a \bg^\m{}_\a{}^\b \psi_\b
\ee
We define the hermitian conjugate
\be
\th^{\dagger\a} = (\th_\a)^*=\th_\a^*
\ee
and the Dirac conjugate
\be
\bar \th = -\th^\dagger\bg^0
\ee
These definitions lead to
\be
\bar \th = \th^*
\ee
The Dirac conjugate is the same as the complex conjugate in our convention.

We define the bosonic spinors
\be\label{bspinors}
u_{\pm\a} = \f{1}{\sqrt{2}} \left( \ba{cc} 1 \\ \mp\ii \ea \right) , ~~
u_{\pm}^{\a} = \f{1}{\sqrt{2}} \left( \mp\ii, -1  \right)
\ee
They satisfy useful identities
\bea
&& u^*_\pm = u_\mp, ~~
   \g_0 u_{\pm} = \pm \ii u_\pm, ~~
   u_{\pm}\g_0 =  \mp \ii u_\pm \nn\\
&& u_+ u_-=-\ii, ~~
   u_- u_+=\ii, ~~
   u_+ u_+=u_-u_-=0
\eea
Introducing
\be
\bg^\pm = \f12(\bg^1 \pm \ii \bg^2)
\ee
we have
\bea
&& \hspace{-8mm}
   \bg^+ u_- = \ii u_+, ~~ u_-\bg^+ = -\ii u_+, ~~
   \bg^- u_+ =  -\ii u_-,  ~~ u_+\bg^- = \ii u_-, ~~
   u_- \bg^+ u_- = u_+ \bg^- u_+ =-1   \nn\\
&& \hspace{-8mm}
   \bg^+ u_+ = u_+\bg^+ = \bg^- u_- = u_-\bg^- = 0, ~~
   u_+ \bg^+ u_- = u_- \bg^+ u_+ =u_+ \bg^- u_- = u_- \bg^- u_+ =0 \nn\\
\eea

\noindent A generic spinor can be written as
\be
\label{spinor1}
\th = u_+ \th_- + u_- \th_+
\ee
with $\th_\pm$ being one-component Grassmann numbers. A similar decomposition holds for its conjugate
\be
\label{spinor2}
\bar\th = u_+ \bar\th_- + u_- \bar\th_+
\ee
Since $u_\pm^*=u_\mp$, we have the following conjugation rule
\be
\bar\th_\pm = (\th_\mp)^*
\ee
Useful identities are
\be
u_+\th = -\ii\th_+, ~~
u_-\th = \ii\th_-, ~~
\th u_+ = \ii\th_+, ~~
\th u_- = -\ii\th_-
\ee
Moreover, the spinor product becomes
\be
\th\psi = \ii(\th_+\psi_- - \th_-\psi_+)
\ee

\section{Infinite mass limit in free field theories}\label{appB}

As in \cite{Lee:2010hk}, we summarize the infinite mass limit in various free field theories. Similar infinite mass limit has also been discussed in \cite{Nakayama:2009cz,Lee:2009mm}.  Note that the fields are not totally free in the sense that they are coupled to an external gauge field.

\subsection{Scalar field}

For a complex massive scalar $d$-dimensional spacetime we have the lagrangian
\be
\mL = -D_\m\bar\Phi D^\m\Phi - m^2 \bar\Phi\Phi
\ee
with covariant derivatives
\be
D_\m\Phi = \p_\m \Phi + \ii A_\m \Phi
\ee
In the $m\to\inf$ limit we can set
\be
\Phi=\f{1}{\sqrt{2m}}\phi\ep^{-\ii m t}
\ee
and get the non-relativistic action
\be
\mL = \ii \bar\phi D_0 \phi
\ee
Alternatively, we can set
\be
\Phi=\f{1}{\sqrt{2m}}\phi\ep^{\ii m t}
\ee
and get
\be
\mL = -\ii \bar\phi D_0 \phi= \ii\phi D_0\bar \phi
\ee
In the second equality we have omitted a total derivative term, as we do in other parts of the paper.

\subsection{Vector field in Maxwell theory}

For a complex vector field in $d$-dimensional Maxwell theory we have the lagrangian
\be
\mL = -\f12 \bar W_{\m\n}W^{\m\n} - m^2 \bar W_\m W^\m
\ee
with
\be
W_{\m\n}=D_\m W_\n - D_\n W_\m, ~~ D_\m W_\n = \p_\m W_\n + \ii A_\m W_\n
\ee
This describes the propagation of  $d-1$ complex degrees of freedom. We can then set
\be
W_\m = \f{1}{\sqrt{2m}}(0,w_1,\cdots, w_{d-1}) \ep^{-\ii mt}
\ee
and we obtain
\be
\mL = \ii \bar w_a D_0 w_a
\ee
where the sum over $a$ is understood.
Alternatively, we can set
\be
W_\m = \f{1}{\sqrt{2m}}(0,w_1,\cdots, w_{d-1}) \ep^{\ii mt}
\ee
and obtain
\be
\mL = -\ii \bar w_a D_0 w_a = \ii w_a D_0 \bar w_a
\ee

\subsection{Vector field in Chern--Simons theory}

For a complex vector field in three dimensions there is the possibility to write the  Chern--Simons lagrangian
\be
\mL = \f{k}{2\pi}\ve^{\m\n\r}\bar W_\m D_\n W_\r - v^2 \bar W_\m W^\m
\ee
where we choose $k>0$.
The vector field has mass
\be
m = \f{2\pi v^2}{k}
\ee
This describes the propagation of one complex degree of freedom.
We have then two options for the choice of massive modes in the lagrangian. One is
\be
 W_\m = \sqrt{\f{\pi}{k}}(0,1,-\ii)w \, \ep^{-\ii m t} \qquad {\rm leading ~to} \qquad \mL = \ii\bar w D_0 w
\ee
whereas the other is
\be
W_\m = \sqrt{\f{\pi}{k}}(0,1,\ii)w\, \ep^{\ii m t}  \qquad {\rm leading ~to} \qquad   \mL = -\ii\bar w D_0 w = \ii w D_0 \bar w
\ee

Similarly, for the lagrangian
\be
\mL = -\f{k}{2\pi}\ve^{\m\n\r}\bar W_\m D_\n W_\r - v^2 \bar W_\m W^\m , \qquad \qquad k > 0
\ee
we can choose
\be
W_\m = \sqrt{\f{\pi}{k}}(0,1,\ii)w\ep^{-\ii m t}  \qquad {\rm leading ~to} \qquad \mL = \ii\bar w D_0 w
\ee
or
\be
W_\m = \sqrt{\f{\pi}{k}}(0,1,-\ii)w\ep^{\ii m t}  \qquad {\rm leading ~to} \qquad \mL = -\ii\bar w D_0 w = \ii w D_0 \bar w
\ee

\subsection{Three-dimensional Dirac field}

Finally, the lagrangian for a three-dimensional Dirac field is
\be
\mL = \ii\bar\Psi\bg^\m D_\m \Psi - \ii m \bar \Psi \Psi
\ee
We can choose the massive modes as
\be
\Psi = u_- \psi \ep^{-\ii m t} \qquad {\rm leading ~to} \qquad  \mL = \ii \bar\psi D_0 \psi
\ee
or
\be
\Psi = u_+ \psi \ep^{\ii m t} \qquad {\rm leading ~to} \qquad  \mL = \ii \bar\psi D_0 \psi = \ii\psi D_0 \bar \psi
\ee

Similarly, for the lagrangian
\be
\mL = \ii\bar\Psi\bg^\m D_\m \Psi + \ii m \bar \Psi \Psi
\ee
we can choose
\be
\Psi = u_+ \psi \ep^{-\ii m t} \qquad {\rm leading ~to} \qquad  \mL = \ii \bar\psi D_0 \psi
\ee
or
\be
\Psi = u_- \psi \ep^{\ii m t} \qquad {\rm leading ~to} \qquad  \mL = \ii \bar\psi D_0 \psi = \ii\psi D_0 \bar \psi
\ee

\section{Killing spinors in AdS$_5\times$S$^5$ spacetime}\label{appC}

Killing spinors in $\gsym$, $\textrm{AdS}_4 \times \textrm{S}^7$, and $\gtzt$ spacetimes have been determined in, for example, \cite{Lu:1998nu,Claus:1998yw,Skenderis:2002vf,Nishioka:2008ib,Drukker:2008zx}. However, since we use different sets of coordinates, we rederive them.  In this appendix we focus on the $\gsym$ case, whereas we devote appendix~\ref{appD} to the calculation for $\textrm{AdS}_4\times\textrm{S}^7$ and appendix~\ref{appF} to $\gtzt$.

In $\gsym$ we assign curved coordinates $x^M=(x^\tm,x^\ti)$, where $x^\tm$ and $x^\ti$ belong to AdS$_5$ and S$^5$, respectively.\footnote{In this paper we use $x^\m=(t,x_1,x_2,x_3)$ to denote the worldvolume coordinates of the stack of D3--branes before taking the near horizon limit, i.e. coordinates of the four--dimensional $\mN=4$ SYM theory. Moreover, we use $x^i=(x_4,x_5,x_6,x_7,x_8,x_9)$ to denote the directions perpendicular to the D3--branes. The $i$ index corresponds to the $SO(6)$ R--symmetry index $I$ in SYM theory.}
In tangent space we use flat coordinates $x^A=(x^{\td a},x^{\td p})$ with $\td a=0,1,2,3,4$ and $\td p=5,6,7,8,9$.

Given the AdS$_5$ and S$^5$ metrics (\ref{ads5}, \ref{s5}), we can easily read the vierbeins
\be
e^0= u dt, ~~ e^1= u d x_1, ~~ e^2= u d x_2, ~~ e^3= u d x_3, ~~ e^4 = \f{du}{u}
\ee
\be
e^5=d\th_1, ~~ e^6=\cos\th_1 d\xi_1, ~~ e^7=\sin\th_1 d\th_2, ~~ e^8=\sin\th_1\cos\th_2 d\xi_2, ~~ e^9=\sin\th_1\sin\th_2 d\xi_3
\nonumber
\ee
The $\gsym$ vierbein components are then given by
\be
E_\tm^{\td a} = R e_\tm^{\td a}, ~~ E_\ti^{\td p} = R e_\ti^{\td p}
\ee
From the constraint $dE+\o \wedge E=0$, we obtain the nonvanishing components of the spin connection
\bea
&& \o^{04}_{t} = \o^{14}_{x_1} = \o^{24}_{x_2} = \o^{34}_{x_3} = u \nn\\
&& \o^{56}_{\xi_1}=\sin\th_1, ~~
   \o^{57}_{\th_2}=-\cos\th_1, ~~
   \o^{58}_{\xi_2}=-\cos\th_1\cos\th_2 \nn\\
&& \o^{78}_{\xi_2}=\sin\th_2, ~~
   \o^{59}_{\xi_3}=-\cos\th_1\sin\th_2, ~~
   \o^{79}_{\xi_3}=-\cos\th_2
\eea

From the SUSY variation of the gravitino in type IIB supergravity we obtain the Killing spinor equation
\be \label{e139}
D_M \e + \f{\ii}{1920}F_{NPQRS}\G^{NPQRS}\G_{M}\e=0
\ee
with $\e$ being a Weyl spinor with positive chirality and $D_M \e=\p_M \e+ \f14 \o^{AB}_M \g_{AB}\e$. Note that we have gamma matrices
\be
\G_M = E_M^A \g_A, ~~ \g_\tm = e_\tm^{\td a} \g_{\td a}, ~~ \g_\ti = e_\ti^{\td p} \g_{\td p}
\ee
and $\G_{\td \m}=R\g_{\td \mu}$, $\G_\ti=R\g_\ti$.
We write the Killing spinor equations (\ref{e139}) as
\be
D_\tm \e = - \f{\ii}{2}\hat\g\g_\tm\e, ~~ D_\ti \e = - \f{\ii}{2}\hat\g\g_\ti\e
\ee
with $\hat \g=\g^{01234}$. Defining $\td \g = \g^{0123}$, they can be rewritten as
\bea
&& \p_{\m} \e = -\f{u}{2}\g_{\m4}(1+\ii\td\g)\e, ~~ \p_u \e = -\f{\ii}{2u} \td\g \e \nn\\
&& \p_{\th_1}\e = -\f{\ii}{2}\hat\g\g_5\e, ~~
   \p_{\th_2}\e = \f{1}{2}\g_{57}\ep^{\ii\th_1\hat\g\g_5}\e, ~~
   \p_{\xi_1}\e = -\f{\ii}2\hat\g\g_{6}\ep^{\ii\th_1\hat\g\g_5}\e \nn\\
&& \p_{\xi_2}\e = \f12 ( \g_{58}\ep^{\ii\th_1\hat\g\g_5}\cos\th_2-\g_{78}\sin\th_2 )\e \nn\\
&& \p_{\xi_3}\e = \f12(\g_{79}\cos\th_2 + \g_{59}\ep^{\ii\th_1\hat\g\g_5}\sin\th_2)\e
\eea
The solution to these equations has been found in \cite{Skenderis:2002vf}. In our conventions it reads
\be\label{ks1}
\e 
   = u^{\f12}h(\e_1+x^{\m}\g_{\m}\e_2) - u^{-\f12} \g_4 h \e_2
\ee
where
\be \label{e96}
h = \ep^{\f{\th_1}{2}\g_{45}} \ep^{\f{\th_2}{2}\g_{57}}
       \ep^{\f{\xi_1}{2}\g_{46}} \ep^{\f{\xi_2}{2}\g_{58}} \ep^{\f{\xi_3}{2}\g_{79}}
\ee
and $\e_1$, $\e_2$ are constant spinors subject to the constraints
\be \label{e95}
\td\g \e_1 = \ii\e_1, ~~ \td\g \e_2 = - \ii \e_2
\ee
Since $\e$ is a positive chirality spinor, i.e. $\g \e=\e$ with $\g=\g_{01\cdots9}$, it follows that  $\g\e_1=\e_1$ and $\g\e_2=-\e_2$. From constraints (\ref{e95}) it follows that the two Weyl spinors $\e_1$, $\e_2$ can be further written in terms of  two Majorana--Weyl spinors $\th$, $\vth$, with respectively positive and negative chiralities
\bea \label{e97}
&& \e_1 = (1-\ii\td\g)\th, ~~ \th = \f12 (\e_1 + \e_1^c) \nn\\
&& \e_2 = (1+\ii\td\g)\vth, ~~ \vth = \f12 (\e_2 + \e_2^c)
\eea
As described in the main text, $\th$ and $\vth$ can be identified respectively with the Poincar\'e supercharges $\th$ and superconformal charges $\vth$ of the four-dimensional SYM theory \cite{Gomis:2006sb}.

\section{Killing spinors in AdS$_4\times$S$^7$ spacetime}\label{appD}

Killing spinors in $\AdS_4\times \rm{S}^7$ spacetime has been obtained in, for example, \cite{Lu:1998nu,Nishioka:2008ib,Drukker:2008zx}. In this appendix we review the derivation in the background
\bea \label{ads4s7}
&& ds^2=R^2 \lt( \frac14 ds^2_{\AdS_4}+ds^2_{ {\rm S}^7} \rt) \nn \\
&& F_{\td\m\td\n\td\r\td\s} = \f{6}{R} \ve_{\td\m\td\n\td\r\td\s}
\eea
with metrics (\ref{ads4}) and (\ref{s7}). Since we use the same $\rmS^7$ coordinates as the ones used in \cite{Drukker:2008zx}, the results therein will be useful to us.

We denote the $\AdS_4\times \rmS^7$ coordinates as $x^M=(x^{\td\m},x^\ti)$, with $x^{\td\m}$ and $x^\ti$ being coordinates of AdS$_4$ and S$^7$, respectively, and tangent space coordinates as $x^A=(x^{\td a},x^{\td p})$ with $\td a=0,1,2,3$ and $\td p=4,5,6,7,8,9,\natural$.\footnote{Before taking the near horizon limit, for the stack of M2--branes we use  worldvolume coordinates $x^\m=(t,x_1,x_2)$ and tangent space coordinates $x^a$ with $a=0,1,2$. For the orthogonal directions we use $x^i=(x_3,x_4,\cdots,x_9,x_\natural)$ and tangent coordinates $x^p$ with $p=3,4,\cdots,9,\natural$. }
For the AdS$_4$ metric (\ref{ads4}) we use the vierbeins
\be
e^0=u dt, ~~ e^1 = u dx_1, ~~ e^2 = u dx_2, ~~ e^3 = \f{d u}{u}
\ee
whereas the vierbeins $e_\ti^{\td p}$ for S$^7$ metric (\ref{s7}) can be found in \cite{Drukker:2008zx} and we avoid rewriting them here.
The vierbeins of the $\AdS_4\times\rmS^7$ background (\ref{ads4s7}) are then given by
\be
E_{\td m}^{\td a} = \f{R}2 e_{\td\m}^{\td a}, ~~ E_\ti^{\td p} = R e_\ti^{\td p}
\ee
The non-vanishing components of the spin connection for AdS$_4$ are
\be
\o^{03}_t = \o^{13}_{x_1} = \o^{23}_{x_2} = u
\ee
and those for S$^7$ can be found in \cite{Drukker:2008zx}.

The Killing spinor equations now read
\be \label{e140}
D^M \e =\f1{288}F_{NPQR}(\G^{MNPQR} - 8G^{MN}\G^{PQR})\e
\ee
with $\e$ being a Majorana spinor.
Note that $\G_{\td \m}=\f{R}{2}\g_{\td \mu}$, $\G_\ti = R\g_\ti$.
We rewrite (\ref{e140}) as
\be
D_{\td\m} \e = \f{1}{2}\hat\g\g_{\td\m}\e, ~~ D_\ti \e = \f{1}{2}\hat\g\g_\ti\e
\ee
with $\hat \g=\g^{0123}$. Defining $\td \g = \g^{012}$,
these equations in the AdS$_4$ directions become
\be
\p_{\m}\e = -\f{u}{2}\g_{\m 3}(1-\td\g)\e, ~~ \p_u \e = \f{1}{2u}\td\g\e
\ee
whereas the ones in the S$^7$ directions can be found in \cite{Drukker:2008zx}.

In our conventions the general solution reads
\be \label{e105}
\e 
   = u^{\f12}h(\e_1+x^{\m}\g_{\m}\e_2) - u^{-\f12} \g_3 h \e_2
\ee
with constant Majorana spinors $\e_1$, $\e_2$ satisfying $\td\g\e_1=\e_1$, $\td\g\e_2=\e_2$, and
\be \label{gh2}
h = \ep^{\frac{\b}{4}(\g_{34}-\g_{7\natural})}
       \ep^{\frac{\theta_1}4 (\g_{35}-\g_{8\natural})}
       \ep^{\frac{\theta_2}4(\g_{46}+\g_{79}) }
       \ep^{\frac{\xi_1}2\g_{3\natural}}
       \ep^{\frac{\xi_2}{2}\g_{58}}
       \ep^{\frac{\xi_3}2\g_{47}}
       \ep^{\frac{\xi_4}2\g_{69}}
\ee
We have in total 32 real degrees of freedom, 16 from $\e_1$ and 16 from $\e_2$.

For our purposes it is convenient to decompose  $\e_1$, $\e_2$ in two different ways.

First, we decompose them in terms of eigenstates of $\g_{03\natural}$, $\g_{058}$, $\g_{047}$, $\g_{069}$. We write \footnote{We note that these equations  are compatible with the Majorana nature of $ \e_i$ \cite{Ouyang:2015ada}.}
\be \label{e107}
\g_{03\natural} \e_i = s_1 \e_i, ~~
\g_{058} \e_i = s_2  \e_i, ~~
\g_{047} \e_i = s_3  \e_i, ~~
\g_{069} \e_i = s_4  \e_i, \qquad i=1,2
\ee
with $s_I = \pm 1$, $I=1,2,3,4$. From the constraint $\td\g\e_i=\e_i$ and the identity $\g_{0123456789\natural}=1$, it follows that
\be
s_1s_2s_3s_4=1
\ee
Therefore, both $\e_1$ and $\e_2$ are decomposed into eight possible states
\bea  \label{e106}
&& (s_1,s_2,s_3,s_4) = (++++), (++--), (+-+-), (+--+), \nn\\
&& \phantom{(s_1,s_2,s_3,s_4) = }
                       (-++-), (-+-+), (--++), (----)
\eea
and each state has two real degrees of freedom.

Alternatively, we can decompose $\e_1, \e_2$ as direct product of Grassmann odd spinors $\th$ and $\vth$ in R$^{1,2}$ and Grassmann even spinors $\eta$ in $\rm{C}^4\cong\rm{R}^8$. Schematically we write
\be \label{e92}
\e_1 \sim \th \otimes \eta \qquad \qquad \e_2 \sim \vth \otimes \eta
\ee
To this end we decompose the eleven-dimensional gamma matrices as
\bea \label{z11}
&& \g_{a} = -\bg_{a} \otimes \bG, ~~ a = 0,1,2 \nn\\
&& \g_{p} = \bo \otimes \bG_{p}, ~~ p = 3,4,5,6,7,8,9,\natural
\eea
where $\bg_a$ are given in (\ref{gm}) (therefore $\td \g = \g^{012} = -\bo \otimes \bG$) and
\be
\bG = \bG_{3456789\natural} = -\bG_{3\natural}\bG_{58}\bG_{47}\bG_{69}
\ee
The $\eta$ spinor can be decomposed in terms of eigenstates
\be \label{z1}
\bG_{3\natural} \eta = \ii t_1 \eta, ~~
\bG_{58} \eta = \ii t_2 \eta, ~~
\bG_{47} \eta = \ii t_3 \eta, ~~
\bG_{69} \eta = \ii t_4 \eta
\ee
with $t_I=\pm$ for $I=1,2,3,4$.
The constraint $\td \g \e_1 = \e_1$ is equivalent to $\bG\eta = -\eta$, and this leads to $t_1t_2t_3t_4=1$.
The $\eta$ spinor is then decomposed into eight states
\bea \label{z6}
&& (t_1,t_2,t_3,t_4) = (++++), (++--), (+-+-), (+--+), \nn\\
&& \phantom{(t_1,t_2,t_3,t_4) = }
                       (-++-), (-+-+), (--++), (----)
\eea
and we name them in the present order as $\eta_i$, $i=1,2,\cdots,8$. Taking the charge conjugate of (\ref{z1}), we obtain
($\bar\eta \equiv \eta^c$, $\bar\eta^i \equiv \eta_i^c$  in $\R^8$)
\be
\bG_{3\natural} \bar\eta = -\ii t_1 \bar\eta, ~~
\bG_{58} \bar\eta = -\ii t_2 \bar\eta, ~~
\bG_{47} \bar\eta = -\ii t_3 \bar\eta, ~~
\bG_{69} \bar\eta = -\ii t_4 \bar\eta
\ee
We normalize $\eta_i$ in such a way that
\be
\bar\eta^1 = \eta_8, ~~ \bar\eta^2 = \eta_7, ~~ \bar\eta^3 = \eta_6, ~~ \bar\eta^4 = \eta_5
\ee
Then we write $\e_1, \e_2$ as
\be
\e_1 = \sum_{i=1}^8 \th^i \otimes \eta_i  \qquad , \qquad \e_2 = \sum_{i=1}^8 \vth^i \otimes \eta_i
\ee
Since they are Majorana spinors, we can define $\bar\th_i=\th^{ic}=(\th^{i})^*$ with the assignment
\be \label{z4}
\bar\th_1 = \th^8, ~~ \bar\th_2 = \th^7, ~~ \bar\th_3 = \th^6, ~~ \bar\th_4 = \th^5
\ee
Finally we can write
\be \label{z2}
\e_1 = \sum_{i=1}^4 ( \th^i \otimes \eta_i + \bar\th_i \otimes \bar\eta^i ) \qquad , \qquad \e_2 = \sum_{i=1}^4 ( \vth^i \otimes \eta_i + \bar\vth_i \otimes \bar\eta^i )
\ee
where the eleven dimensional Majorana spinors have been expressed in terms of four independent Dirac spinors in three dimensions.

\section{Higgsing procedure in general ${\cal N}=4$ SCSM theories} \label{appE}

In this appendix we give details about the Higgsing procedure for Wilson loops $W_1$, $W_1^{(\ell)}$, $\td W_1$ ,$\td W_1^{(\ell)}$ in a general $\mN = 4$ SCSM theory with alternating levels.

As can be inferred from the quiver diagram, Figure~\ref{necklace}, in a general $\mN=4$ SCSM theory with alternating levels we have gauge fields $A_\m^{(2\ell-1)}$, $B_\m^{(2\ell)}$ and bi--fundamental matter fields $\phi_\hi^{(2\ell-1)}$, $\psi^\hi_{(2\ell-1)}$, $\bar\phi^\hi_{(2\ell-1)} = (\phi_\hi^{(2\ell-1)})^\dagger$, $\bar\psi_\hi^{(2\ell-1)} = (\psi^\hi_{(2\ell-1)})^\dagger$, $\phi_i^{(2\ell)}$, $\psi^i_{(2\ell)}$, $\bar\phi^i_{(2\ell)} = (\phi_i^{(2\ell)})^\dagger$, $\bar\psi_i^{(2\ell)} = (\psi^i_{(2\ell)})^\dagger$ that couple to them.  Here $\ell=1,2,\cdots,r$ with identifications $(2r+1)=(1)$, $(2r)=(0)$, and  $i=1,2$, $\hi=\ho,\hw$.

We write the lagrangian  as a sum of four terms
\be \label{LagN4}
\mL = \mL_{CS} + \mL_{k} +\mL_{p} + \mL_{Y}
\ee
Explicitly, the Chern--Simons part is given  by
{\small \be \hspace{-2mm}
\mL_{CS} =\f{k}{4\pi} \sum_{\ell=1}^{r} \ve^{\m\n\r} \Tr \Big(
A_\m^{(2\ell-1)}\p_\n A_\r ^{(2\ell-1)} + \f{2\ii}{3}A_\m^{(2\ell-1)}A_\n^{(2\ell-1)}A_\r^{(2\ell-1)}
- B_\m^{(2\ell)}\p_\n B_\r ^{(2\ell)} - \f{2\ii}{3}B_\m^{(2\ell)}B_\n^{(2\ell)}B_\r^{(2\ell)} \Big)
\ee}%
The kinetic part of the scalars and fermions is
{\small \be \hspace{-1mm}
\mL_{k} = \sum_{\ell=1}^{r} \Tr \Big(
- D_\m \bar\phi^\hi_{(2\ell-1)} D^\m \phi_\hi ^{(2\ell-1)}
+ \ii  \bar\psi_i^{(2\ell-1)} \bg^\m D_\m  \psi^i _{(2\ell-1)}
-D_\m \bar\phi^i_{(2\ell)} D^ \m \phi_i ^{(2\ell)}
+ \ii  \bar\psi_\hi^{(2\ell)} \bg^\m D_\m  \psi^\hi _{(2\ell)}  \Big)
\ee}%
with covariant derivatives being
\bea
&& D_\m \phi_\hi^{(2\ell-1)} =\p_\m \phi_\hi^{(2\ell-1)} +\ii A_\m^{(2\ell-1)} \phi_\hi^{(2\ell-1)} -\ii \phi_\hi^{(2\ell-1)} B_\m ^{(2\ell)} \nn\\
&& D_\m \bar\phi^\hi_{(2\ell-1)} =\p_\m \bar\phi^\hi_{(2\ell-1)} +\ii B_\m ^{(2\ell)} \bar\phi^\hi_{(2\ell-1)} -\ii \bar\phi^\hi_{(2\ell-1)}A_\m^{(2\ell-1)} \nn\\
&& D_\m \psi^i_{(2\ell-1)} =\p_\m \psi^i_{(2\ell-1)} +\ii A_\m^{(2\ell-1)} \psi^i_{(2\ell-1)} -\ii \psi^i_{(2\ell-1)} B_\m ^{(2\ell)} \nn\\
&& D_\m \phi_i^{(2\ell)} = \p_\m \phi_i^{(2\ell)}
                             +\ii A_\m^{(2\ell+1)} \phi_i^{(2\ell)}
                             -\ii \phi_i^{(2\ell)} B_\m ^{(2\ell)} \nn\\
&& D_\m \bar\phi^i_{(2\ell)} = \p_\m \bar\phi^i_{(2\ell)}
                                +\ii B_\m ^{(2\ell)} \bar\phi^i_{(2\ell)}
                                -\ii \bar\phi^i_{(2\ell)}A_\m^{(2\ell+1)} \nn\\
&& D_\m \psi^\hi_{(2\ell)} = \p_\m \psi^\hi_{(2\ell)}
                               +\ii A_\m^{(2\ell+1)} \psi^\hi_{(2\ell)}
                               -\ii \psi^\hi_{(2\ell)} B_\m ^{(2\ell)}
\eea
The  potential part is
\be
\mL_{p} = \f{4\pi^2}{3k^2}\sum_{\ell=1}^{r} \Tr \Big(\mL_{p}^{(2\ell-1)} + \mL_{p}^{(2\ell)}  \Big)
\ee
with{\small
\bea
&& \hspace{-4mm}
  \mL_{p}^{(2\ell-1)} = \phi_\hi ^{(2\ell-1)}\bar\phi^\hi_{(2\ell-1)}\phi_\hj^{(2\ell-1)}\bar\phi^\hj_{(2\ell-1)}\phi_\hk ^{(2\ell-1)} \bar\phi^\hk_{(2\ell-1)} + \phi_\hi ^{(2\ell-1)}\bar\phi^\hj_{(2\ell-1)}\phi_\hj^{(2\ell-1)} \bar\phi^\hk_{(2\ell-1)}\phi_\hk ^{(2\ell-1)}\bar\phi^\hi_{(2\ell-1)} \nn\\
&&  \hspace{-4mm} \phantom{\mL_{p}^{(2\ell-1)}=}+4\phi_\hi ^{(2\ell-1)}\bar\phi^\hj_{(2\ell-1)}\phi_\hk^{(2\ell-1)}\bar\phi^\hi_{(2\ell-1)}\phi_\hj^{(2\ell-1)}\bar\phi^\hk_{(2\ell-1)} - 6\phi_\hi^{(2\ell-1)} \bar\phi^\hj_{(2\ell-1)}\phi_\hj^{(2\ell-1)}\bar\phi^\hi_{(2\ell-1)}\phi_\hk^{(2\ell-1)} \bar\phi^\hk_{(2\ell-1 )} \nn\\
&& \hspace{-4mm} \phantom{\mL_{p}^{(2\ell-1)}=}+ 3\phi_\hi ^{(2\ell-1)}\bar\phi^\hi_{(2\ell-1)}\phi_\hj^{(2\ell-1)}\bar\phi^\hj_{(2\ell-1)}\phi_k ^{(2\ell-2)} \bar\phi^k_{(2\ell-2)} + 3\phi_i ^{(2\ell)}\bar\phi^\hj_{(2\ell-1)}\phi_\hj^{(2\ell-1)}\bar\phi^\hk_{(2\ell-1)}\phi_\hk ^{(2\ell-1)} \bar\phi^i_{(2\ell)} \nn\\
&& \hspace{-4mm} \phantom{\mL_{p}^{(2\ell-1)}=}+ 12\phi_i ^{(2\ell)}\bar\phi^\hj_{(2\ell-1)}\phi_\hk^{(2\ell-1)}\bar\phi^i_{(2\ell)}\phi_\hj ^{(2\ell+1)} \bar\phi^\hk_{(2\ell+1)} - 6\phi_\hi ^{(2\ell-1)}\bar\phi^\hj_{(2\ell-1)}\phi_\hj^{(2\ell-1)}\bar\phi^\hi_{(2\ell-1)}\phi_k ^{(2\ell-2)} \bar\phi^k_{(2\ell-2)}\nn\\
&& \hspace{-4mm} \phantom{\mL_{p}^{(2\ell-1)}=} - 6\phi_i^{(2\ell)}\bar\phi^\hj_{(2\ell-1)}\phi_\hj^{(2\ell-1)}\bar\phi^i_{(2\ell)}\phi_\hk^{(2\ell+1)}\bar\phi^\hk_{(2\ell+1)} - 6\phi_\hi^{(2\ell-1)}\bar\phi^j_{(2\ell)}\phi_j^{(2\ell)}\bar\phi^\hi_{(2\ell-1)}\phi_\hk^{(2\ell-1)}\bar\phi^\hk_{(2\ell-1)}
\eea}%
and
\bea
&& \hspace{-8mm}
  \mL_{p}^{(2\ell)} = \phi_i^{(2\ell)}\bar\phi^i_{(2\ell)}\phi_j^{(2\ell)}\bar\phi^j_{(2\ell)}\phi_k ^{(2\ell)}\bar\phi^k_{(2\ell)} + \phi_i^{(2\ell)}\bar\phi^j_{(2\ell)}\phi_j^{(2\ell)} \bar\phi^k_{(2\ell)}\phi_k^{(2\ell)}\bar\phi^i_{(2\ell)} \nn\\
&& \hspace{-8mm} \phantom{\mL_{p}^{(2\ell)}=}+ 4\phi_i^{(2\ell)}\bar\phi^j_{(2\ell)}\phi_k^{(2\ell)}\bar\phi^i_{(2\ell)}\phi_j^{(2\ell)}\bar\phi^k_{(2\ell)} - 6\phi_i^{(2\ell)}\bar\phi^j_{(2\ell)}\phi_j^{(2\ell)}\bar\phi^i_{(2\ell)}\phi_k ^{(2\ell)}\bar\phi^k_{(2\ell)}\nn\\
&& \hspace{-8mm} \phantom{\mL_{p}^{(2\ell)}=}+ 3\phi_i ^{(2\ell)}\bar\phi^i_{(2\ell)}\phi_j^{(2\ell)}\bar\phi^j_{(2\ell2)}\phi_\hk ^{(2\ell+1)} \bar\phi^\hk_{(2\ell+1)} + 3\phi_\hi ^{(2\ell-1)}\bar\phi^j_{(2\ell)}\phi_j^{(2\ell)}\bar\phi^k_{(2\ell)}\phi_k ^{(2\ell)} \bar\phi^\hi_{(2\ell-1)} \nn\\
&& \hspace{-8mm} \phantom{\mL_{p}^{(2\ell)}=}+ 12\phi_i ^{(2\ell)}\bar\phi^j_{(2\ell)}\phi_\hk^{(2\ell+1)}\bar\phi^i_{(2\ell+2)}\phi_j ^{(2\ell+2)} \bar\phi^\hk_{(2\ell+1)} - 6\phi_i ^{(2\ell)}\bar\phi^j_{(2\ell)}\phi_j^{(2\ell)}\bar\phi^i_{(2\ell)}\phi_\hk ^{(2\ell+1)} \bar\phi^\hk_{(2\ell+1)}\nn\\
&& \hspace{-8mm} \phantom{\mL_{p}^{(2\ell)}=} - 6\phi_\hi^{(2\ell+1)}\bar\phi^j_{(2\ell+2)}\phi_j^{(2\ell+2)}\bar\phi^\hi_{(2\ell+1)}\phi_k^{(2\ell)}\bar\phi^k_{(2\ell)} - 6\phi_i^{(2\ell)}\bar\phi^\hj_{(2\ell-1)}\phi_\hj^{(2\ell-1)}\bar\phi^i_{(2\ell)}\phi_k^{(2\ell)}\bar\phi^k_{(2\ell)}
\eea
The part containing Yukawa couplings is
\bea
&&\mL_{Y} = \f{2\pi \ii}{k} \sum_{\ell=1}^{r} \Tr \Big(
                       \phi_i^{(2\ell)}\bar\phi^i_{(2\ell)}\psi^j_{(2\ell+1)}\bar\psi_j^{(2\ell+1)}
                       + \phi_\hi^{(2\ell+1)}\bar\phi^\hi_{(2\ell+1)}\psi^j_{(2\ell+1)}\bar\psi_j^{(2\ell+1)}  \nn\\
&&\phantom{\mL_{Y} =} +\phi_i^{(2\ell)}\bar\phi^i_{(2\ell)}\psi^\hj_{(2\ell)}\bar\psi_\hj^{(2\ell)}
                      + \phi_\hi^{(2\ell+1)}\bar\phi^\hi_{(2\ell+1)}\psi^\hj_{(2\ell)}\bar\psi_\hj^{(2\ell)} \nn\\
&&\phantom{\mL_{Y} =} -2 \phi_i^{(2\ell)}\bar\phi^j_{(2\ell)}\psi^i_{(2\ell+1)}\bar\psi_j^{(2\ell+1)}
                      -2 \phi_\hi^{(2\ell-1)}\bar\phi^j_{(2\ell)}\psi^\hi_{(2\ell)}\bar\psi_j^{(2\ell-1)} \nn\\
&&\phantom{\mL_{Y} =} -2 \phi_i^{(2\ell)}\bar\phi^\hj_{(2\ell-1)}\psi^i_{(2\ell-1)}\bar\psi_\hj^{(2\ell)}
                      -2  \phi_\hi^{(2\ell+1)}\bar\phi^\hj_{(2\ell+1)}\psi^\hi_{(2\ell)}\bar\psi_\hj^{(2\ell)} \nn\\
&&\phantom{\mL_{Y} =} - \bar\phi^i_{(2\ell)}\phi_i^{(2\ell)}\bar\psi_j^{(2\ell+1)}\psi^j_{(2\ell+1)}
                      - \bar\phi^\hi_{(2\ell+1)}\phi_\hi^{(2\ell+1)}\bar\psi_j^{(2\ell+1)}\psi^j_{(2\ell+1)} \nn\\
&&\phantom{\mL_{Y} =} - \bar\phi^i_{(2\ell)}\phi_i^{(2\ell)}\bar\psi_\hj^{(2\ell)}\psi^\hj_{(2\ell)}
                      - \bar\phi^\hi_{(2\ell+1)}\phi_\hi^{(2\ell+1)}\bar\psi_\hj^{(2\ell)}\psi^\hj_{(2\ell)} \nn\\
&&\phantom{\mL_{Y} =} + 2 \bar\phi^i_{(2\ell)}\phi_j^{(2\ell)}\bar\psi_i^{(2\ell+1)}\psi^j_{(2\ell+1)}
                      + 2 \bar\phi^i_{(2\ell)}\phi_\hj^{(2\ell-1)}\bar\psi_i^{(2\ell-1)}\psi^\hj_{(2\ell)} \nn\\
&&\phantom{\mL_{Y} =} + 2 \bar\phi^\hi_{(2\ell-1)}\phi_j^{(2\ell)}\bar\psi_\hi^{(2\ell)}\psi^j_{(2\ell-1)}
                      +2 \bar\phi^\hi_{(2\ell+1)} \phi_\hj^{(2\ell+1)}\bar\psi_\hi^{(2\ell)}\psi^\hj_{(2\ell)} \nn\\
&&\phantom{\mL_{Y} =} + 2 \ve^{ij}\ve^{\hk\hl} \phi_i^{(2\ell)} \bar\psi_j^{(2\ell-1)} \phi_\hk^{(2\ell-1)} \bar\psi_\hl^{(2\ell)}
                      - \ve^{ij}\ve^{\hk\hl} \phi_i^{(2\ell)} \bar\psi_\hk^{(2\ell)} \phi_j^{(2\ell)} \bar\psi_\hl^{(2\ell)} \nn\\
&&\phantom{\mL_{Y} =} + 2 \ve^{\hi\hj}\ve^{kl} \phi_\hi^{(2\ell+1)} \bar\psi_k^{(2\ell+1)} \phi_l^{(2\ell)} \bar\psi_\hj^{(2\ell)}
                      - \ve^{\hi\hj}\ve^{kl} \phi_\hi^{(2\ell+1)} \bar\psi_k^{(2\ell+1)} \phi_\hj^{(2\ell+1)} \bar\psi_l^{(2\ell+1)} \nn\\
&&\phantom{\mL_{Y} =} - 2 \ve_{ij}\ve_{\hk\hl} \bar\phi^i_{(2\ell)}\psi^j_{(2\ell+1)} \bar\phi^\hk_{(2\ell+1)}\psi^\hl_{(2\ell)}
                      + \ve_{ij}\ve_{\hk\hl} \bar\phi^i_{(2\ell)}\psi^\hk_{(2\ell)} \bar\phi^j_{(2\ell)}\psi^\hl_{(2\ell)} \nn\\
&&\phantom{\mL_{Y} =} -2 \ve_{\hi\hj}\ve_{kl} \bar\phi^\hi_{(2\ell-1)}\psi^k_{(2\ell-1)}\bar\phi^l_{(2\ell)}\psi^\hj_{(2\ell)}
                      + \ve_{\hi\hj}\ve_{kl} \bar\phi^\hi_{(2\ell-1)}\psi^k_{(2\ell-1)}\bar\phi^\hj_{(2\ell-1)}\psi^l_{(2\ell-1)} \Big)
\eea
with $\ve_{ij}$, $\ve_{\hi\hj}$, $\ve^{ij}$, $\ve^{\hi\hj}$ being antisymmetric and $\ve_{12}=\ve^{12}=\ve_{\hat{1}\hat{2}}=\ve^{\hat{1}\hat{2}}=1$. \\[0.2cm]
The lagrangian \eqref{LagN4} is invariant under the following SUSY transformations:
\noindent \\[0.15cm]
-- Gauge vectors
{\small \bea
&&  \hspace{-2mm}
  \d A_\m^{(2\ell-1)}=-\f{2\pi}{k} \Big[ \Big( \phi_i^{(2\ell-2)}\bar\psi_\hi^{(2\ell-2)}
                                              - \phi_\hi^{(2\ell-1)}\bar\psi_i^{(2\ell-1)} \Big) \bg_\m \e^{i\hi}
                   +\bar\e_{i\hi}\bg_\m \Big( \psi^\hi_{(2\ell-2)}\bar\phi^i_{(2\ell-2)}
                                            - \psi^i_{(2\ell-1)}\bar\phi^\hi_{(2\ell-1)} \Big) \Big]    \nn\\
&&  \hspace{-2mm}
  \d B_\m^{(2\ell)}=-\f{2\pi}{k} \Big[ \Big( \bar\psi_i^{(2\ell-1)}\phi_\hi^{(2\ell-1)}
                                              - \bar\psi_\hi^{(2\ell)}\phi_i^{(2\ell)} \Big) \bg_\m\e^{i\hi}
                   +\bar\e_{i\hi}\bg_\m \Big( \bar\phi^i_{(2\ell)}\psi^\hi_{(2\ell)}
                                            - \bar\phi^\hi_{(2\ell-1)}\psi^i_{(2\ell-1)} \Big) \Big]
\eea}%
-- Scalar fields
{\small \be
\d\phi_\hi^{(2\ell-1)} = -\ii\bar\e_{i\hi}\psi^i_{(2\ell-1)}, ~~
\d\bar\phi^\hi_{(2\ell-1)} = -\ii\bar\psi_i^{(2\ell-1)}\e^{i\hi} , ~~
\d\phi_i^{(2\ell)} = \ii\bar\e_{i\hi}\psi^\hi_{(2\ell)}, ~~
\d\bar\phi^i_{(2\ell)} = \ii\bar\psi_\hi^{(2\ell)}\e^{i\hi}
\ee}%
-- Fermion fields
{\small \bea
&& \hspace{-5mm}
   \d\psi^i_{(2\ell-1)}= \bg^\m\e^{i\hi} D_\m\phi_\hi^{(2\ell-1)} +  \vth^{i\hi} \phi_\hi^{(2\ell-1)}
                      -\f{4\pi}{k}\e^{j\hj} \Big(  \phi_\hj^{(2\ell-1)}\bar\phi^i_{(2\ell)}\phi_j^{(2\ell)}
                                                  -\phi_j^{(2\ell-2)}\bar\phi^i_{(2\ell-2)}\phi_\hj^{(2\ell-1)} \Big) \nn\\
&& \hspace{-5mm}\phantom{\d\psi^i_{(2\ell-1)}=}
                      +\f{2\pi}{k}\e^{i\hi} \Big( \phi_\hi^{(2\ell-1)}\bar\phi^\hj_{(2\ell-1)}\phi_\hj^{(2\ell-1)}
                                                 +\phi_\hi^{(2\ell-1)}\bar\phi^j_{(2\ell)}\phi_j^{(2\ell)} \nn\\
&& \hspace{-5mm}\phantom{\d\psi^i_{(2\ell-1)}=}
                                                 -\phi_\hj^{(2\ell-1)}\bar\phi^\hj_{(2\ell-1)}\phi_\hi^{(2\ell-1)}
                                                 -\phi_j^{(2\ell-2)}\bar\phi^j_{(2\ell-2)}\phi_\hi^{(2\ell-1)}      \Big) \nn\\
&& \hspace{-5mm}
   \d\bar\psi_i^{(2\ell-1)}= -\bar\e_{i\hi}\bg^\m D_\m\bar\phi^\hi_{(2\ell-1)} + \bar\vth_{i\hi} \bar\phi^\hi_{(2\ell-1)}
                      +\f{4\pi}{k}\bar\e_{j\hj} \Big(  \bar\phi^\hj_{(2\ell-1)}\phi_i^{(2\ell-2)}\bar\phi^j_{(2\ell-2)}
                                                      -\bar\phi^j_{(2\ell)}\phi_i^{(2\ell)}\bar\phi^\hj_{(2\ell-1)} \Big) \nn\\
&& \hspace{-5mm}\phantom{\d\bar\psi_i^{(2\ell-1)}=}
                      -\f{2\pi}{k}\bar\e_{i\hi} \Big(  \bar\phi^\hi_{(2\ell-1)}\phi_\hj^{(2\ell-1)}\bar\phi^\hj_{(2\ell-1)}
                                                      +\bar\phi^\hi_{(2\ell-1)}\phi_j^{(2\ell-2)}\bar\phi^j_{(2\ell-2)} \nn\\
&& \hspace{-5mm}\phantom{\d\bar\psi_i^{(2\ell-1)}=}
                                                      -\bar\phi^\hj_{(2\ell-1)}\phi_\hj^{(2\ell-1)}\bar\phi^\hi_{(2\ell-1)}
                                                      -\bar\phi^j_{(2\ell)}\phi_j^{(2\ell)}\bar\phi^\hi_{(2\ell-1)} \big) \nn\\
&& \hspace{-5mm}
   \d\psi^\hi_{(2\ell)}=-\bg^\m\e^{i\hi} D_\m\phi_i^{(2\ell)} - \vth^{i\hi} \phi_i^{(2\ell)}
                      -\f{4\pi}{k}\e^{j\hj} \Big(  \phi_\hj^{(2\ell+1)}\bar\phi^\hi_{(2\ell+1)}\phi_j^{(2\ell)}
                                                  -\phi_j^{(2\ell)}\bar\phi^\hi_{(2\ell-1)}\phi_\hj^{(2\ell-1)} \Big) \nn\\
&& \hspace{-5mm}\phantom{\d\psi^\hi_{(2\ell)}=}
                      -\f{2\pi}{k}\e^{i\hi} \Big(  \phi_i^{(2\ell)}\bar\phi^\hj_{(2\ell-1)}\phi_\hj^{(2\ell-1)}
                                                  +\phi_i^{(2\ell)}\bar\phi^j_{(2\ell)}\phi_j^{(2\ell)}  \nn\\
&& \hspace{-5mm}\phantom{\d\psi^\hi_{(2\ell)}=}
                                                  -\phi_\hj^{(2\ell+1)}\bar\phi^\hj_{(2\ell+1)}\phi_i^{(2\ell)}
                                                  -\phi_j^{(2\ell)}\bar\phi^j_{(2\ell)}\phi_i^{(2\ell)}  \Big)   \nn\\
&& \hspace{-5mm}
   \d\bar\psi_\hi^{(2\ell)} = \bar\e_{i\hi}\bg^\m D_\m\bar\phi^i_{(2\ell)} - \bar\vth_{i\hi}\bar\phi^i_{(2\ell)}
                      +\f{4\pi}{k}\bar\e_{j\hj} \Big(  \bar\phi^\hj_{(2\ell-1)}\phi_\hi^{(2\ell-1)}\bar\phi^j_{(2\ell)}
                                                      -\bar\phi^j_{(2\ell)}\phi_\hi^{(2\ell+1)}\bar\phi^\hj_{(2\ell+1)} \Big) \nn\\
&& \hspace{-5mm}\phantom{\d\psi_\hi^{(2\ell)}=}
                      +\f{2\pi}{k}\bar\e_{i\hi} \Big(  \bar\phi^i_{(2\ell)}\phi_\hj^{(2\ell+1)}\bar\phi^\hj_{(2\ell+1)}
                                                       +\bar\phi^i_{(2\ell)}\phi_j^{(2\ell)}\bar\phi^j_{(2\ell)}  \nn\\
&& \hspace{-5mm}\phantom{\d\psi_\hi^{(2\ell)}=}
                                                       -\bar\phi^\hj_{(2\ell-1)}\phi_\hj^{(2\ell-1)}\bar\phi^i_{(2\ell)}
                                                       -\bar\phi^j_{(2\ell)}\phi_j^{(2\ell)}\bar\phi^i_{(2\ell)} \Big)
\eea}
Here we have the SUSY parameters $\e^{i\hi} = \th^{i\hi} + x^\m \bg_\m \vth^{i\hi}$, $\bar\e_{i\hi} = \bar\th_{i\hi} - \bar\vth_{i\hi} x^\m \bg_\m$. The parameters $\th^{i\hi}$, $\bar\th_{i\hi}$ are Poincar\'e supercharges, and $\vth^{i\hi}$, $\bar\vth_{i\hi}$ are superconformal charges, and they are Dirac spinors subject to the following constraints
\bea
&& (\th^{i\hi})^*=\bar \th_{i\hi} ~~, ~~ \bar\th_{i\hi}=\ve_{ij}\ve_{\hi\hj}\th^{j\hj} \nn\\
&& (\vth^{i\hi})^*=\bar \vth_{i\hi} ~~, ~~ \bar\vth_{i\hi}=\ve_{ij}\ve_{\hi\hj}\vth^{j\hj}
\eea
In analogy with what has been done for the orbifold ABJM theory (see section~\ref{sec5}), we consider the 1/2 BPS Wilson loop $W_1$ defined as the holonomy of the superconnection
\be \label{j76}
L_1=\diag (L_1^{(1)},L_1^{(2)}, \cdots , L_1^{(r)})
\ee
with
\bea \label{j77}
&& L_1^{(\ell)}=\lt( \ba{cc} \mA^{(2\ell-1)} &  f_1^{(2\ell-1)} \\ f_2^{(2\ell-1)} & \mB^{(2\ell)} \ea \rt), ~~
   f_1^{(2\ell-1)}=\sr{\f{4\pi}{k}}\psi^1_{(2\ell-1)+}, ~~  f_2^{(2\ell-1)}=\sr{\f{4\pi}{k}}\bar\psi_{1-}^{(2\ell-1)}  \nn\\
&& \mA^{(2\ell-1)}=A_0^{(2\ell-1)} +\f{2\pi}{k} \lt(-\phi_1^{(2\ell-2)}\bar\phi^1_{(2\ell-2 )} + \phi_2^{(2\ell-2)}\bar\phi^2_{(2\ell-2)} + \phi_\hi^{(2\ell-1)}\bar\phi^\hi_{(2\ell-1)} \rt) \nn\\
&& \mB^{(2\ell)}=B_0^{(2\ell)} +\f{2\pi}{k} \lt(-\bar\phi^1_{(2\ell)}\phi_1^{(2\ell)} + \bar\phi^2_{(2\ell)}\phi_2^{(2\ell)} + \bar\phi^\hi_{(2\ell-1)}\phi_\hi^{(2\ell-1)} \rt)
\eea
Given the diagonal nature of superconnection (\ref{j76}), we can write $W_1 = \sum_{\ell} W_1^{(\ell)}$ where $W_1^{(\ell)}$ operators are nothing but WLs associated to $L_1^{(\ell)}$ superconnections.
$W_1$ and $W_1^{(\ell)}$ operators all preserve half of the supercharges
\be
\th_+^{1 \hi}, ~~ \th_-^{2 \hi}, ~~ \bar\th_{1 \hi -}, ~~ \bar\th_{2 \hi +}
\ee
Wilson loops $W_2$, $W_2^{(\ell)}$, $W_\ho$, $W_\ho^{(\ell)}$, $W_\hw$, $W_\hw^{(\ell)}$ and their preserved supercharges can be obtained by acting with R--symmetry rotations on $W_1$, $W_1^{(\ell)}$ and the corresponding supercharges.

We can also introduce the 1/2 BPS Wilson loop $\td W_1$ defined in terms of the  superconnection
\be \label{j91}
\td L_1=\diag (\td L_1^{(1)},\td L_1^{(2)}, \cdots , \td L_1^{(r)})
\ee
where we have defined
\bea \label{j92}
&& \td L_1^{(\ell)}=\lt( \ba{cc} \td \mA^{(2\ell-1)} &  \td f_1^{(2\ell-1)} \\ \td f_2^{(2\ell-1)} & \td \mB^{(2\ell)} \ea \rt), ~~
   \td f_1^{(2\ell-1)}=\sr{\f{4\pi}{k}}\psi^1_{(2\ell-1)-}, ~~  \td f_2^{(2\ell-1)}=-\sr{\f{4\pi}{k}}\bar\psi_{1+}^{(2\ell-1)}  \nn\\
&& \td \mA^{(2\ell-1)}=A_0^{(2\ell-1)} +\f{2\pi}{k} \lt( \phi_1^{(2\ell-2)}\bar\phi^1_{(2\ell-2 )}
                                                       - \phi_2^{(2\ell-2)}\bar\phi^2_{(2\ell-2)}
                                                       + \phi_\hi^{(2\ell-1)}\bar\phi^\hi_{(2\ell-1)} \rt) \nn\\
&& \td \mB^{(2\ell)}=B_0^{(2\ell)} +\f{2\pi}{k} \lt( \bar\phi^1_{(2\ell)}\phi_1^{(2\ell)}
                                                   - \bar\phi^2_{(2\ell)}\phi_2^{(2\ell)}
                                                   + \bar\phi^\hi_{(2\ell-1)}\phi_\hi^{(2\ell-1)} \rt)
\eea
Again, we can write $\td W_1 = \sum_{\ell} \td W_1^{(\ell)}$ where $\td W_1^{(\ell)}$ is the WL associated to the  $\td L_1^{(\ell)}$ superconnection.
All these WLs preserve half of the supercharges
\be
\th_-^{1 \hi}, ~~ \th_+^{2 \hi}, ~~ \bar\th_{1 \hi +}, ~~ \bar\th_{2 \hi -}
\ee
Wilson loops $\td W_2$, $\td W_2^{(\ell)}$, $\td W_\ho$, $\td W_\ho^{(\ell)}$, $\td W_\hw$, $\td W_\hw^{(\ell)}$ and their preserved supercharges can be obtained by applying R--symmetry rotations.

\vskip 10pt

We break the gauge group $\prod_{\ell=1}^{r}[U(N_{2\ell-1}+1)\times U(N_{2\ell}+1)]$ to $\prod_{\ell=1}^{r}[U(N_{2\ell-1})\times U(N_{2\ell})]$ by Higgsing with the ansatz
\bea \label{N4hig}
A_\m^{(2\ell-1)}&=&\lt( \ba{cc} A_\m^{(2\ell-1)} & W_\m^{(2\ell-1)} \\ \bar{W}_\m^{(2\ell-1)} & 0 \ea \rt) ~~~~~~
   B_\m^{(2\ell)}=\lt( \ba{cc} B_\m^{(2\ell)} & Z_\m^{(2\ell)} \\ \bar{Z}_\m^{(2\ell)} & 0 \ea \rt)   \nn\\\
 \phi_i^{(2\ell)} &=&\lt( \ba{cc} \phi_i^{(2\ell)} &  R_i^{(2\ell)} \\ \bar{S}_i^{(2\ell)} & v_i \ea \rt) \qquad \qquad  ~~
   \bar\phi^i_{(2\ell)}=\lt( \ba{cc} \bar\phi^i_{(2\ell)} &  S^i_{(2\ell)} \\ \bar{R}^i_{(2\ell)} & \bar{v}^i \ea \rt)   \nn\\
\phi_\hi^{(2\ell-1)}&=&\lt( \ba{cc} \phi_\hi^{(2\ell-1)} &  R_\hi^{(2\ell-1)} \\ \bar{S}_\hi^{(2\ell-1)} & 0 \ea \rt) ~~~~~~
   \bar\phi^\hi_{(2\ell-1)}=\lt( \ba{cc} \bar\phi^\hi_{(2\ell-1)} &  S^\hi_{(2\ell-1)} \\ \bar{R}^\hi_{(2\ell-1)} & 0 \ea \rt)  \nn\\
\psi^i_{(2\ell-1)}&=&\lt( \ba{cc} \psi^i_{(2\ell-1)} &  \O^i_{(2\ell-1)} \\ \bar\S^i_{(2\ell-1)} & 0 \ea \rt)  ~~~~ ~~
   \bar\psi_i^{(2\ell-1)}=\lt( \ba{cc} \bar\psi_i^{(2\ell-1)} &  \S_i^{(2\ell-1)} \\ \bar\O_i^{(2\ell-1)} & 0 \ea \rt) \nn\\
\psi^\hi_{(2\ell)}&=&\lt( \ba{cc} \psi^\hi_{(2\ell)} &  \O^\hi_{(2\ell)} \\ \bar\S^\hi_{(2\ell)} & 0 \ea \rt)  \qquad \qquad ~~~
   \bar\psi_\hi^{(2\ell)}=\lt( \ba{cc} \bar\psi_\hi^{(2\ell)} &  \S_\hi^{(2\ell)} \\ \bar\O_\hi^{(2\ell)} & 0 \ea \rt)
\eea
where $v_i = v \delta_i^1$, $v>0$, and $v \ra \inf$. We work in the unitary gauge where $R_1^{(2\ell)} = \bar{R}^1_{(2\ell)} = S^1_{(2\ell)} = \bar{S}_1^{(2\ell)} = 0$. Inserting ansatz \eqref{N4hig} into the lagrangian (\ref{LagN4}) we can read the terms that are relevant for the dynamics of massive particles in the $v\to\inf$ limit. Explicitly, for the vector part we have
\bea
&& \mL_v = \sum_{\ell=1}^{r}\Big[ \f{k}{2\pi} \ve^{\m\n\r} \Big(\bar{W}_\m^{(2\ell-1)}D_\n W_\r^{(2\ell-1)}
                                                             - \bar{Z}_\m^{(2\ell-2)}D_\n Z_\r^{(2\ell-2)} \Big)  \nn\\
&&\phantom{\mL_s = }
     - \bar{W}_\m^{(2\ell-1)}\Big(v^2 + \phi_i^{(2\ell-2)}\bar\phi^i_{(2\ell-2)}
                                      + \phi_\hi^{(2\ell-1)}\bar\phi^\hi_{(2\ell-1)} \Big) W_\r^{(2\ell-1)} \nn\\
&& \phantom{\mL_s =  }
    - \bar{Z}_\m^{(2\ell-2)}\Big(v^2 + \bar\phi^i_{(2\ell-2)}\phi_i^{(2\ell-2)}
                                     + \bar\phi^\hi_{(2\ell-3)}\phi_\hi^{(2\ell-3)} \Big) Z_\r^{(2\ell-2)}  \nn\\
&& \phantom{\mL_s =  }
         + 2v  \bar{W}_\m^{(2\ell-1)}\phi_1^{(2\ell-2)}Z^\m_{(2\ell-2)}
         + 2v  \bar{Z}_\m^{(2\ell-2)} \bar\phi^1_{(2\ell-2)}W^\m_{(2\ell-1)} \Big]
\eea
while for the scalar part
{\small \bea
&&  \hspace{-4mm}
  \mL_s = \sum_{\ell=1}^{r} \Big\{ -D_\m \bar{R}^2_{(2\ell)} D^\m R_2^{(2\ell)}
                                  -D_\m \bar{R}^\hi_{(2\ell-1)} D^\m R_\hi^{(2\ell-1)}
                                  - D_\m\bar{S}_2^{(2\ell)} D^\m S^2_{(2\ell)}
                                  - D_\m\bar{S}_\hi^{(2\ell+1)} D^\m S^\hi_{(2\ell+1)}  \nn\\
&&\hspace{-4mm} \phantom{\mL_s = }
   - \f{4\pi v^4}{k^2}\Big( \bar{R}^2_{(2\ell)} R_2^{(2\ell)}
                         + \bar{R}^\hi_{(2\ell-1)}R_\hi^{(2\ell-1)}
                         + \bar{S}_2^{(2\ell)} S^2_{(2\ell)}
                         + \bar{S}_\hi^{(2\ell+1)}  S^\hi_{(2\ell+1)}  \Big) \nn\\
&&\hspace{-4mm} \phantom{\mL_s = }
   - \f{4\pi v^2}{k^2}\Big[ 2\bar{R}^2_{(2\ell)} \Big(-\phi_1^{(2\ell)}\bar\phi^1_{(2\ell)}
                                                    + \phi_2^{(2\ell)}\bar\phi^2_{(2\ell)}
                                                    + \phi_\hi^{(2\ell+1)}\bar\phi^\hi_{(2\ell+1)} \Big) R_2^{(2\ell)}
   - \bar{R}^2_{(2\ell)}\phi_2^{(2\ell)}\bar\phi^2_{(2\ell)} R_2^{(2\ell)}  \nn\\
&&\hspace{-4mm} \phantom{\mL_s = - \f{4\pi v^2}{k^2}}
   +2\bar{R}^\hi_{(2\ell-1)} \Big(- \phi_1^{(2\ell-2)}\bar\phi^1_{(2\ell-2)}
                                 + \phi_2^{(2\ell-2)}\bar\phi^2_{(2\ell-2)}
                                 + \phi_\hj^{(2\ell-1)}\bar\phi^\hj_{(2\ell-1)} \Big) R_\hi^{(2\ell-1)}  \nn\\
&&\hspace{-4mm} \phantom{\mL_s = - \f{4\pi v^2}{k^2}}
                                 - \bar{R}^\hi_{(2\ell-1)}\phi_\hi^{(2\ell-1)}\bar\phi^\hj_{(2\ell-1)} R_\hj^{(2\ell-1)}
                                 - \bar{R}^2_{(2\ell)}\phi_2^{(2\ell)}\bar\phi^\hi_{(2\ell-1)} R_\hi^{(2\ell-1)}
                                 - \bar{R}^\hi_{(2\ell-1)}\phi_\hi^{(2\ell-1)}\bar\phi^2_{(2\ell)} R_2^{(2\ell)} \Big] \nn\\
&&\hspace{-4mm} \phantom{\mL_s = }
   - \f{4\pi v^2}{k^2}\Big[ 2\bar{S}_2^{(2\ell)}\Big(-\bar\phi^1_{(2\ell)}\phi_1^{(2\ell)}
                                                   + \bar\phi^2_{(2\ell)}\phi_2^{(2\ell)}
                                                   + \bar\phi^\hi_{(2\ell-1)}\phi_\hi^{(2\ell-1)} \Big)  S^2_{(2\ell)}
                          - \bar{S}_2^{(2\ell)}\bar\phi^2_{(2\ell)}\phi_2^{(2\ell)} S^2_{(2\ell)}  \nn\\
&&\hspace{-4mm} \phantom{\mL_s = - \f{4\pi v^2}{k^2}}
   +2\bar{S}_\hi^{(2\ell+1)}\Big(-\bar\phi^1_{(2\ell+2)}\phi_1^{(2\ell+2)}
                               + \bar\phi^2_{(2\ell+2)}\phi_2^{(2\ell+2)}
                               + \bar\phi^\hj_{(2\ell+1)}\phi_\hj^{(2\ell+1)} \Big)  S^\hi_{(2\ell+1)} \\ 
&&\hspace{-4mm} \phantom{\mL_s = - \f{4\pi v^2}{k^2}}
        - \bar{S}_\hi^{(2\ell+1)}\bar\phi^\hi_{(2\ell+1)}\phi_\hj^{(2\ell+1)} S^\hj_{(2\ell+1)}
        - \bar{S}_2^{(2\ell)}\bar\phi^2_{(2\ell)}\phi_\hi^{(2\ell+1)} S^\hi_{(2\ell+1)}
        - \bar{S}_\hi^{(2\ell+1)}\bar\phi^\hi_{(2\ell+1)}\phi_2^{(2\ell)} S^2_{(2\ell)} \Big] \Big\} \nn
\eea}%
The fermion part is further split into a sum of three parts,  $\mL_f=\mL_{f1}+\mL_{f2}+\mL_{f3}$ with
{\small \bea
&&\hspace{-8mm} \mL_{f1} = \sum_{\ell=1}^{r} \Big[
    \ii\bar\O_1^{(2\ell-1)}\bg^\m D_\m\O^1_{(2\ell-1)}
    + \f{2\pi\ii v^2}{k}\bar\O_1^{(2\ell-1)}\O^1_{(2\ell-1)}
    +\ii\bar\S^1_{(2\ell+1)}\bg^\m D_\m\S_1^{(2\ell+1)}
    - \f{2\pi\ii v^2}{k}\bar\S^1_{(2\ell+1)}\S_1^{(2\ell+1)}  \nn\\
&& \hspace{-8mm}\phantom{ \mL_{f1}=}
    + \f{2\pi\ii}{k}\bar\O_1^{(2\ell-1)} \Big(-\phi_1^{(2\ell)}\bar\phi^1_{(2\ell)}
                                            +\phi_2^{(2\ell)}\bar\phi^2_{(2\ell)}
                                            +\phi_\hi^{(2\ell-1)}\bar\phi^\hi_{(2\ell-1)}\Big) \O^1_{(2\ell-1)} \nn\\
&& \hspace{-8mm}\phantom{ \mL_{f1}=}
    + \bar\O_i^{(2\ell-1)}\bg^\m\psi^i_{(2\ell-1)} Z_\m^{(2\ell)}
    + \bar\O_\hi^{(2\ell)}\bg^\m\psi^\hi_{(2\ell)} Z_\m^{(2\ell)}
    + \bar Z_\m^{(2\ell)} \bar\psi_i^{(2\ell-1)}\bg^\m\O^i_{(2\ell-1)}
    + \bar Z_\m^{(2\ell)} \bar\psi_\hi^{(2\ell)}\bg^\m\O^\hi_{(2\ell)}  \nn\\
&&\hspace{-8mm} \phantom{ \mL_{f1}=}
    -\f{2\pi\ii}{k}\bar\S^1_{(2\ell+1)} \Big(-\bar\phi^1_{(2\ell+2)}\phi_1^{(2\ell+2)}
                                             + \bar\phi^2_{(2\ell+2)}\phi_2^{(2\ell+2)}
                                             + \bar\phi^\hi_{(2\ell+1)}\phi_\hi^{(2\ell+1)}\Big) \S_1^{(2\ell+1)}  \\
&& \hspace{-8mm}\phantom{ \mL_{f1}=}
     + \bar\S^i_{(2\ell+1)}\bg^\m\bar\psi_i^{(2\ell+1)} W_\m^{(2\ell+1)}+ \bar\S^\hi_{(2\ell)}\bg^\m\bar\psi_\hi^{(2\ell)} W_\m^{(2\ell)} + \bar W_\m^{(2\ell+1)} \psi^i_{(2\ell+1)}\bg^\m\S_i^{(2\ell+1)}+ \bar W_\m^{(2\ell)} \psi^\hi_{(2\ell)}\bg^\m\S_\hi^{(2\ell)} \Big] \nn
\eea
\bea
&&\hspace{-6mm} \mL_{f2}=\sum_{\ell=1}^{r} \Big\{ \ii\bar\O_2^{(2\ell-1)}\bg^\m D_\m\O^2_{(2\ell-1)} + \ii\bar\O_\hi^{(2\ell)}\bg^\m D_\m\O^\hi_{(2\ell)} - \f{2\pi\ii v^2}{k}\Big( \bar\O_2^{(2\ell-1)}\O^2_{(2\ell-1)} + \bar\O_\hi^{(2\ell)}\O^\hi_{(2\ell)} \Big)  \nn\\
&&\hspace{-6mm}\phantom{ \mL_{f1}=} +\f{2\pi\ii}{k}\Big[ \bar\O_2^{(2\ell-1)}\Big(\phi_i^{(2\ell)}\bar\phi^i_{(2\ell)}+\phi_\hi^{(2\ell-1)}\bar\phi^\hi_{(2\ell-1)}\Big)\O^2_{(2\ell-1)} + \bar\O_\hi^{(2\ell)}\Big(\phi_j^{(2\ell)}\bar\phi^j_{(2\ell)}+\phi_\hj^{(2\ell+1)}\bar\phi^\hj_{(2\ell+1)}\Big)\O^\hi_{(2\ell)}  \nn\\
&&\hspace{-6mm}\phantom{ \mL_{f1}=+\f{2\pi\ii}{k}} - 2\bar\O_2^{(2\ell-1)}\phi_2^{(2\ell)}\bar\phi^2_{(2\ell)}\O^2_{(2\ell-1)}- 2\bar\O_2^{(2\ell-1)}\phi_\hi^{(2\ell-1)}\bar\phi^2_{(2\ell)}\O^\hi_{(2\ell)}- 2\bar\O_\hi^{(2\ell)}\phi_2^{(2\ell)}\bar\phi^\hi_{(2\ell-1)}\O^2_{(2\ell-1)}  \nn\\
&&\hspace{-6mm}\phantom{ \mL_{f1}=+\f{2\pi\ii}{k}}  - 2\bar\O_\hi^{(2\ell)}\phi_\hj^{(2\ell+1)}\bar\phi^\hi_{(2\ell+1)}\O^\hj_{(2\ell)} +2v\bar\O_2^{(2\ell-1)}\psi^1_{(2\ell-1)}S^2_{(2\ell)} +2v\bar\O_\hi^{(2\ell)}\psi^1_{(2\ell+1)}S^\hi_{(2\ell+1)} \nn\\
&&\hspace{-6mm}\phantom{ \mL_{f1}=+\f{2\pi\ii}{k}}  +2v\bar S_2^{(2\ell)}\bar\psi_1^{(2\ell-1)}\O^2_{(2\ell-1)} +2v\bar S_\hi^{(2\ell+1)}\bar\psi_1^{(2\ell+1)}\O^\hi_{(2\ell)} \Big] \nn\\
&&\hspace{-6mm}\phantom{ \mL_{f1}=} +\ii\bar\S^2_{(2\ell+1)}\bg^\m D_\m\S_2^{(2\ell+1)} + \ii\bar\S^\hi_{(2\ell)}\bg^\m D_\m\S_\hi^{(2\ell)}  + \f{2\pi\ii v^2}{k}\Big( \bar\S^2_{(2\ell+1)}\S_2^{(2\ell+1)} + \bar\S^\hi_{(2\ell)}\S_\hi^{(2\ell)}\Big) \nn\\
&&\hspace{-6mm}\phantom{\mL_{f1}=} -\f{2\pi\ii}{k} \Big[\bar\S^2_{(2\ell+1)} \Big(\bar\phi^i_{(2\ell+2)}\phi_i^{(2\ell+2) } + \bar\phi^\hi_{(2\ell+1)}\phi_\hi^{(2\ell+1)}\Big)\S_2^{(2\ell+1)}  + \bar\S^\hi_{(2\ell)}\Big( \bar\phi^j_{(2\ell)}\phi_j^{(2\ell) } + \bar\phi^\hj_{(2\ell-1)}\phi_\hj^{(2\ell-1)} \Big)\S_i^{(2\ell)}  \nn\\
&&\hspace{-6mm}\phantom{ \mL_{f1}=-\f{2\pi\ii}{k}} - 2\bar\S^2_{(2\ell+1)} \bar\phi^2_{(2\ell+2)}\phi_2^{(2\ell+2)}\S_2^{(2\ell+1)}- 2\bar\S^2_{(2\ell+1)} \bar\phi^\hi_{(2\ell+1)}\phi_2^{(2\ell)}\S_\hi^{(2\ell)}- 2\bar\S^\hi_{(2\ell)} \bar\phi^2_{(2\ell)}\phi_\hi^{(2\ell+1)}\S_2^{(2\ell+1)} \nn\\
&&\hspace{-6mm}\phantom{ \mL_{f1}=-\f{2\pi\ii}{k}} - 2\bar\S^\hi_{(2\ell)} \bar\phi^\hj_{(2\ell-1)}\phi_\hi^{(2\ell-1)}\S_\hj^{(2\ell)} + 2v\bar\S^2_{(2\ell+1)}\bar\psi_1^{(2\ell+1)}R_2^{(2\ell)}  + 2v\bar\S^\hi_{(2\ell)}\bar\psi_1^{(2\ell-1)}R_\hi^{(2\ell-1)} \nn\\
&&\hspace{-6mm}\phantom{ \mL_{f1}=-\f{2\pi\ii}{k}} + 2v\bar R^2_{(2\ell)}\psi^1_{(2\ell+1)}\S_2^{(2\ell+1)} + 2v\bar R^\hi_{(2\ell-1)}\psi^1_{(2\ell-1)}\S_\hi^{(2\ell)} \Big] \nn\\
&&\hspace{-6mm}\phantom{ \mL_{f1}=} +\f{4\pi\ii v}{k} \Big[ \ve^{\hi\hj}\Big(\bar\O_2^{(2\ell-1)}\phi_\hi^{(2\ell-1)}\S_\hj^{(2\ell)} -\bar\O_\hi^{(2\ell)}\phi_2^{(2\ell)}\S_\hj^{(2\ell)} + \bar\O_\hi^{(2\ell)}\phi_\hj^{(2\ell+1)}\S_2^{(2\ell+1)} \Big)  \nn\\
&&\hspace{-6mm}\phantom{ \mL_{f1}= +\f{4\pi\ii v}{k}}  -\ve_{\hi\hj} \Big(\bar\S^2_{(2\ell+1)}\bar\phi^\hi_{(2\ell+1)}\O^\hj_{(2\ell)} - \bar\S^\hi_{(2\ell)}\bar\phi^2_{(2\ell)}\O^\hj_{(2\ell)} + \bar\S^\hi_{(2\ell)}\bar\phi^\hj_{(2\ell-1)}\O^2_{(2\ell-1)} \Big) \Big] \Big\}
\eea
\bea
&&\hspace{-2mm}\mL_{f3}=-\f{4\pi\ii}{k}\sum_{\ell=1}^{r} \Big( \bar\O_1^{(2\ell-1)}\phi_2^{(2\ell)}\bar\phi^1_{(2\ell)}\O^2_{(2\ell-1)} + \bar\O_1^{(2\ell-1)}\phi_\hi^{(2\ell-1)}\bar\phi^1_{(2\ell)}\O^\hi_{(2\ell)} + \bar\O_2^{(2\ell-1)}\phi_1^{(2\ell)}\bar\phi^2_{(2\ell)}\O^1_{(2\ell-1)}  \nn\\
&&\hspace{-2mm}\phantom{\mL_{f3}=} + \bar\O_\hi^{(2\ell)}\phi_1^{(2\ell)}\bar\phi^\hi_{(2\ell-1)}\O^1_{(2\ell-1)} - \bar\S^1_{(2\ell+1)}\bar\phi^2_{(2\ell+2)}\phi_1^{(2\ell+2)}\S_2^{(2\ell+1)} - \bar\S^1_{(2\ell+1)}\bar\phi^\hi_{(2\ell+1)}\phi_1^{(2\ell)}\S_\hi^{(2\ell)} \nn\\
&&\hspace{-2mm}\phantom{\mL_{f3}=}  - \bar\S^2_{(2+1\ell)}\bar\phi^1_{(2\ell+2)}\phi_2^{(2\ell+2)}\S_1^{(2\ell+1)}- \bar\S^\hi_{(2\ell)}\bar\phi^1_{(2\ell)}\phi_\hi^{(2\ell+1)}\S_1^{(2\ell+1)} \Big)
\eea}%
It is convenient to redefine the bosonic fields as
\bea
&& W_\m^{(2\ell-1)} \ra  \Big(1 + \f{\phi_1^{(2\ell-2)}\bar\phi^1_{(2\ell-2)}}{2v^2} \Big) W_\m^{(2\ell-1)}
                       + \f{\phi_1^{(2\ell-2)}}{v}Z_\m^{(2\ell-2)} \nn\\
&& Z_\m^{(2\ell)} \ra  \Big(1 + \f{\bar\phi^1_{(2\ell)}\phi_1^{(2\ell)}}{2v^2} \Big)Z_\m^{(2\ell)}
                     + \f{\bar\phi^1_{(2\ell)}}{v}W_\m^{(2\ell+1)}  \nn\\
&& R_2^{(2\ell)} \ra  R_2^{(2\ell)}
                    + \f{\phi_2^{(2\ell)}\bar\phi^2_{(2\ell)}}{2v^2} R_2^{(2\ell)}
                    + \f{\phi_2^{(2\ell)}\bar\phi^\hi_{(2\ell-1)}}{2v^2}R_\hi^{(2\ell-1)} \nn\\
&& R_\hi^{(2\ell-1)} \ra    R_\hi^{(2\ell-1)}
                          + \f{\phi_\hi^{(2\ell-1)}\bar\phi^\hj_{(2\ell-1)} }{2v^2}R_\hj^{(2\ell-1)}
                          + \f{\phi_\hi^{(2\ell-1)}\bar\phi^2_{(2\ell)}}{2v^2} R_2^{(2\ell)}  \nn\\
&& S^2_{(2\ell)} \ra   S^2_{(2\ell)}
                     + \f{\bar\phi^2_{(2\ell)}\phi_2^{(2\ell)}}{2v^2} S^2_{(2\ell)}
                     + \f{\bar\phi^2_{(2\ell)}\phi_\hi^{(2\ell+1)}}{2v^2} S^\hi_{(2\ell+1)} \\
&& S^\hi_{(2\ell-1)}\ra   S^\hi_{(2\ell-1)}
                        + \f{\bar\phi^\hi_{(2\ell-1)}\phi_\hj^{(2\ell-1)} }{2v^2}S^\hj_{(2\ell-1)}
                        + \f{\bar\phi^\hi_{(2\ell-1)}\phi_2^{(2\ell-2)}}{2v^2} S^2_{(2\ell-2)} \nn
\eea
and the fermion fields as
{\small \bea
&& \O^1_{(2\ell-1)} \to \O^1_{(2\ell-1)}, ~~
   \S_1^{(2\ell+1)} \to \S_1^{(2\ell+1)} \nn\\
&& \O^2_{(2\ell-1)} \to \O^2_{(2\ell-1)}
                        + \f{1}{v}\ve^{\hi\hj}\phi_\hi^{(2\ell-1)}\S_\hj^{(2\ell)}
                        + \f{1}{2v^2}\phi_\hi^{(2\ell-1)}\lt(  \bar\phi^2_{(2\ell)}\O^\hi_{(2\ell)}
                                                              -\bar\phi^\hi_{(2\ell-1)}\O^2_{(2\ell)}\rt)  \nn\\
&& \O^\hi_{(2\ell)} \to \O^\hi_{(2\ell)}
                      + \f{1}{v}\ve^{\hi\hj}\lt(\phi_\hj^{(2\ell+1)}\S_2^{(2\ell+1)}
                      - \phi_2^{(2\ell)}\S_\hj^{(2\ell)}\rt) \nn\\
&& \phantom{\O^\hi_{(2\ell)} \to}
             + \f{1}{2v^2}\lt[ \phi_\hj^{(2\ell+1)}\lt(\bar\phi^\hi_{(2\ell+1)}\O^\hj_{(2\ell)}
                              -\bar\phi^\hj_{(2\ell+1)}\O^\hi_{(2\ell)} \rt)
                              + \phi_2^{(2\ell)}\lt( \bar\phi^\hi_{(2\ell-1)}\O^2_{(2\ell-1)}
                                                     -\bar\phi^2_{(2\ell)}\O^\hi_{(2\ell-1)}\rt) \rt]  \nn\\
&& \S_2^{(2\ell-1)} \to  \S_2^{(2\ell-1)}
                       + \f{1}{v}\ve_{\hi\hj}\bar\phi^\hi_{(2\ell-1)}\O^\hj_{(2\ell-2)}
                       + \f{1}{2v^2}\bar\phi^\hi_{(2\ell-1)}\lt( \phi_2^{(2\ell-2)}\S_\hi^{(2\ell-2)}
                                                                -\phi_\hi^{(2\ell-1)}\S_2^{(2\ell-1)} \rt) \nn\\
&& \S_\hi^{(2\ell)} \to  \S_\hi^{(2\ell)}
                       + \f{1}{v}\ve_{\hi\hj}\lt( \bar\phi^\hj_{(2\ell-1)}\O^2_{(2\ell-1)}
                                               - \bar\phi^2_{(2\ell)}\O^\hj_{(2\ell)}\rt)  \\
&& \phantom{\S_\hi^{(2\ell)} \to }
     + \f{1}{2v^2}\lt[\bar\phi^\hj_{(2\ell-1)}\lt( \phi_\hi^{(2\ell-1)}\S_\hj^{(2\ell)}
                                                  -\phi_\hj^{(2\ell-1)}\S_\hi^{(2\ell)} \rt)
     + \bar\phi^2_{(2\ell)}\lt( \phi_\hi^{(2\ell+1)}\S_2^{(2\ell+1)}
                               -\phi_2^{(2\ell)}\S_\hi^{(2\ell)} \rt) \rt] \nn
\eea}%
If we now choose the particle modes
\bea
W_\m^{(2\ell-1)} = \sr{\f{k}{\pi}}\lt(0,1,-\ii\rt)w^{(2\ell-1)} \ep^{-\ii mt}, && Z_\m^{(2\ell)} = \sr{\f{k}{\pi}}\lt(0,1,\ii\rt)z^{(2\ell)} \ep^{-\ii mt}\nn\\
\O^2_{(2\ell-1)} = u_- \o^2_{(2\ell-1)} \ep^{-\ii mt},                         && \S_2^{(2\ell-1)} = u_+ \s_2^{(2\ell-1)} \ep^{-\ii mt} \nn\\
\O^\hi_{(2\ell)} = u_- \o^\hi_{(2\ell)} \ep^{-\ii mt},                         && \S_\hi^{(2\ell)} = u_+ \s_\hi^{(2\ell)} \ep^{-\ii mt} \nn\\
R_2^{(2\ell)}= \sr{\f{k}{4\pi}}\f{1}{v}r_2^{(2\ell)} \ep^{-\ii mt},            && R_\hi^{(2\ell-1)}= \sr{\f{k}{4\pi}}\f{1}{v}r_\hi^{(2\ell-1)} \ep^{-\ii mt} \nn\\
S^2_{(2\ell)}= \sr{\f{k}{4\pi}}\f{1}{v}s^2_{(2\ell)} \ep^{-\ii mt},            && S^\hi_{(2\ell-1)}= \sr{\f{k}{4\pi}}\f{1}{v}s^\hi_{(2\ell-1)} \ep^{-\ii mt} \nn\\
\O^1_{(2\ell-1)} = u_+ \o^1_{(2\ell-1)} \ep^{-\ii mt},                         && \S_1^{(2\ell-1)} = u_- \s_1^{(2\ell-1)} \ep^{-\ii mt}
\eea
combine them into the following supermatrices
{\small \be
\Psi_1^{(\ell)}=\lt( \ba{cc} w^{(2\ell-1)} & \o^1_{(2\ell-1)} \\ \s_1^{(2\ell-1)} & z^{(2\ell)} \ea \rt), ~~
\Psi_2^{(\ell)}=\lt( \ba{cc} r_2^{(2\ell-2)} &  -\o^2_{(2\ell-1)} \\ -\s_2^{(2\ell-1)} & s^2_{(2\ell)} \ea \rt), ~~ \Psi_\hi^{(\ell)}=\lt( \ba{cc} r_\hi^{(2\ell-1)} &  -\o^\hi_{(2\ell-2)} \\ -\s_\hi^{(2\ell)} & s^\hi_{(2\ell-1)} \ea \rt)
\ee}%
and define
\be
\Psi_i= \diag( \Psi_i^{(1)}, \Psi_i^{(2)}, \cdots, \Psi_i^{(r)}), ~~
\Psi_\hi= \diag( \Psi_\hi^{(1)}, \Psi_\hi^{(2)}, \cdots, \Psi_\hi^{(r)})
\ee
the non-relativistic lagrangian can be put in the following form
\be
\mL = \ii \sum_{\ell=1}^{r} \Big( \Tr \bar\Psi^i_{(\ell)} \mf{D}_0  \Psi_i^{(\ell)}
                               + \Tr \bar\Psi^\hi_{(\ell)} \mf{D}_0  \Psi_\hi^{(\ell)}  \Big)
= \ii \Big( \Tr \bar\Psi^i \mf{D}_0  \Psi_i
          + \Tr \bar\Psi^\hi \mf{D}_0  \Psi_\hi \Big)
\ee
Here covariant derivatives are defined as
\bea
\mf D_0 \Psi_i^{(\ell)} &=& \p_0 \Psi_i^{(\ell)}+ \ii L_1^{(\ell)} \Psi_i^{(\ell)} \qquad , \qquad \mf D_0 \Psi_\hi^{(\ell)} = \p_0 \Psi_\hi^{(\ell)}+ \ii L_1^{(\ell)} \Psi_\hi^{(\ell)} \nn \\
\mf D_0 \Psi_i &=& \p_0 \Psi_i+ \ii L_1 \Psi_i \qquad \qquad , \qquad \mf D_0 \Psi_\hi = \p_0 \Psi_\hi+ \ii L_1 \Psi_\hi
\eea
where $L_1^{(\ell)}$ are exactly the connections in (\ref{j77}) that define Wilson loops $W_1^{(\ell)}$, and $L_1$ is connection (\ref{j76}) that defines the $W_1$ operator.

Alternatively, we choose the antiparticle modes
\bea
W_\m^{(2\ell-1)} = \sr{\f{k}{\pi}}\lt(0,1,\ii\rt)w^{(2\ell-1)} \ep^{\ii mt},  && Z_\m^{(2\ell)} = \sr{\f{k}{\pi}}\lt(0,1,-\ii\rt)z^{(2\ell)} \ep^{\ii mt}\nn\\
\O^2_{(2\ell-1)} = u_+ \o^2_{(2\ell-1)} \ep^{\ii mt},                         && \S_2^{(2\ell-1)} = u_- \s_2^{(2\ell-1)} \ep^{\ii mt} \nn\\
\O^\hi_{(2\ell)} = u_+ \o^\hi_{(2\ell)} \ep^{\ii mt},                         && \S_\hi^{(2\ell)} = u_- \s_\hi^{(2\ell)} \ep^{\ii mt} \nn\\
R_2^{(2\ell)}= \sr{\f{k}{4\pi}}\f{1}{v}r_2^{(2\ell)} \ep^{\ii mt},            && R_\hi^{(2\ell-1)}= \sr{\f{k}{4\pi}}\f{1}{v}r_\hi^{(2\ell-1)} \ep^{\ii mt} \nn\\
S^2_{(2\ell)}= \sr{\f{k}{4\pi}}\f{1}{v}s^2_{(2\ell)} \ep^{\ii mt},            && S^\hi_{(2\ell-1)}= \sr{\f{k}{4\pi}}\f{1}{v}s^\hi_{(2\ell-1)} \ep^{\ii mt}\nn\\
\O^1_{(2\ell-1)} = u_- \o^1_{(2\ell-1)} \ep^{\ii mt},                         && \S_1^{(2\ell-1)} = u_+ \s_1^{(2\ell-1)} \ep^{\ii mt}
\eea
combined into the supermatrices
{\small \be
\Psi_1^{(\ell)}=\lt( \ba{cc} w^{(2\ell-1)} & -\o^1_{(2\ell-1)} \\ \s_1^{(2\ell-1)} & z^{(2\ell)} \ea \rt), ~~
\Psi_2^{(\ell)}=\lt( \ba{cc} r_2^{(2\ell-2)} &  -\o^2_{(2\ell-1)} \\ \s_2^{(2\ell-1)} & s^2_{(2\ell)} \ea \rt), ~~ \Psi_\hi^{(\ell)}=\lt( \ba{cc} r_\hi^{(2\ell-1)} &  -\o^\hi_{(2\ell-2)} \\ \s_\hi^{(2\ell)} & s^\hi_{(2\ell-1)} \ea \rt)
\ee}%
With the further definition
\be
\Psi_i= \diag( \Psi_i^{(1)}, \Psi_i^{(2)}, \cdots, \Psi_i^{(r)}), ~~
\Psi_\hi= \diag( \Psi_\hi^{(1)}, \Psi_\hi^{(2)}, \cdots, \Psi_\hi^{(r)})
\ee
the non-relativistic lagrangian becomes
\be
\mL = \ii \sum_{\ell=1}^{r} \Big( \Tr \Psi_i^{(\ell)} \mf{D}_0 \bar\Psi^i_{(\ell)}
                               + \Tr \Psi_\hi^{(\ell)} \mf{D}_0 \bar\Psi^\hi_{(\ell)} \Big)
= \ii \Big( \Tr \Psi_i \mf{D}_0 \bar\Psi^i
          + \Tr \Psi_\hi \mf{D}_0 \bar\Psi^\hi \Big)
\ee
with covariant derivatives
\bea
\mf D_0 \bar\Psi^i_{(\ell)} &=& \p_0 \bar\Psi^i_{(\ell)}- \ii \bar\Psi^i_{(\ell)}\td L_1^{(\ell)} \qquad , \qquad \mf D_0 \bar\Psi^\hi_{(\ell)} = \p_0 \bar\Psi^\hi_{(\ell)} - \ii \bar\Psi^\hi_{(\ell)}\td L_1^{(\ell)} \nn \\
\mf D_0 \bar\Psi^i &=& \p_0 \bar\Psi^i - \ii \bar\Psi^i \td L_1 \qquad \qquad , \qquad \mf D_0 \bar\Psi^\hi = \p_0 \bar\Psi^\hi - \ii \bar\Psi^\hi \td L_1
\eea
Here $\td L_1^{(\ell)}$ are connections (\ref{j92}) that define Wilson loops $\td W_1^{(\ell)}$, and $\td L_1$ is connection (\ref{j91}) that defines $\td W_1$.

The other $6(r+1)$ 1/2 BPS Wilson loops $W_2$, $W_2^{(\ell)}$, $W_\ho$, $W_\ho^{(\ell)}$, $W_\hw$, $W_\hw^{(\ell)}$, $\td W_2$, $\td W_2^{(\ell)}$, $\td W_\ho$, $\td W_\ho^{(\ell)}$, $\td W_\hw$, $\td W_\hw^{(\ell)}$ can be obtained with an identical procedure that we will not repeat here.

\section{M2--branes in AdS$_7\times$S$^4$ spacetime}\label{appF}

The six-dimensional (2,0) superconformal field theory is supposed to be dual to M--theory in AdS$_7\times$S$^4$ spacetime with a four-form flux turned on in S$^4$ \cite{Maldacena:1997re}
\bea \label{ads7s4}
&& ds^2 = R^2 (ds^2_{\AdS_7} + \f14 ds^2_{\rmS^4}) \nn \\
&& F_{\ti\tj\td k\td l} = \f{6}{R} \ve_{\ti\tj\td k\td l}
\eea
where $\ve_{\ti\tj\td k\td l}$ is the volume form of S$^4$.

Although in the (2,0) theory a 1/2 BPS Wilson surface could be defined in terms of the two--form field and possibly other bosonic and fermionic fields \cite{Maldacena:1998im}, in the absence of an explicit lagrangian and SUSY transformations we cannot construct it explicitly. Still we can investigate their possible gravity duals, and the corresponding preserved supercharges.
We look for the 1/2 BPS M2--brane configurations in AdS$_7\times$S$^4$ spacetime, and also discuss briefly M5--brane solutions at the end of this appendix.\footnote{Wilson surfaces in six-dimensional (2,0) superconformal field theory and their gravity duals have been also discussed in, for examples, \cite{Ganor:1996nf,Berenstein:1998ij,Corrado:1999pi,Gustavsson:2004gj,Lunin:2007ab,Chen:2007ir}.}

For AdS$_7$ we use the metric
\be \label{ads7}
ds^2_{\AdS_7} = u^2( -dt^2 + dx_1^2 + dx_2^2 + dx_3^2 + dx_4^2 + dx_5^2 ) + \f{du^2}{u^2}
\ee
We embed S$^4$ in $\rm{R}^5\cong \rm{R}\times\rm{C}^2$ by parameterizing
\bea
&& z_0 = \cos\th_1 = x_6 \nn\\
&& z_1 = \sin\th_1 \cos\th_2 \, \ep^{\ii\xi_1} = x_7 + \ii x_9 \nn\\
&& z_2 = \sin\th_1 \sin\th_2 \, \ep^{\ii\xi_2} = x_8 + \ii x_\natural
\eea
with $\th_1 \in [0,\pi]$, $\th_2 \in [0,\pi/2]$, $\xi_{1,2} \in [0,2\pi]$. This leads to the S$^4$ metric
\be \label{s4}
ds^2_{\rmS^4} = d\th_1^2 + \sin^2\th_1 ( d \th_2^2 + \cos^2\th_2 \, d\xi_1^2 + \sin^2\th_2 \, d\xi_2^2 )
\ee

We begin by deriving Killing spinors in $\gtzt$. Given coordinates $x^M=(x^{\td\m},x^\ti)$, with $x^{\td\m}$, and $x^\ti$ being coordinates of AdS$_7$ and S$^4$ respectively, and tangent space coordinates $x^A=(x^{\td a},x^{\td p})$ with $\td a=0,1,2,3,4,5,6$ and $\td p=7,8,9,\natural$, for the AdS$_7$ metric (\ref{ads7}) we use the vierbeins
\be
e^0= u dt, ~~ e^1 = u dx_1, ~~  e^2 = u dx_2, ~~ e^3 = u dx_3, ~~ e^4 = u dx_4, ~~ e^5 = u dx_5, ~~ e^6 = \f{du}{u}
\ee
whereas for S$^4$ metric (\ref{s4}) we use
\be
e^7=d\th_1, ~~ e^8=\sin\th_1 d\th_2, ~~ e^9=\sin\th_1\cos\th_2 d\xi_1, ~~ e^\natural=\sin\th_1\sin\th_2 d\xi_2
\ee
The vierbeins of $\gtzt$ (\ref{ads7s4}) are then given by
\be
E_{\td\m}^{\td a} = R e_{\td\m}^{\td a}, ~~ E_\ti^{\td p} = \f{R}2 e_\ti^{\td p}
\ee
The non-vanishing components of the spin connection are
\bea
&& \o^{06}_t=\o^{16}_{x_1}=\o^{26}_{x_2}=\o^{36}_{x_3}=\o^{46}_{x_4}=\o^{56}_{x_5}=u   \nn\\
&& \o^{78}_{\th_2}=-\cos\th_1, ~~
   \o^{79}_{\xi_1}=-\cos\th_1\cos\th_2, ~~
   \o^{89}_{\xi_1}=\sin\th_2   \nn\\
&&
   \o^{7\natural}_{\xi_2}=-\cos\th_1\sin\th_2, ~~
   \o^{8\natural}_{\xi_2}=-\cos\th_2
\eea

The Killing spinor equations read
\be
D^M \e =\f1{288}F_{NPQR}(\G^{MNPQR} - 8G^{MN}\G^{PQR})\e
\ee
with $\e$ being a Majorana spinor.
Note that $\G_{\td \m}=R\g_{\td \mu}$, $\G_\ti=\f{R}{2}\g_\ti$.
We rewrite them as
\be
D_{\td\m} \e = \f{1}{2}\hat\g\g_{\td\m}\e, ~~ D_\ti \e = \f{1}{2}\hat\g\g_\ti\e
\ee
with $\hat \g=\g^{789\natural}$. We define $\td \g = \g^{6789\natural}$,
and the explicit Killing spinor equations are
\bea
&& \p_{\m} \e = -\f{u}{2}\g_{\m6}(1-\td\g)\e, ~~ \p_u \e = \f{1}{2u} \td\g \e \nn\\
&& \p_{\th_1}\e = \f{1}{2}\hat\g\g_7\e, ~~
   \p_{\th_2}\e = \f{1}{2}\g_{78}\ep^{-\th_1\hat\g\g_7}\e \nn\\
&& \p_{\xi_1}\e = \f12 ( \g_{79}\ep^{-\th_1\hat\g\g_7}\cos\th_2-\g_{89}\sin\th_2 )\e \nn\\
&& \p_{\xi_2}\e = \f12 ( \g_{8\natural}\cos\th_2+\g_{7\natural}\ep^{-\th_1\hat\g\g_7}\sin\th_2 )\e
\eea
The general solution reads
\be
\e 
   = u^{\f12}h(\e_1+x^{\m}\g_{\m}\e_2) - u^{-\f12} \g_6 h \e_2
\ee
where
\be
h= \ep^{\f{\th_1}{2}\td\g\g_{67}} \ep^{\f{\th_2}{2}\g_{78}}
   \ep^{\f{\xi_1}{2}\g_{79}} \ep^{\f{\xi_2}{2}\g_{8\natural}}
\ee
and constant Majorana spinors $\e_1$, $\e_2$ satisfying $\td\g\e_1=\e_1$, $\td\g\e_2=-\e_2$.

\vskip 10pt
Now we consider a M2--brane described by worldvolume coordinates $\s^{0,1,2}$, embedded in $\gtzt$ as
\be
t=\s^0, ~~ x_1=\s^1, ~~x_2=x_3=x_4=x_5=0, ~~ u=\s^2
\ee
The brane is localized on the compact space $\rmS^4$ that is specified by coordinates $(\a,\b,\xi_1,\xi_2)$.
The supercharges preserved by the M2--brane are given by the condition
\be
\g_{016}\e=\e
\ee
which is equivalent to
\be
h^{-1} \g_{016} h \e_1 =\e_1, ~~ h^{-1} \g_{016} h \e_2 =\e_2
\ee
It turns out that
\be
h^{-1} \g_{016} h = \g_{01I}n^I \qquad \quad I=6,7,8,9,\natural
\ee
where $n^I$ is  the unit vector in $\rm{R}^5$
\be
n^I = ( \cos\th_1, \sin\th_1\cos\th_2\cos\xi_1, \sin\th_1\sin\th_2\cos\xi_2, \sin\th_1\cos\th_2\sin\xi_1, \sin\th_1\sin\th_2\sin\xi_2 )
\ee
The supercharges preserved  by the M2--brane are then
\be
\g_{01I}n^I \e_1 = \e_1, ~~ \g_{01I}n^I \e_2 = \e_2
\ee

In order to discuss possible overlapping of the spectrum of preserved supercharges, we consider ten different M2--brane configurations $M_2^{(i)}$, $i = 1, \cdots ,10$, localized at different positions in the compact S$^4$ space. They are listed in table~\ref{tab6}, together with their positions and the corresponding supercharges. Since the five matrices $\g_{016}$, $\g_{017}$, $\g_{019}$, $\g_{018}$, $\g_{01\natural}$ do not commute with each other, there is no supercharge overlapping among the M2--branes. We expect them to be dual to non-degenerate Wilson surfaces in the six-dimensional (2,0) superconformal field theory.

Similarly, we can consider ten anti--M2--brane configurations $\bar{M}_2^{(i)}$, localized at the points listed in table~\ref{tab6}. The corresponding preserved supercharges can be determined by
\be
\g_{01I}n^I \e_1 = -\e_1, ~~ \g_{01I}n^I \e_2 = -\e_2
\ee
An anti--M2--brane preserves a set of supercharges that is complementary to the one of the corresponding M2--brane localized at the same position.

\begin{table}[htbp]\centering\begin{tabular}{|c|l|l|l|}\hline
      brane    & \multicolumn{2}{c|}{position} & \multicolumn{1}{c|}{preserved supercharges} \\ \hline

  $M_2^{(1)}$  & $z_0=1$    & $\th_1=0$                       & $\g_{016}\e_1=\e_1$, $\g_{016}\e_2=\e_2$ \\ \hline
  $M_2^{(2)}$  & $z_0=-1$   & $\th_1=\pi$                     & $\g_{016}\e_1=-\e_1$, $\g_{016}\e_2=-\e_2$ \\ \hline

  $M_2^{(3)}$  & $z_1=1$    & $\th_1=\pi/2,\th_2=\xi_1=0$        & $\g_{017}\e_1=\e_1$, $\g_{017}\e_2=\e_2$ \\ \hline
  $M_2^{(4)}$  & $z_1=\ii$  & $\th_1=\pi/2,\th_2=0,\xi_1=\pi/2$  & $\g_{019}\e_1=\e_1$, $\g_{019}\e_2=\e_2$ \\ \hline
  $M_2^{(5)}$  & $z_1=-1$   & $\th_1=\pi/2,\th_2=0,\xi_1=\pi$    & $\g_{017}\e_1=-\e_1$, $\g_{017}\e_2=-\e_2$ \\ \hline
  $M_2^{(6)}$  & $z_1=-\ii$ & $\th_1=\pi/2,\th_2=0,\xi_1=3\pi/2$ & $\g_{019}\e_1=-\e_1$, $\g_{019}\e_2=-\e_2$ \\ \hline

  $M_2^{(7)}$  & $z_2=1$    & $\th_1=\th_2=\pi/2,\xi_2=0$        & $\g_{018}\e_1=\e_1$, $\g_{018}\e_2=\e_2$ \\ \hline
  $M_2^{(8)}$  & $z_2=\ii$  & $\th_1=\th_2=\pi/2,\xi_2=\pi/2$    & $\g_{01\natural}\e_1=\e_1$, $\g_{01\natural}\e_2=\e_2$ \\ \hline
  $M_2^{(9)}$  & $z_2=-1$   & $\th_1=\th_2=\pi/2,\xi_2=\pi$      & $\g_{018}\e_1=-\e_1$, $\g_{018}\e_2=-\e_2$ \\ \hline
  $M_2^{(10)}$ & $z_2=-\ii$ & $\th_1=\th_2=\pi/2,\xi_2=3\pi/2$   & $\g_{01\natural}\e_1=-\e_1$, $\g_{01\natural}\e_2=-\e_2$ \\ \hline
  \end{tabular}
\caption{Ten different M2--branes placed at different positions, and their preserved supercharges.}\label{tab6}
\end{table}

However, it is easy to realize that
there are ten pairs of brane and anti--brane located at opposite points of S$^4$ that preserve the same set of supercharges. These are for instance
$(M_2^{(1)}, \bar{M}_2^{(2)})$,
$(M_2^{(2)}, \bar{M}_2^{(1)})$,
$(M_2^{(3)}, \bar{M}_2^{(5)})$,
$(M_2^{(4)}, \bar{M}_2^{(6)})$,
$(M_2^{(5)}, \bar{M}_2^{(3)})$,
$(M_2^{(6)}, \bar{M}_2^{(4)})$, etc.
It is then natural to expect that a similar degeneracy occurs also in the spectrum of Wilson surfaces in SCFT. At this stage it is impossible to establish whether this degeneracy is trivial as in the four-dimensional ${\cal N}=4$ SYM case, or it actually signals the existence of two different Wilson surfaces preserving the same supercharges as in the three-dimensional ${\cal N}=4$ SCSM theories. However, given that the dual picture mostly resembles the case of four-dimensional ${\cal N}=4$ SYM, we are tempted to believe that M2--brane pairs describe the same operator.

A possible M5--brane configuration dual to a 1/2 BPS Wilson surface needs necessarily to wrap along three directions in the compact space S$^4$. However, since there are no non--contractible three--cycles in S$^4$, this configuration could not be stable or BPS, unless we consider some quotient of S$^4$ and/or turn on some flux in the M5--brane worldvolume. The reader can find more details in \cite{Lunin:2007ab,Chen:2007ir}. We do not further investigate this problem here.

\providecommand{\href}[2]{#2}\begingroup\raggedright\endgroup


\end{document}